\documentclass[aps,amsmath,amssymb,aps,showpacs,twocolumn]{revtex4-1}

\usepackage{graphicx}
\usepackage{dcolumn}
\usepackage{bm}
\usepackage{balance}
\usepackage{amsmath}
\usepackage{amsfonts}
\usepackage[applemac]{inputenc}
\usepackage{bbm}
\usepackage{braket}
\usepackage{float}
\usepackage{enumitem}  
\setlist[itemize]{leftmargin=*}
\setlist[enumerate]{leftmargin=*}
\usepackage{titlesec}
\usepackage{mathtools}

\begin{document}
\title{Classification of topological ladder models}

\author{Carlos G. Velasco}
\author{Bel\'en Paredes}

\affiliation{Arnold Sommerfeld Center for Theoretical Physics	\\		
Ludwig-Maximilians-Universit\"at M\"unchen, 80333 M\"unchen, Germany}

\begin{abstract} 
Ladder architectures are fruitful systems to realize topological phases of matter. Here we present a classification of ladder models giving rise to topological insulators. We identify six different types of topological ladder models, three in the BDI symmetry class, and three in the AIII symmetry class. They correspond to six distinct configurations of Wilson fermions. The six types are manifested in distinctive momentum distributions of the corresponding topological edge modes. The number of Wilson fermions, their chirality and mass, are directly manifested in the number, momentum and height of the peaks of the momentum distribution of the corresponding topological edge modes. We  identify a canonical ladder geometry, the {\em bowtie ladder}, from which any other topological ladder model can be obtained by a unitary transformation. We identify, classify and list all possible topological ladder geometries, determining the parameter regimes in which each of the six types of topological edge modes can be realized.  Our results open a route for the experimental realization and detection of topological insulators in novel symmetry classes with ladder architectures.

\end{abstract}


\maketitle

\section{Introduction}

The understanding of topological many-body phases and the synthesis of novel topological materials is one of 
most important and difficult challenges of modern theoretical and experimental quantum physics. With ultracold atoms in optical lattices, the recent realization of ladder architectures with artificial gauge fields has opened a fascinating route towards the realization and detection of artificially created one-dimensional topological phases.
Ladder systems are fruitful one-dimensional systems \cite{Huegel2014} where to explore the physics of two-dimensional topological phases. By combining a superlattice structure together with laser-assisted tunneling \cite{Aidelsburger2011, Aidelsburger2013}, ladders in the presence of effective magnetic fluxes habe been realized, and chiral currents were experimentally observed \cite{Atala2014}.
Additionally, ladder systems with effective magnetic fluxes have also been created by using a synthetic dimension \cite{Boada2012, Celi2014, Mancini2015, Stuhl2015, Livi2016, Kolkowitz2017, An2017, Kang2018, Han2019, Kang2019}, where each ladder leg corresponds to a different internal atomic degree of freedom. Furthermore, interesting instances of interacting topological phases have been recently theoretically investigated in ladder systems
\cite{T1.2, Juenemann2017, Meier2018, Leseleuc2018, Gholizadeh2018}.

Here, we aim at the investigation of ladder architectures for the realization of novel symmetry classes of topological insulators. Specially, we focus in the realization of the AIII symmetry class \cite{Ryu2010, Altland1997}, where the topological insulator breaks both time reversal and charge-conjugation symmetry, while preserving chiral symmetry.
In contrast to the BDI symmetry class, represented by the seminal model of Su, Schrieffer, and Heeger \cite{Su1979} (the SSH model), which has been realized and probed with ultracold atoms \cite{Atala2013, Meier2016, Leder2016}, the AIII symmetry class lacks to our knowledge experimental realization and has been only rarely theoretically explored. AIII topological insulators are believed to open a physical pathway to Riemann's conjecture, for one-dimensional models realizing the Riemann zeros seem to belong to the AIII symmetry class \cite{Sierra2014}. 

We present a classification of ladder models giving rise to topological insulators. We identify six different types of topological ladder models, three in the BDI symmetry class, and three in the AIII symmetry class. They correspond to six distinct configurations of Wilson fermions. The six types are manifested in distinctive momentum distributions of the corresponding topological edge modes. The number of Wilson fermions, their chirality and mass, are directly manifested in the number, momentum and height of the peaks of the momentum distribution of the corresponding topological edge modes. We  identify a canonical ladder geometry, the {\em bowtie ladder}, from which any other topological ladder model can be obtained by a unitary transformation. We identify, classify and list all possible topological ladder geometries, determining the parameter regimes in which each of the six types of topological edge modes can be realized. Our results open a route for the experimental realization and detection of the six types of topological insulators in a ladder architecture.

\section{Symmetry classes of 1D topological insulators}

\begin{figure}[t]
  \centering
    \includegraphics[width=0.485\textwidth]{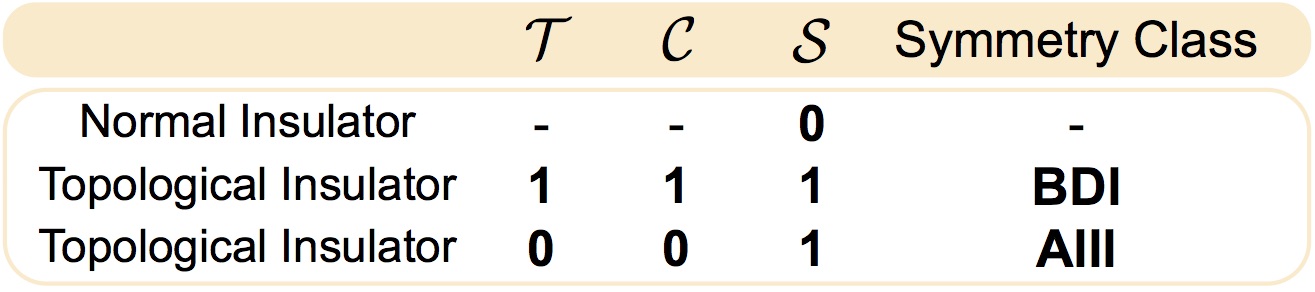}
    \caption[Topological classes of 1D insulators.]{\textbf{Topological classes of 1D insulators.} Symmetry classes for one-dimensional insulators according to the presence (1) or absence (0) of time reversal ($\mathcal{T}$), charge conjugation ($\mathcal{C}$) and chiral ($\mathcal{S}$) symmetries.}
    \label{tab:1DTopInsulatorClasses} 
\end{figure}

Topological insulators are classified according to their symmetry properties under time reversal, charge conjugation and chiral transformations \cite{Schnyder2008,Ryu2010,Altland1997}.
In one dimension, the presence or absence of chiral symmetry determines completely the topological or trivial character of an insulator. Since chiral symmetry is the composition of time reversal and charge conjugation symmetries, thus either both time reversal and charge conjugation symmetries are present or none of them is. These two cases correspond, respectively, to the two distinct symmetry classes of one-dimensional topological insulators: the BDI class and the AIII class (see Fig.~\ref{tab:1DTopInsulatorClasses}).

\begin{figure}[t]
  \centering
    \includegraphics[width=0.375\textwidth]{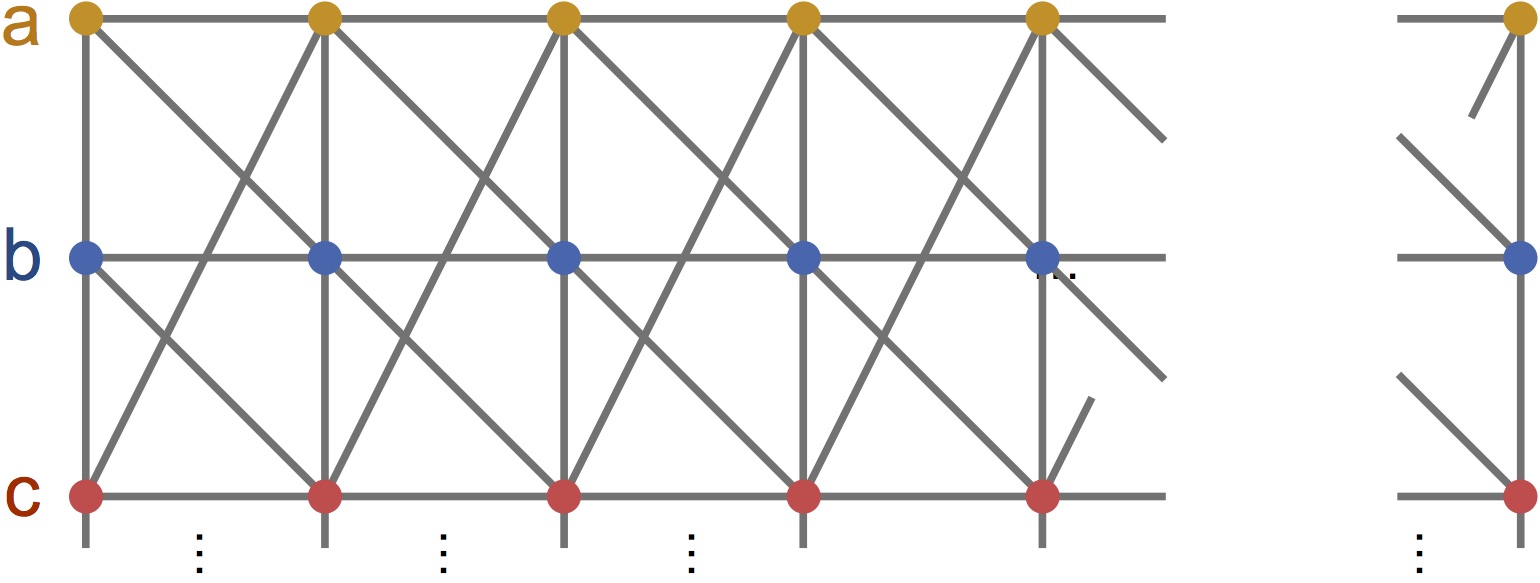}
    \caption[1D topological insulator.] {\textbf{1D topological insulator.} A model for a 1D topological insulator with translational invariance can be represented as a ladder in which each leg corresponds to a different type of site within every unit cell.}
    \label{fig:Esquema1DInsulator} 
\end{figure}

We consider a Hamiltonian $H$ for a one-dimensional non-interacting topological insulator with translational invariance. This system can be illustrated as a ladder with a leg for each different type of site per unit cell, where every hopping term is represented by a line (see.~Fig.~\ref{fig:Esquema1DInsulator}). The translational invariance allows us to write the Hamiltonian using the momentum representation as:
\begin{equation}\label{eq:GeneralHamiltonianMomentumBasis}
H=-\sum_{k}\psi^{\dagger}_{k}\,M(k)\,\psi^{}_{k},
\end{equation}
being $\psi^{\dagger}_{k}$ a vector that contains all creation modes with Bloch momentum $k$, $\psi^{\dagger}_{k}=( \hat{a}_{k}^{\dagger}\quad \hat{b}_{k}^{\dagger}\,\dots\,)$, and $M(k)$ a hermitian matrix called the \textit{Hamiltonian matrix}.\\
\indent In this context time reversal, charge conjugation and chiral symmetries are transformations acting on the spinor $\psi_{k}$ that produce a particular change in the Hamiltonian.
Time reversal symmetry is an antiunitary transformation $\mathcal{T}$ that leaves the Hamiltonian invariant, charge conjugation symmetry is an antinunitary transformation $\mathcal{C}$ under which the Hamiltonian changes its sign and, finally, chiral symmetry is a unitary transformation $\mathcal{S}$ that changes the sign of the Hamiltonian. That is:
\begin{align}
&\mathcal{T}\,\psi_{k}=U_{T}\,(\psi_{k})^{*},\qquad\,\mathcal{T}H\mathcal{T}^{-1}=H\\
&\mathcal{C}\,\psi_{k}=U_{C}\,(\psi_{k})^{*},\qquad\,\, \mathcal{C}H\mathcal{C}^{-1}=-H\\
&\mathcal{S}\,\psi_{k}=U_{S}\,\psi_{k},\qquad\quad\,\,\, \mathcal{S}H\mathcal{S}^{-1}=-H.
\end{align}
Where $U_{T}, U_{C}$ and $U_{S}$ are unitary constant matrices. Therefore, time reversal symmetry implies that the eigenstates of the Hamiltonian come in pairs with the same energy but opposite momenta, charge conjugation symmetry implies that they come in pairs with opposite energies and momenta, and finally chiral symmetry implies that they come in pairs with opposite energies and same momentum.\\
\indent From Eq.~(\ref{eq:GeneralHamiltonianMomentumBasis}) it follows that these symmetry operations acting on the Hamiltonian are equivalent to a transformation of the Hamiltonian matrix.
Therefore, the conditions for time reversal, charge conjugation and chiral symmetry can be expressed as:
\begin{align}
&\mathcal{T}:\,U_{T}\,|\quad U_{T}\,M^{*}(-k)\,U_{T}^{\dagger}=M(k) \label{eq:ConditionT}\\
&\mathcal{C}:\,U_{C}\,|\quad U_{C}\,M^{*}(-k)\,U_{C}^{\dagger}=-M(k) \label{eq:ConditionC}\\
&\mathcal{S}:\,U_{S}\,|\quad U_{S}\,M(k)\,U_{S}^{\dagger}=-M(k), \label{eq:ConditionS}
\end{align}
being $U_{T}, U_{C}$ and $U_{S}$ unitary matrices with no dependence on the momentum.\\

\section{Hamiltonian characterization of topological ladder models}

\begin{figure}[t]
  \centering
    \includegraphics[width=0.35\textwidth]{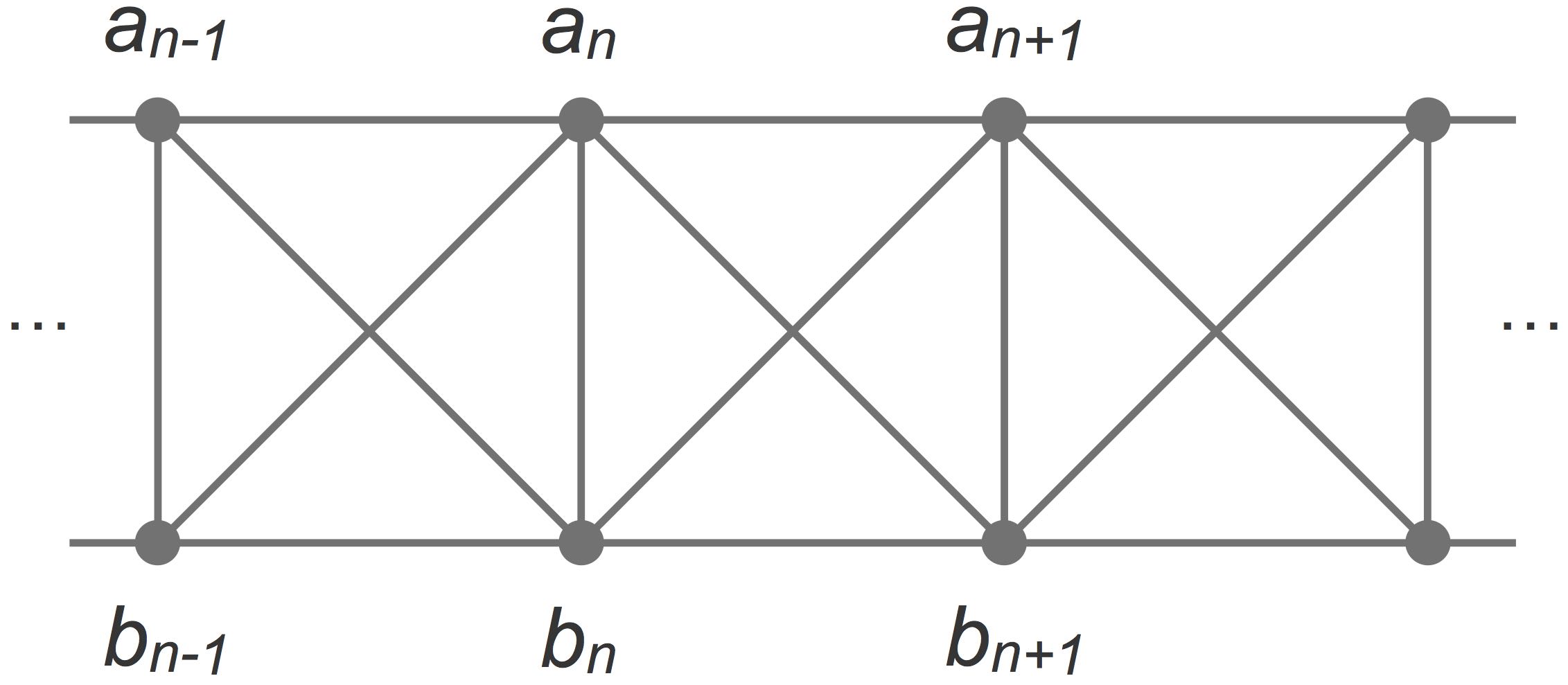}
    \caption[Ladder model.]{\textbf{Ladder model.} Schematic illustration of a ladder model. Each leg corresponds to one of the two different sites per unit cell and all possible couplings are represented by lines.}
    \label{fig:FiguraLadder01} 
\end{figure}

In this work we study models for non-interacting one-dimensional topological insulators with two different sites per unit cell, translational symmetry and in which only couplings that change the lattice index by one unit at most are allowed. That is what we call here a ladder model (see Fig.~\ref{fig:FiguraLadder01}).

We denote as $\hat{a}_{n}^{\dagger}$ and $\hat{b}_{n}^{\dagger}$ the operators that create a particle in the sublattice site $a_{n}$ and $b_{n}$ of the $n$-th lattice cell, respectively.  The most general ladder model corresponds to the following Hamiltonian:
\begin{equation}\label{eq:GeneralLadderHamiltonian}
H=-\sum_{n=1}^{N}\left(\psi^{\dagger}_{n}\,C\,\psi^{}_{n}+\psi^{\dagger}_{n+1}\,T\,\psi^{}_{n}+\psi^{\dagger}_{n}\,T^{\dagger}\,\psi^{}_{n+1}\right),
\end{equation}
where $\psi^{\dagger}_{n}=\begin{pmatrix}
\hat{a}^{\dagger}_{n} & \hat{b}^{\dagger}_{n}
\end{pmatrix}$, $N$ is the total number of unit cells in the lattice, $C$ is a $2\times2$ hermitian matrix and $T$ a general complex $2\times2$ matrix. Both matrices have no dependence on the lattice index $n$, as we are considering the model to be translational invariant. We change to the momentum representation and write the Hamiltonian $H$ as:
\begin{equation}
H=-\sum_{k}\psi^{\dagger}_{k}\,M(k)\,\psi_{k},
\end{equation}
where:
\begin{equation}\label{eq:MatrizM1}
M(k)=C+e^{-ik}\,T+e^{ik}\,T^{\dagger}
\end{equation}
is the Hamiltonian matrix. As any $2\times2$ hermitian matrix, it can be written as:
\begin{equation}\label{eq:GeneralMatrixComponentes} 
M(k)=\lambda(k)\mathbb{I}+\rho(k)\,\bm{n}(k)\cdot\bm{\sigma},
\end{equation}
being $\mathbb{I}$ the identity and $\bm{\sigma}$ a vector containing the three Pauli matrices. Here $\lambda(k)$ and $\rho(k)$ are real functions and $\bm{n}(k)$ is a real unit vector in the 3-dimensional space. The Hamiltonian matrix depends on the momentum only through the sine and cosine functions [see Eq.~(\ref{eq:MatrizM1})] and therefore the vector $\bm{n}(k)$ can be decomposed as:
\begin{equation}
\rho(k)\,\bm{n}(k)=\bm{n}_{0}+\bm{n}_{c}\cos k+\bm{n}_{s}\sin k,\label{eq:DescomposicionN}
\end{equation}
being $\bm{n}_{0}$, $\bm{n}_{c}$ and $\bm{n}_{s}$ three constant vectors.\\
\indent In the following we show how the symmetry class of a ladder Hamiltonian is determined by the properties of these three vectors. We also identify the general form of a ladder Hamiltonian in the BDI symmetry class and a ladder Hamiltonian in the AIII symmetry class.
For that, we first impose chiral symmetry and obtain the most general Hamiltonian of a topologically nontrivial ladder model. Afterwards, we impose time reversal symmetry and distinguish between ladder Hamiltonian in the BDI class and a ladder Hamiltonian in the AIII class.

\subsection{Ladder Hamiltonian with chiral symmetry}

We consider the Hamiltonian matrix of a general ladder model, Eq.~(\ref{eq:GeneralMatrixComponentes}). It can be proved that it has chiral symmetry if and only if $\lambda(k)=0$ and the vector $\bm{n}(k)$ lives in a plane that crosses the origin (see Appendix A for the details). That is, the conditions for chiral symmetry are:
\begin{align} 
&M(k)=\rho(k)\,\bm{n}(k)\cdot\bm{\sigma},\label{eq:ChiralCondition1}\\
&\left\{\bm{n}_{0},\,\bm{n}_{c},\,\bm{n}_{s}\right\}\,\text{are linearly dependent.}\label{eq:ChiralCondition2}
\end{align}
In other words, the Hamiltonian matrix has no component proportional to the identity and the three components of $\bm{n}(k)$ lie in a common plane. As a consequence, we can choose a particular orthonormal basis within that plane made by just two vectors $\bm{n}_1$ and $\bm{n}_2$ and write the Hamiltonian matrix as a combination of $\sigma_1=\bm{n}_1\cdot\bm{\sigma}$ and $\sigma_2=\bm{n}_2\cdot\bm{\sigma}$, being $\left\{\sigma_{1},\sigma_{2}\right\}=0$.\\
\indent In conclusion, the most general ladder model for a topological insulator corresponds to a Hamiltonian matrix of the form:
\begin{equation}\label{eq:GeneralChiralMatrix}
M(k)=f_1(k)\,\sigma_{1}+f_2(k)\,\sigma_{2},
\end{equation}
with $\left\{\sigma_{1},\sigma_{2}\right\}=0$ and being $f_{1}(k)$ and $f_{2}(k)$ two real functions that depend on the momentum through the sine and cosine functions. Defining:
\begin{equation}
\sigma_3=-(i/2)\left[ \sigma_1,\sigma_2 \right],
\end{equation}
it follows that $\sigma_3M(k)\sigma_3=-M(k)$. Therefore we can identify the orthogonal direction to the plane where $\bm{n}(k)$ lives with the unitary operator that fulfils the chiral condition (\ref{eq:ConditionS}), that is: $U_{S}=\sigma_{3}$.

\subsection{Ladder Hamiltonian in the BDI class}

We consider a general Hamiltonian matrix with chiral symmetry, written as in Eq.~(\ref{eq:ChiralCondition1}). By imposing the condition for time reversal symmetry in Eq.~(\ref{eq:ConditionT}), it can be proved (see Appendix A for the details) that the Hamiltonian has time reversal symmetry if and only if:
 \begin{align}
& \bm{n}_{s}\cdot\bm{n}_{0}=0\label{eq:GeneralTRCondition1}\\
& \bm{n}_{s}\cdot\bm{n}_{c}=0.\label{eq:GeneralTRCondition2}
 \end{align} 
In other words, the even and odd components of the Hamiltonian matrix must be perpendicular to each other. Combining this with the second condition for chiral symmetry, Eq.~(\ref{eq:ChiralCondition2}), it follows that the Hamiltonian matrix of a BDI model must be a combination of two Pauli matrices along perpendicular directions, corresponding each to an even and an odd function of the momentum, respectively. This means that there exist real parameters $\alpha$, $\beta$ and $\gamma$ and a particular choice of the orthonormal basis $\left\{\bm{n}_{1}, \bm{n}_{2}\right\}$ such that:
\begin{align}
&f_{1}(k)=\alpha+\beta\cos k\\
&f_{2}(k)=\gamma\sin k.
\end{align}
Therefore, the most general Hamiltonian for a ladder model in the BDI symmetry class corresponds to a matrix of the form:
\begin{equation}\label{eq:GeneralBDIMatrix}
M(k)=(\alpha+\beta\cos k)\,\sigma_1+\gamma\sin k\,\sigma_2,
\end{equation}
where $\alpha$, $\beta$ and $\gamma$ are real parameters and $\left\{\sigma_{1},\sigma_{2}\right\}=0$. A Hamiltonian of this form has chiral symmetry because it is a particular case of Eq.~(\ref{eq:GeneralChiralMatrix}), and it has time reversal symmetry since:
\begin{equation}
\sigma_2\sigma_yM^{*}(-k)\sigma_y\sigma_2=M(k).
\end{equation}
Chiral and time reversal symmetries imply charge conjugation symmetry, therefore the condition for charge conjugation symmetry is also fulfilled. We can identify the time reversal and charge conjugation transformations that satisfy the respective symmetry conditions, Eq.~(\ref{eq:ConditionT}) and Eq.~(\ref{eq:ConditionC}), as: $U_{T}=\sigma_{2}\sigma_{y}$ and $U_{C}=\sigma_{1}\sigma_{y}$.

\begin{figure}[t]
  \centering
    \includegraphics[width=0.485\textwidth]{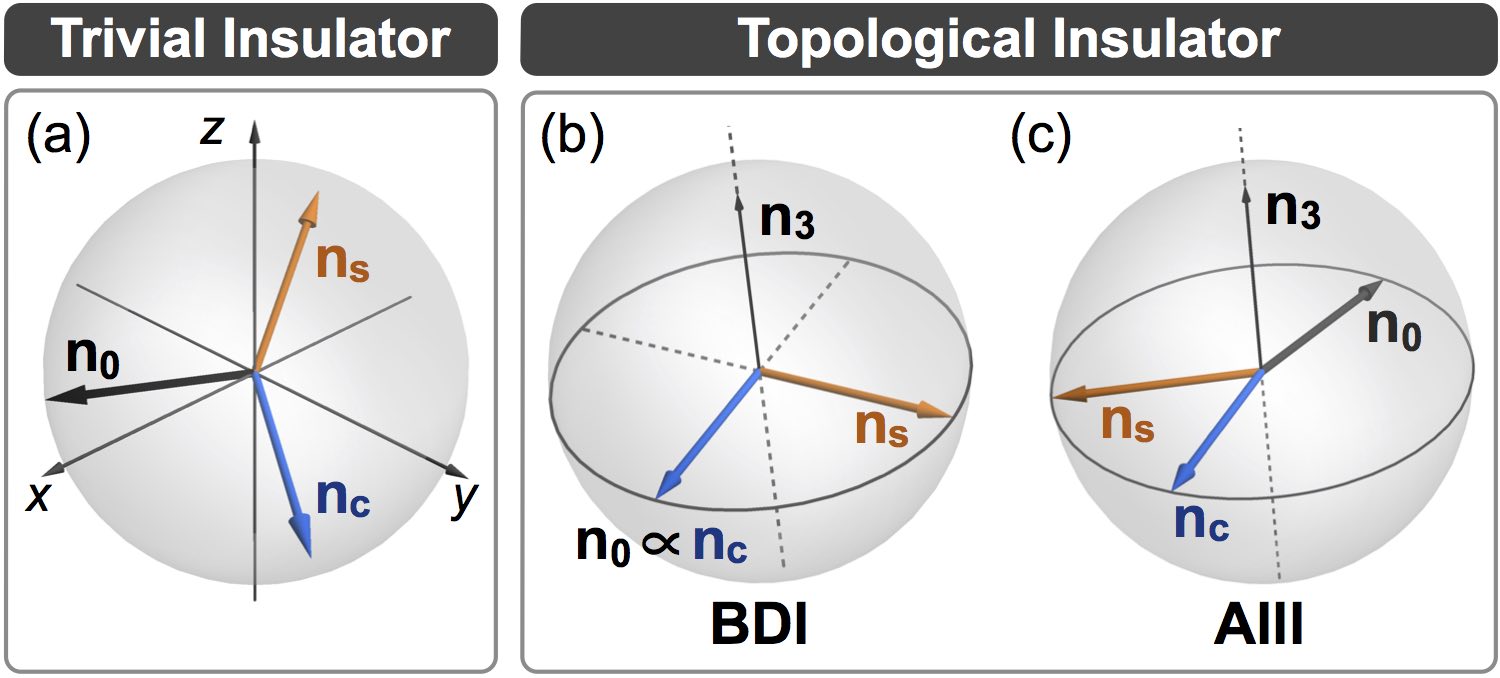}
    \caption[Geometrical classification of ladder Hamiltonians.]{\textbf{Geometrical classification of ladder Hamiltonians.} The symmetry class of a ladder model is determined by the geometry of its Hamiltonian matrix $M(k)=\left(\bm{n}_{0}+\bm{n}_{c}\cos k+\bm{n}_{s}\sin k \right)\cdot\bm{\sigma}$. In the most general case it corresponds to a normal insulator (a). When all the components lie in a common plane the system has chiral symmetry and, thus, exhibits topological properties. The orthogonal direction to that plane, denoted by the unit vector $\bm{n}_{3}$, corresponds to the chiral symmetry operator: $U_{S}=\sigma_{3}=\bm{n}_{3}\cdot\bm{\sigma}$. If the odd component is normal to the even ones, the system is in the BDI class (b), whereas it belongs to the AIII class otherwise (c). For the shake of clarity all vectors are represented with the same modulus, although they are in general different.}
    \label{fig:FiguraLadder02} 
\end{figure}

\subsection{Ladder Hamiltonian in the AIII class}

A model in the AIII symmetry class has chiral symmetry but is not time reversal symmetric. Therefore, it corresponds to a Hamiltonian matrix of the form in Eq.~(\ref{eq:GeneralChiralMatrix}) that cannot be written as in Eq.~(\ref{eq:GeneralBDIMatrix}).

In addition to the form in Eq.~(\ref{eq:GeneralChiralMatrix}), the Hamiltonian matrix of a chiral model can be written in an alternative way, using the basis formed by $\bm{n}_{c}$ and $\bm{n}_{s}$. Since the components of the Hamiltonian matrix of a chiral model lie in a common plane, there exist real parameters $\alpha$, $\beta$, $\eta$ and $\gamma$ such that:
\begin{equation}\label{eq:GeneralAIIIMatrix}
M(k)=(\alpha+\beta\cos k)\,\sigma_c+(\eta+\gamma\sin k)\,\sigma_s,
\end{equation}
with $\sigma_{c}=(\bm{n}_{c}\cdot\bm{\sigma})/|\bm{n}_{c}|$ and $\sigma_{s}=(\bm{n}_{s}\cdot\bm{\sigma})/|\bm{n}_{s}|$.

Comparing this expression to the general Hamiltonian matrix of a BDI model, Eq.~(\ref{eq:GeneralBDIMatrix}), we conclude that the most general Hamiltonian for a ladder model in the AIII class corresponds to a Hamiltonian matrix of the form in Eq.~(\ref{eq:GeneralAIIIMatrix}) with $\eta\neq0$ and/or $\left\{\sigma_{c},\sigma_{s}\right\}\neq0$. In other words, the model must break at least one of the two conditions $\eta=0$ and $\left\{\sigma_{c},\sigma_{s}\right\}=0$ in order to be in the AIII class, which are equivalent to the time reversal conditions in Eq.~(\ref{eq:GeneralTRCondition1}) and Eq.~(\ref{eq:GeneralTRCondition2}).\\

In this way, we use two different representations for the Hamiltonian matrix of a chiral ladder. The first one, Eq.~(\ref{eq:GeneralChiralMatrix}), uses an orthonormal basis $\left\{\bm{n}_{1},\bm{n}_{2}\right\}$ and, thus, two anticommuting Pauli matrices $\bm{\sigma}_{1}$ and $\bm{\sigma}_{2}$. The second one, Eq.~(\ref{eq:GeneralAIIIMatrix}), in contrast, uses two Pauli matrices $\bm{\sigma}_{c}$ and $\bm{\sigma}_{s}$ whose anticommutator is in general different from zero.  They are combinations of  $\bm{\sigma}_{1}$ and $\bm{\sigma}_{2}$ and the four of them lie in the same plane. The first representation has the advantage of using an orthonormal basis, which is easier to deal with. On the other hand, in the second representation the functions $\sin k$ and $\cos k$ are separated in different components, which is more convenient in some situations. Only for the BDI class these two representations coincide and the even and odd components of the Hamiltonian matrix can be separated in orthogonal directions.

In conclusion, the symmetry class of a ladder model for a topological insulator is determined by the geometry of the Hamiltonian matrix. Decomposing it as $M(k)=\left(\bm{n}_{0}+\bm{n}_{c}\cos k+\bm{n}_{s}\sin k \right)\cdot\bm{\sigma}$, the model is topologically non-trivial only when all components lie in a common plane, corresponding to a topologically trivial insulator otherwise [Fig.~\ref{fig:FiguraLadder02}(a)]. In the former case, the system belongs to the BDI class if the even and odd components are perpendicular to each other [Fig.~\ref{fig:FiguraLadder02}(b)], whereas it is in the AIII class in the most general situation [Fig.~\ref{fig:FiguraLadder02}(c)].


\section{6 types os topological ladder models and 6 types of Wilson fermion configurations}

\subsection{Topological characterization of a ladder model.}

As we explained in Chapter 6, a ladder model with chiral symmetry corresponds to a Hamiltonian matrix that can be written using two different representations. On one hand:
\begin{equation}
M(k)=\rho(k)\,\bm{n}(k)\cdot\bm{\sigma},
\end{equation}
being the vector $\bm{n}(k)$ confined in a particular plane; that is, there exists a unitary vector $\bm{n}_{3}$ such that $\bm{n}_{3}\cdot\bm{n}(k)=0$ $\forall k\in 1^{st}\mathcal{BZ}$. On the other hand, we can define another two unitary vectors $\bm{n}_{1}$ and $\bm{n}_{2}$ such that the three of them, $\{\bm{n}_{1}, \bm{n}_{2}, \bm{n}_{3}\}$, form an orthonormal basis. Therefore, the Hamiltonian matrix can also be written as:
\begin{equation}
M(k)=f_{1}(k)\,\sigma_{1}+f_{2}(k)\,\sigma_{2},
\end{equation}
with $\{\sigma_{1},\sigma_{2}\}=0$, being $\sigma_{j}=\bm{n}_{j}\cdot\bm{\sigma},\quad j=1,2$ and $f_{1}(k)$ and $f_{2}(k)$ two real functions of the momentum.

Both ways of writing the Hamiltonian matrix can be related by defining the complex function $z(k)$ and the real function $\varphi(k)$ as:
\begin{equation}\label{eq:DefinitionRho}
z(k)=\rho(k)\,e^{i\varphi(k)}=f_{1}(k)+if_{2}(k),
\end{equation}
so that $\bm{n}(k)=\cos\varphi(k)\,\bm{n}_{1}+\sin\varphi(k)\,\bm{n}_{2}$. In this way $\varphi(k)$ is the azimuth angle of the vector $\bm{n}(k)$ when using spherical coordinates if we choose the three cartesian coordinates $x$, $y$ and $z$ in the directions of $\bm{n}_{1}$, $\bm{n}_{2}$ and $\bm{n}_{3}$, respectively (see Fig.~\ref{fig:FiguraLadder03}).
Hence, the eigenvectors of the Hamiltonian matrix are easy to obtain, as they correspond to the vectors $\pm\bm{n}(k)$ in the Bloch sphere representation. That is:
\begin{figure}[t]
  \centering
    \includegraphics[width=0.45\textwidth]{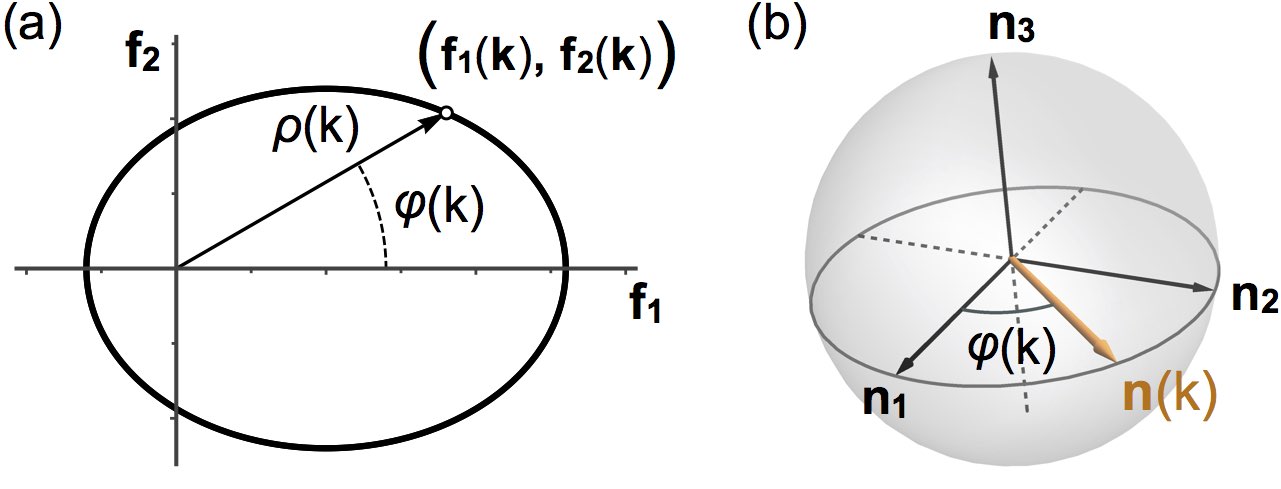}
    \caption[Hamiltonian matrix curve of a ladder model.]{\textbf{Hamiltonian matrix curve of a ladder model.} (a) A ladder model with chiral symmetry is characterized by the functions $f_{1}(k)$ and $f_{2}(k)$, which describe a closed curve in the plane. This curve determines the two energy bands of the system as $E_{\pm}(k)=\mp\rho(k)$, being $\rho(k)$ the distance from each point to the origin. (b) In addition, each point in the curve determines also an angle $\varphi(k)$, which defines an isospin vector $\bm{n}(k)$ that characterizes each eigenstate. The presence of chiral symmetry makes the vector $\bm{n}(k)$ be constrained to the plane generated by the two vectors $\bm{n}_{1}$ and $\bm{n}_{2}$.}
    \label{fig:FiguraLadder03} 
\end{figure}
\begin{figure}[t]
  \centering
    \includegraphics[width=0.485\textwidth]{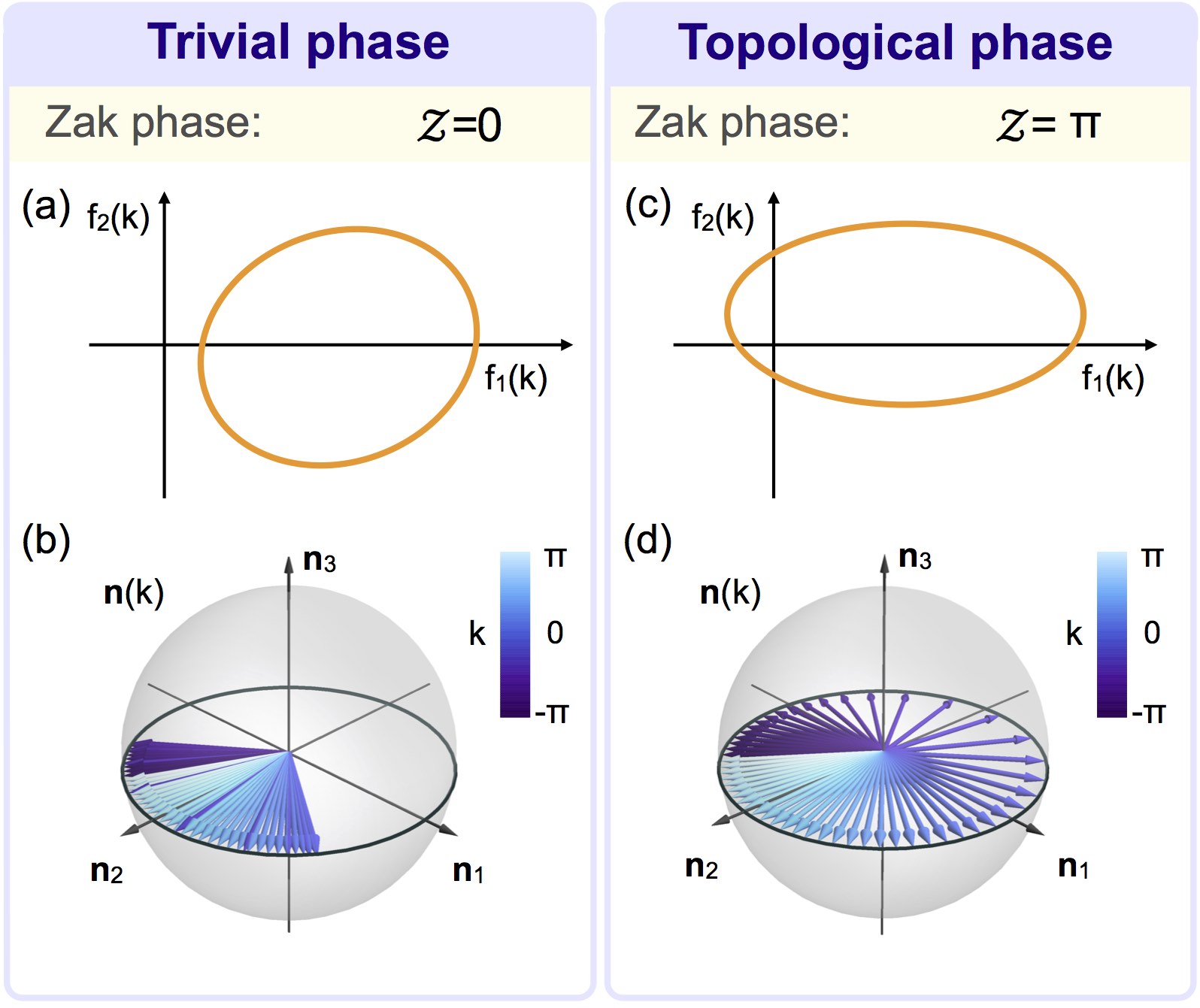}
    \caption[Topological characterization of a ladder model.]{\textbf{Topological characterization of a ladder model.} The trivial and topological phases of a ladder model are characterized and distinguished by the Zak phase. In the trivial phase the curve drawn by the Hamiltonian matrix in the complex plane, $z(k)=f_{1}(k)+if_{2}(k)$, does not enclose the origin (a), so that the vector $\bm{n}(k)$ on the Bloch sphere describes a circular arc as the momentum varies across the Brillouin zone (b). This situation corresponds to a Zak phase $\mathcal{Z}=0$. On the other hand, when the system is in the topologically non-trivial phase, $z(k)$ encloses the origin (c), and the vector $\bm{n}(k)$ describes a closed circumference (d). In this case the Zak phase takes the value $\mathcal{Z}=\pi$.}
    \label{fig:FiguraLadder04} 
\end{figure}
\begin{equation}\label{eq:Isospin}
\pm\bm{n}(k)\longrightarrow
U\begin{pmatrix}
1\\
\pm e^{i\varphi(k)}
\end{pmatrix},
\end{equation}
being $U$ the rotation that transforms $\sigma_{x}(\sigma_{y})$ into $\sigma_{1}(\sigma_{2})$ and corresponding each of the two possibilities $\pm$ to each of the two different energy bands. The eigenmodes of the Hamiltonian under periodic boundary conditions are therefore:
\begin{equation}\label{eq:BlochEigenstates}
\ket{k}_{\pm}=\begin{pmatrix}
\hat{a}_{k}^{\dagger} & \hat{b}_{k}^{\dagger}
\end{pmatrix}
\frac{1}{\sqrt{2}}\,U\begin{pmatrix}
1\\
\pm e^{i\varphi(k)}
\end{pmatrix}\ket{0},
\end{equation}
whose corresponding energies are:
\begin{equation}\label{eq:BlochEigenenergy}
E_{\pm}(k)=\mp\rho(k)=\mp\sqrt{f_{1}^{2}(k)+f_{2}^{2}(k)}.
\end{equation}
Once we have obtained the Bloch modes, i.e. the eigenstates of the Hamiltonian for periodic boundary conditions, we can compute the Zak phase, which characterizes the topological nature of the system \cite{Zak1989}. It consists of the geometrical phase acquired by a particle when completing an adiabatic path across the whole first Brillouin zone, that is:
\begin{equation}
\mathcal{Z}=-i\oint_{1^{st}BZ}dk\,\langle k|_{\pm}\,\partial_{k}\,\ket{k}_{\pm}=\frac{\Delta\varphi}{2}.
\end{equation}
As we see, the Zak phase is quantized according to the winding number around the origin of $z(k)$ in the complex plane and, equivalently, to the winding number of the vector $\bm{n}(k)$ around the origin in the plane generated by the two vectors $\bm{n}_{1}$ and $\bm{n}_{2}$ (see Fig.~\ref{fig:FiguraLadder04}).
When the curve encloses the origin (winding number $1$) the system is in a topological phase and correspodns to a Zak phase $\mathcal{Z}=\pi$. On the contrary, if the curve does not enclose the origin (winding number $0$) the system is in a trivial phase and $\mathcal{Z}=0$.\\

In this way, the presence of chiral symmetry makes the Hamiltonian matrix live on the plane generated by $\bm{n}_{1}$ and $\bm{n}_{2}$, where the functions $f_{1}(k)$ and $f_{2}(k)$ define a closed curve parametrized by the momentum.
These two functions completely characterize the ladder model, being the particular directions of the vectors $\bm{n}_{1}$ and $\bm{n}_{2}$ irrelevant as they can be changed by performing a global unitary transformation.
On one hand, the winding number of the curve determines weather or not the system is found to be in its topological phase.
On the other hand, each point in the curve, given by a particular value of the momentum $k$, determines an angle $\varphi(k)$ and a distance to the origin $\rho(k)$. The first function defines an isospin vector $\bm{n}(k)$, which corresponds through the Bloch sphere representation to the particular superposition between the modes $\hat{a}^{\dagger}_{k}$ and $\hat{b}^{\dagger}_{k}$ that constitutes an eigenstate of the Hamiltonian, Eq.~(\ref{eq:BlochEigenstates}), while the second function gives us the energy of such modes, Eq.~(\ref{eq:BlochEigenenergy}).

%
%

\subsection{6 types of topological ladder models}

In this section we analyse the different types of energy bands that a topological ladder model can exhibit. We pay special attention to the number of energy gaps between the two bands, as well as their width and location in the first Brillouin zone. These aspects are quite relevant since, as we show in Sec.~$6$, they are directly related to the properties of the symmetry protected edge modes that a topological ladder model give rise to. In this way, we have classified all possible topological ladder models into six different types, three of them belonging to the BDI symmetry class, and another three in the AIII symmetry class.

\subsubsection{3 types of topological ladder models in the BDI class}

As we concluded in Sec.~$3$, the Hamiltonian matrix of a topological ladder model in the BDI symmetry class has the following form:
\begin{equation}
M(k)=f_{1}(k)\,\sigma_{1}+f_{2}(k)\,\sigma_{2},
\end{equation}
with $\{\sigma_{1}, \sigma_{2}\}=0$ and:
\begin{equation}
\begin{cases}
f_{1}(k)=\alpha+\beta\,\cos k\\
f_{2}(k)=\gamma\,\sin k
\end{cases},
\end{equation}
being $\alpha$, $\beta$ and $\gamma$ three real parameters. Therefore the curve drawn by the Hamiltonian matrix in the complex plane, $z(k)=f_{1}(k)+if_{2}(k)$, is an ellipse with its center located at the point $(\alpha, 0)$ and whose horizontal and vertical axes are given by $|\beta|$ and $|\gamma|$, respectively. That is:
\begin{equation}
\left[ \frac{f_{1}(k)-\alpha}{\beta} \right]^{2}+\left[ \frac{f_{2}(k)}{\gamma} \right]^{2}=1.
\end{equation}
The two energy bands of such topological ladder model are:
\begin{equation}\label{eq:BDIEnergyBands}
E_{\pm}(k)=\mp\sqrt{\alpha^{2}+\gamma^{2}+2\alpha\beta\cos k+(\beta^{2}-\gamma^{2})\cos^{2}k}.
\end{equation}
The energy bands are distinguished by a global sign, $E_{\pm}(k)=\mp\rho(k)$, so that the energy gaps will be located at the minima of the function $\rho(k)$. Being $q$ a certain minimum of such function, the width of the corresponding energy gap will be $E_{gap}=2\rho(q)$.\\

Taking into account the location and width of the energy gaps we can distinguish three types of topological ladder models in the BDI symmetry class. On one hand, the number of energy gaps cannot be larger than two, as the curve described by the Hamiltonian matrix in the complex plane is an ellipse. On the other hand, the presence of time reversal symmetry makes the energy bands be symmetric with respect to the momentum, that is: $E_{\pm}(-k)=E_{\pm}(k)$. Therefore there are three distinct possibilities, namely: \textit{i)} a single energy gap located at momentum $q=0$ or $q=\pi$ (as the momenta $k=\pi$ and $k=-\pi$ are identified), \textit{ii)} two energy gaps of the same width located at opposite momenta $q_{1}=q$ and $q_{2}=-q$, with $q\neq0,\pi$ , and \textit{iii)} two energy gaps with, in general, different widths and located at the momentum values $q_{1}=0$ and $q_{2}=\pi$. We call these three cases the \textit{SSH-like} model, the \textit{balanced} BDI model and the \textit{imbalanced} BDI model, respectively.\\

\textbf{\textit{i)} SSH-like model}\\

\begin{figure*}
  \centering
    \includegraphics[width=0.85\textwidth]{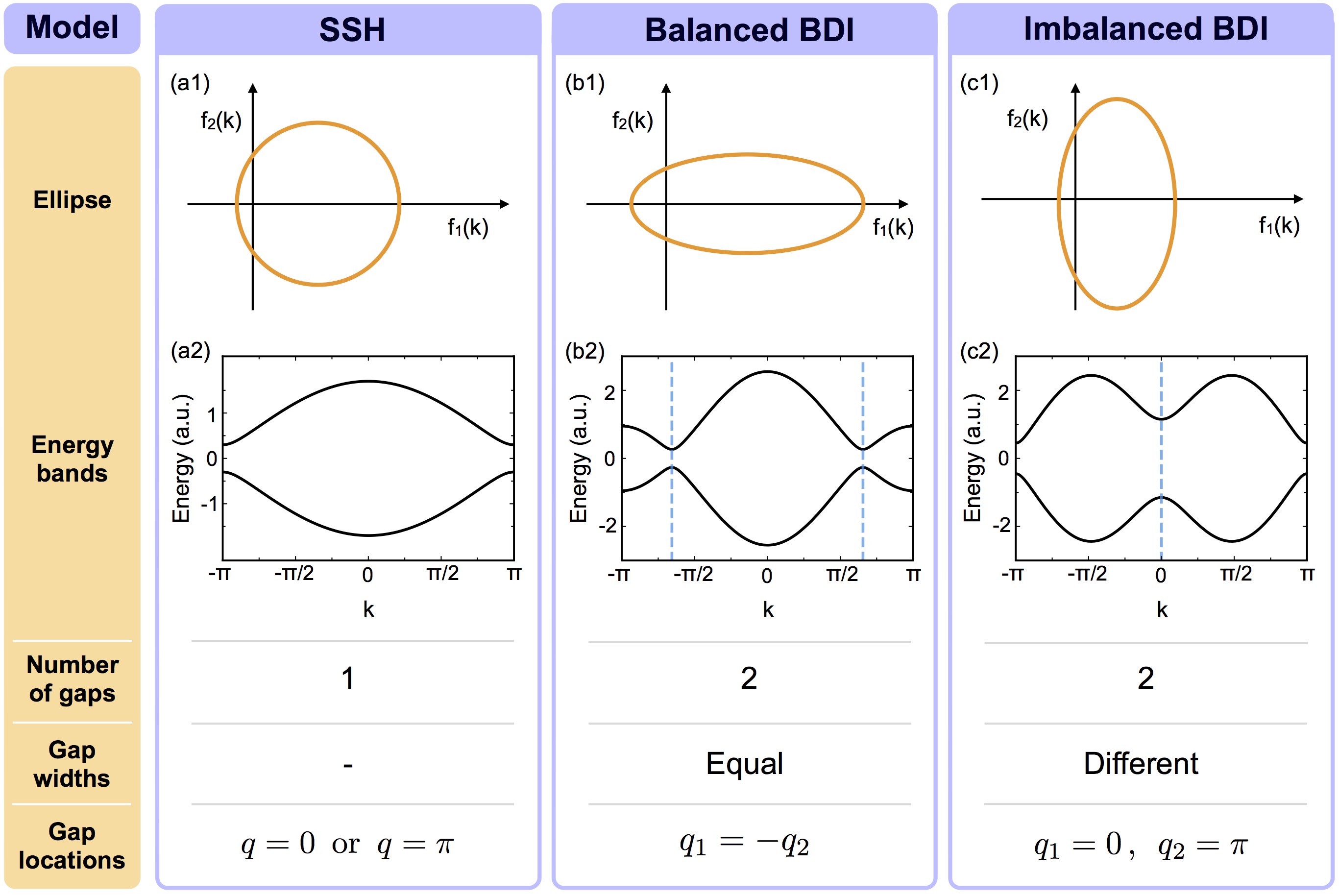}
    \caption[3 types of topological ladder models in the BDI class.]{\textbf{3 types of topological ladder models in the BDI class.} There are three types of topological ladder models in the BDI symmetry class, which are distinguished and characterized by their energy gap configurations. (a) The SSH model (as the representative of a slightly more general model, the SSH-like model) corresponds to a circumference in the $(f_{1}, f_{2})$-plane and a single energy gap between the two energy bands of the system. This gap is located at $q=0$ or $q=\pi$, depending on the relative sign between the Hamiltonian parameters $\alpha$ and $\beta$ (see text). (b) The balanced BDI model corresponds to an ellipse in the $(f_{1}, f_{2})$-plane whose horizontal axis is bigger than the vertical one, so that there are two energy gaps of the same width and located at opposite momenta $q_{1}=q$ and $q_{2}=-q$, with $\cos q=\alpha\beta/(\gamma^{2}-\beta^{2})$. (c) The imbalanced BDI model is characterized by a Hamiltonian matrix that draws an ellipse in the $(f_{1}, f_{2})$-plane with a vertical axis bigger than the horizontal one, and thus there are two gaps between the energy bands, located at momentum $q_{1}=0$ and $q_{2}=\pi$. In general, each gap has a different width.}
    \label{fig:FiguraLadder05} 
\end{figure*}

The SSH model corresponds to the particular case in which $|\beta|=|\gamma|$, so that the ellipse that characterizes the Hamiltonian matrix in the complex plane collapses into a circle [see Fig.~\ref{fig:FiguraLadder05}(a1)]. As a consequence the function $\rho(k)$ has just one minimum and, therefore, the energy bands show a single energy gap, which can be located at momentum $q=0$ or $q=\pi$ [see Fig.~\ref{fig:FiguraLadder05}(a2)]. For $\alpha\beta>0$, the energy gap is found to be at $q=0$, whereas it is located at $q=\pi$ for $\alpha\beta<0$.

In the case in which $|\beta|\neq|\gamma|$, so that the two axes of the ellipse in the complex plane are different from each other, there is a region in the parameter space in which the corresponding ellipse has an eccentricity low enough for the bands to have a single energy gap, located at the same position as in the SSH model. This more general model is what we call the SSH-like model and is defined by the condition $|\alpha\beta|\geq|\beta^{2}-\gamma^{2}|$.
To see this, we consider the first derivative with respect to the momentum of the lower energy band of a general BDI ladder model [see Eq.~(\ref{eq:BDIEnergyBands})]:
\begin{equation}
\rho^{\prime}(k)=-\frac{1}{\rho(k)}\sin k\left[ \alpha\beta+(\beta^{2}-\gamma^{2})\cos k \right].
\end{equation}
It is straightforward to conclude that a momentum $q$ needs to fulfil one of the following conditions in order to be a minimum (or a maximum) of the band:
\textit{i)} $\sin q=0$,
or
\textit{ii)} $\cos q=-\alpha\beta/(\beta^{2}-\gamma^{2})$.
In the case in which $|\alpha\beta|>|\beta^{2}-\gamma^{2}|$ condition \textit{ii)} does not define a real value for $q$ and, therefore, there are only two solutions: $q=0$ and $q=\pi$, one will correspond to a maximum and the other to a minimum. This is precisely the case for the SSH-like model. In the particular case in which $|\alpha\beta|=|\beta^{2}-\gamma^{2}|$ condition \textit{ii)} can be indeed satisfied by a real value of the momentum, however the solution coincides with one of the solutions of condition \textit{i)} and thus this situation represents another case of SSH-like model. Finally, when $|\alpha\beta|<|\beta^{2}-\gamma^{2}|$, there are four different solutions. Among them, two will be minima and the other two maxima. This last situation includes the two other types of BDI ladder models different from the SSH-like model.

Nevertheless, the edge modes of the SSH-like model are not significantly different from the ones that appear in the SSH model (see Chapter $9$), which is a simpler and widely known model. Therefore, we will often use the SSH model as a representative of the first type of topological ladder model in the BDI class, instead of using the slightly more general SSH-like model.\\

\textbf{\textit{ii)} Balanced BDI model}\\

This model corresponds to the situation in which $|\alpha\beta|<|\beta^{2}-\gamma^{2}|$, what implies
that the upper band has two maxima and two minima, and $|\beta|>|\gamma|$, so that the horizontal axis of the ellipse in the $(f_{1}, f_{2})$-plane is bigger than the vertical one [see Fig.~\ref{fig:FiguraLadder05}(b1)]. In this situation, the minima are located at $q_{1}=q$ and $q_{2}=-q$ with $\cos q=-\alpha\beta/(\beta^{2}-\gamma^{2})$ and the maxima are found at $k=0$ and $k=\pi$. In this way, the energy gaps between the two bands are located at opposite momenta $\pm q$ and, as a consequence of time reversal symmetry, their widths are the same [see Fig.~\ref{fig:FiguraLadder05}(b2)].\\

\textbf{\textit{iii)} Imbalanced BDI model}\\

The third type of BDI ladder model corresponds to the last possibility: $|\alpha\beta|<|\beta^{2}-\gamma^{2}|$ and $|\beta|<|\gamma|$, so that the ellipse in the $(f_{1}, f_{2})$-plane has a vertical axis bigger
\begin{figure}[H]
  \centering
    \includegraphics[width=0.485\textwidth]{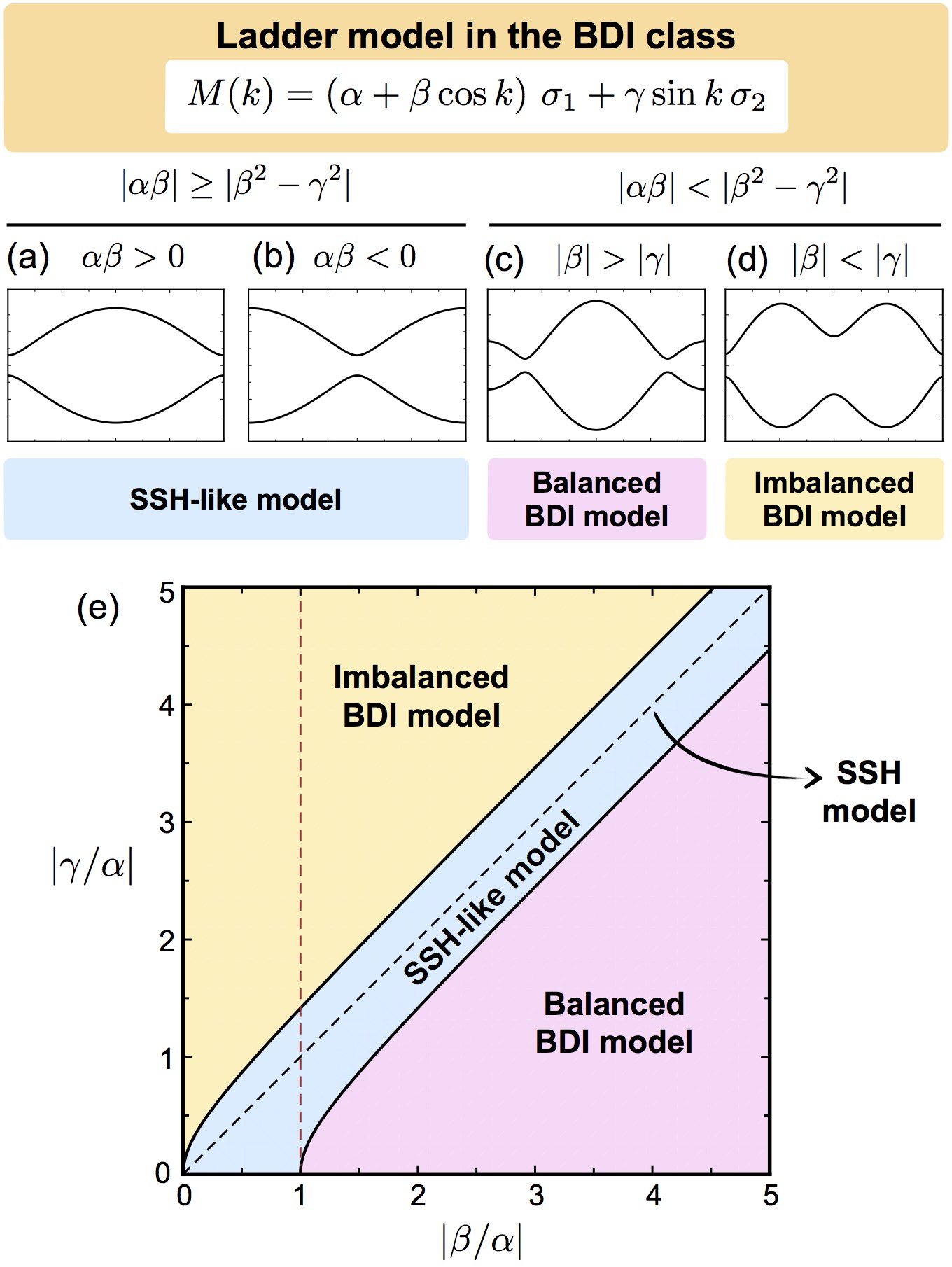}
    \caption[Classification of BDI ladder models.]{\textbf{Classification of BDI ladder models.} A general ladder model in the BDI symmetry class can be classified into one of three different types according to the parameters of its Hamiltonian matrix $M(k)=(\alpha+\beta\cos k)\,\sigma_{1}+\gamma\sin k\,\sigma_{2}$. For $|\alpha\beta|\geq|\beta^{2}-\gamma^{2}|$, it corresponds to the SSH-like model, which is characterized by (a) a single energy gap located at $q=0$ if $\alpha\beta>0$ or (b) a single energy gap located at $q=\pi$ for $\alpha\beta<0$. The SSH model is a representative of this type and corresponds to the particular case in which $|\beta|=|\gamma|$. On the other hand, when $|\alpha\beta|<|\beta^{2}-\gamma^{2}|$, the system corresponds to the balanced BDI model for $|\beta|>|\gamma|$ and to the imbalanced BDI model for $|\beta|<|\gamma|$. The former one is characterized through its energy bands by (c) the presence of two energy gaps of the same width located at opposite momenta, whereas the last one gives rise to (d) a pair of energy gaps of the different widths at momenta $q=0$ and $q=\pi$. (e) Parameter space of a general BDI ladder model, where the two axes correspond to the quantities $|\beta/\alpha|$ and $|\gamma/\alpha|$. The blue region corresponds to the SSH-like model, the pink one to the balance BDI model and the yellow one to the imbalance BDI model. The dashed black line corresponds to the SSH model and the dashed red line indicates the critical point that separates the trivial phase, $|\alpha|>|\beta|$, from the topological one, $|\alpha|<|\beta|$.}
    \label{fig:FiguraLadder06} 
\end{figure}
\noindent than its horizontal axis [see Fig.~\ref{fig:FiguraLadder05}(c1)]. As in the balanced BDI model, the imbalanced BDI model is characterized by two minima and two maxima in its energy bands. However, their locations are interchanged, being the minima located at $q_{1}=0$ and $q_{2}=\pi$, whereas the maxima take place at $\pm q$ with $\cos q=-\alpha\beta/(\beta^{2}-\gamma^{2})$. Therefore, while the locations of the energy gaps are fixed at $q_{1}=0$ and $q_{2}=\pi$, their widths can be in general different [see Fig.~\ref{fig:FiguraLadder05}(c2)].\\

It is illuminating to draw the parameter space of a general topological ladder model in the BDI symmetry class and identify the region corresponding to each one of the three types of BDI ladder models. The condition $|\alpha\beta|\geq|\beta^{2}-\gamma^{2}|$, that defines the SSH-like model, can be written in terms of the ratios $|\beta/\alpha|$ and $|\gamma/\alpha|$. If we define $x=|\beta/\alpha|$ and $y=|\gamma/\alpha|$, the SSH-like model corresponds to the region whose boundaries are given by the following functions:
\begin{align}
&y=\sqrt{x^{2}-x},\\
&y=\sqrt{x^{2}+x},
\end{align}
which are asymptotically close to the lines $y=x-1/2$ and $y=x+1/2$, respectively. Therefore, the SSH-like model region in the parameter space corresponds approximately to a stripe of width $1/\sqrt{2}$ whose center is given by the line $|\beta/\alpha|=|\gamma/\alpha|$ (SSH model), corresponding the two remaining regions to the balanced and imbalanced BDI models [see Fig.~\ref{fig:FiguraLadder06}(e)].

\begin{figure*}
  \centering
    \includegraphics[width=0.85\textwidth]{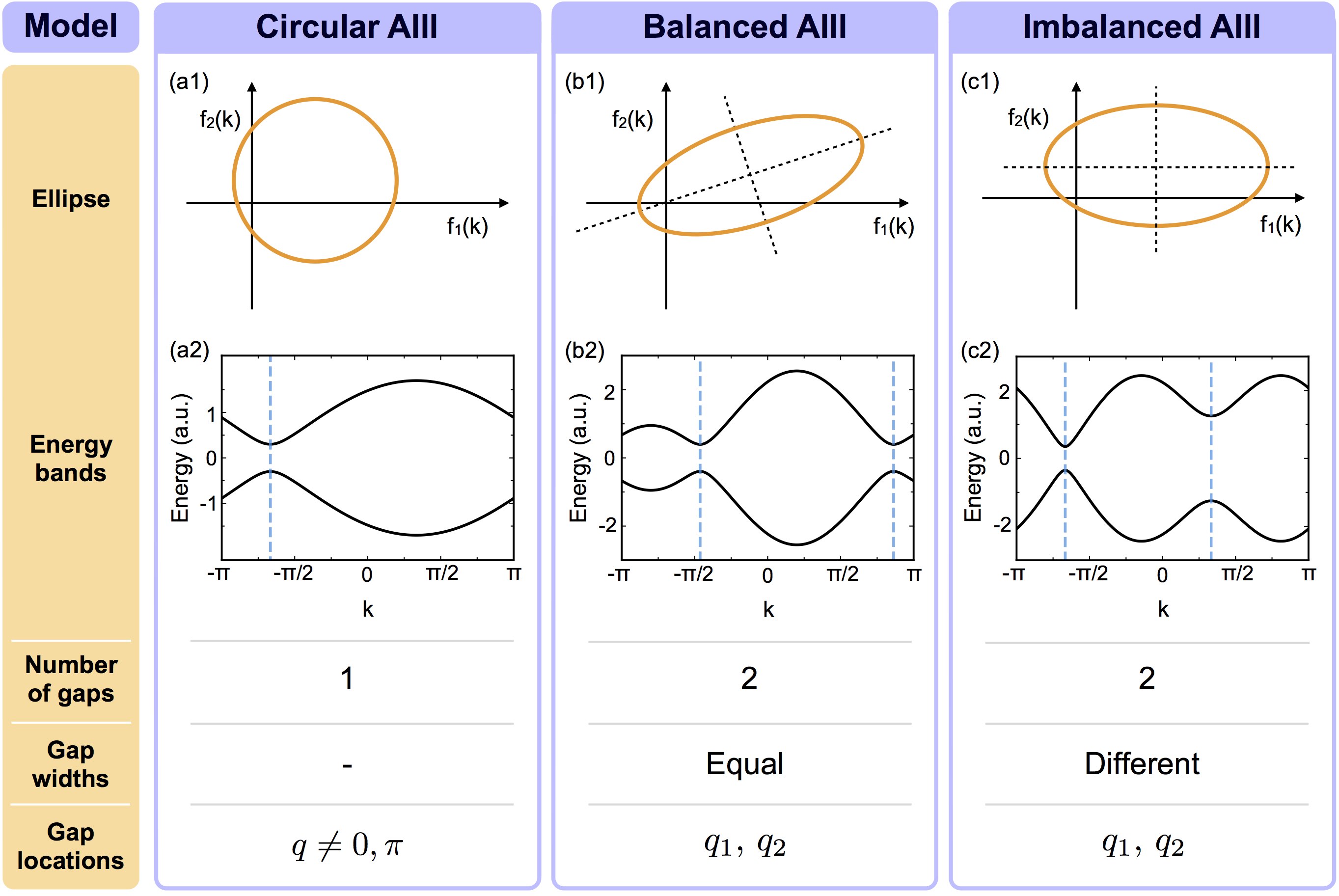}
    \caption[3 types of topological ladder models in the AIII class.]{\textbf{3 types of topological ladder models in the AIII class.} There are three types of topological ladder models in the AIII symmetry class, characterized by distinc energy gap configurations. (a) The circular AIII model corresponds to an ellipse in the $(f_{1}, f_{2})$-plane whose eccentricity is low enough for the energy bands to have a single gap. Due to the lack of time reversal symmetry, this energy gap is located at $q\neq0,\pi$. In this figure we show the particular case in which the Hamiltonian matrix ellipse is actually a circumference, which represents the simplest circular AIII model. (b) The balanced AIII model corresponds to a Hamiltonian matrix ellipse such that one of its axis crosses the origin, so that there are two energy gaps of the same width. The lack of time reversal symmetry implies that they are located at not opposite momenta: $q_{1}$ and $q_{2}$ with $q_{1}\neq-q_{2}$. (c) The imbalanced AIII model corresponds to the most general situation. Therefore, it shows two energy gaps of different widths and located at any positions in momentum space; excluding the case in which $q_{1}=0$ and $q_{2}=\pi$, as it would correspond to the imbalanced BDI model.}
    \label{fig:FiguraLadder07} 
\end{figure*}

\subsubsection{3 types of topological ladder models in the AIII class}

From Chapter $6$, we know that a topological ladder model in the AIII symmetry class is characterized by a Hamiltonian matrix of the form:
\begin{equation}
M(k)=(\alpha+\beta\cos k)\,\sigma_c+(\eta+\gamma\sin k)\,\sigma_s,
\end{equation}
being $\eta\neq0$ and/or $\{ \sigma_{c}, \sigma_{s}\}\neq0$. This implies that, if we write the Hamiltonian matrix using two anti-commuting Pauli matrices $\sigma_{1}$ and $\sigma_{2}$ as:
\begin{equation}
M(k)=f_{1}(k)\,\sigma_{1}+f_{2}(k)\,\sigma_{2},
\end{equation}
the two functions $f_{1}(k)$ and $f_{2}(k)$ describe an ellipse such that, in contrast to the BDI case, its center can be located anywhere in the plane and its axes can be rotated with respect to the abscises and ordinates axes. As a consequence, the energy gap configurations that a topological ladder model in the AIII class can exhibit are richer than those in the BDI class, as they are not constrained by the presence time reversal symmetry. Analogously to the classification of BDI ladder models, we can distinguish three distinct kinds of topological ladder models in the AIII symmetry class, characterized by a different energy gap configuration, namely: \textit{i)} a single energy gap located at momentum $q\neq0, \pi$, \textit{ii)} two energy gaps of the same width located at \textit{not opposite momenta} $q_{1}$ and $q_{2}$, and \textit{iii)} two energy gaps with different widths and located at any momentum values $q_{1}$ and $q_{2}$, excluding the case in which $q_{1}=0$ and $q_{2}=\pi$. We call these three cases the \textit{circular} AIII model, the \textit{balanced} AIII model and the \textit{imbalanced} AIII model, respectively.\\

\textbf{\textit{i)} Circular AIII model}\\

In the case in which $|\beta|=|\gamma|$ the curve described by the Hamiltonian matrix in the complex plane is a circle [see Fig.~\ref{fig:FiguraLadder07}(a1)]. Therefore the energy bands give rise to a single energy gap, which is located at any momentum $q$ different from $0$ and $\pi$ [see Fig.~\ref{fig:FiguraLadder07}(a2)], as these two particular cases take place only in the BDI class. Nevertheless, not only the specific case in which $|\beta|=|\gamma|$ corresponds to this gap configuration. Analogously to the SSH model and the SSH-like model, the case $|\beta|=|\gamma|$ is just a particular model within a larger region in parameter space in which the Hamiltonian matrix ellipse has an eccentricity low enough for the energy bands to have a single energy gap. For the sake of simplicity, we refer to this type of AIII ladder models as the circular AIII model, since the case $|\beta|=|\gamma|$ (the only situation in which the Hamiltonian matrix curve is truly a circle) is the simplest representative of this family of models and there is no significant distinction between any of them.\\

\textbf{\textit{ii)} Balanced AIII model}\\

This model corresponds to the case in which the Hamiltonian matrix ellipse is such that one of its axes crosses the origin [see Fig.~\ref{fig:FiguraLadder07}(b1)], so that the points associated to the minima of the upper band are at the same distance from the origin and, thus, there are two energy gaps of the same width between the two energy bands [see Fig.~\ref{fig:FiguraLadder07}(b2)]. These energy gaps cannot be located at opposite momenta, as this would correspond to the BDI class. That is, the gaps are located at $q_{1}$ and $q_{2}$ with $q_{1}\neq-q_{2}$.\\
\begin{figure*}
  \centering
    \includegraphics[width=1\textwidth]{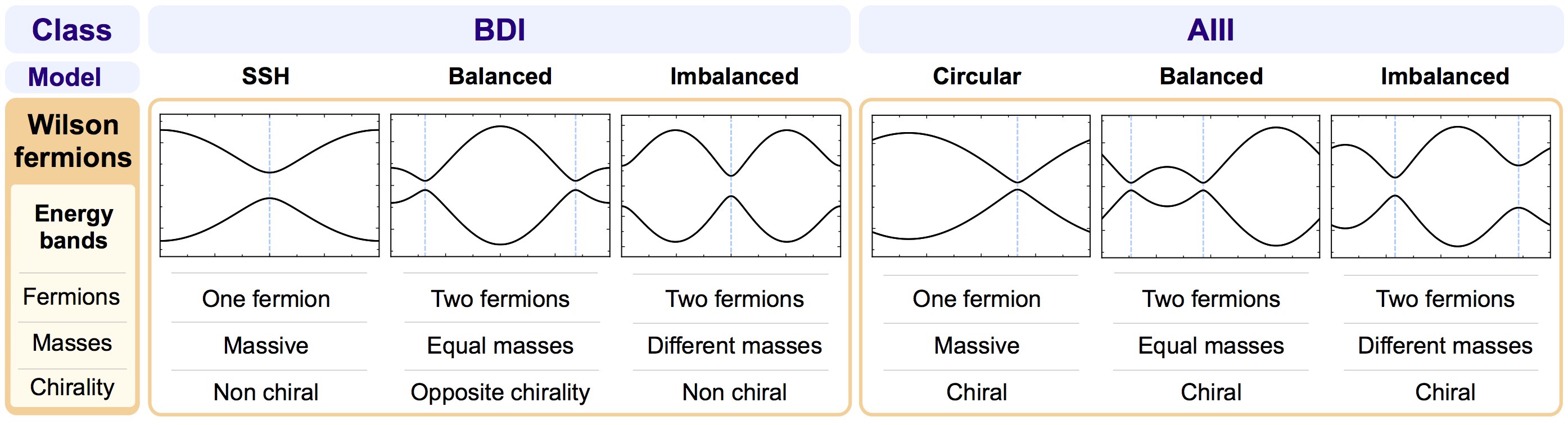}
    \caption[Wilson fermion configurations of a ladder model.]{\textbf{Wilson fermion configurations of a ladder model.} There are three different configurations of Wilson fermions that a ladder model in the BDI class can describe, and another three in the AIII class. Each of this six possibilities is characterized by three aspects. First, the number of fermions, which can be one or two. Second, in case there are two fermions, weather they have the same mass or not. Third, the chirality of the fermions, which is related to the symmetry class of the ladder model (see text).}
    \label{fig:FiguraLadder08} 
\end{figure*}

\textbf{\textit{iii)} Imbalanced AIII model}\\

Finally, the imbalanced AIII model consists of the more general case, in which the Hamiltonian matrix ellipse does not correspond to any of the previous situations [see Fig.~\ref{fig:FiguraLadder07}(c1)]. In this case the energy bands show a pair of energy gaps with different widths and located at any momenta $q_{1}$ and $q_{2}$ [see Fig.~\ref{fig:FiguraLadder07}(c2)], excluding the case $q_{1}=0$ and $q_{2}=\pi$ as it corresponds to the imbalanced BDI model.

\subsection{6 types of Wilson fermion configurations}

A ladder model with chiral symmetry and a Fermi energy between the two bands can be interpreted as a discretized description of a fermion, where a whole in the filled sea of negative energy states represents an antiparticle.

When there is a gap between the two bands the fermion is massive, whereas it is massless at the critical point, where the two bands cross each other.
The simplest ladder model describing a fermion with a kinetic and a mass term is called the naive lattice transcription of the Dirac equation \cite{Creutz1994}.
Each of the two momentum values at which the upper band reaches a local minimum corresponds to a pole of the propagator and thus to a different specie of fermion \cite{Creutz2001}.
In the context of particle physics this was called the doubling problem and was solved by Wilson \cite{Wilson1977}, who added another kinetic term so that one of the fermions gets heavier, leaving only one light fermion.

In the context of condensed matter, the naive Dirac fermion corresponds to a trivial insulator, whereas the Wilson fermion corresponds to a topological ladder in the BDI symmetry class, more precisely to the Creutz ladder \cite{Creutz1999}. It describes two Wilson fermions of different masses localized at momenta $k=0$ and $k=\pi$. After considering the most general ladder model Hamiltonian we have found more possible Wilson fermion configurations, as well as a relation between the symmetry class of a topological ladder model and the set of Wilson fermions that it describes.\\

There are six distinct Wilson fermion configurations that a ladder model can exhibit, and they correspond to the six different types of ladder models we showed in the previous section. The SSH model corresponds to just one fermion, which is massive, and appears at momentum $k=0$ or $k=\pi$. Therefore we say that it is a non chiral Wilson fermion. The balanced BDI model shows two fermions of the same mass, since both energy gaps in this kind of ladder model have the same width. The energy gaps appear at opposite momenta and, thus, the two corresponding fermions have opposite chirality. The imbalance BDI model describes two fermions with different masses, being both of them non chiral, as the energy gaps are located in momentum space at $k=0$ and $k=\pi$. The circular AIII model shows a single Wilson fermion, which is massive and chiral, as the energy gap is located at any momentum different from $0$ and $\pi$. The balanced AIII model describes two chiral fermions of the same mass, and whose chiralities are not opposite to each other. Finally, the imbalanced AIII model corresponds to the most general case in which there are two chiral fermions of different masses [see Fig.~\ref{fig:FiguraLadder08}].

As we see, there are two variables: the number of Wilson fermions and their masses. A topological ladder model can describe one or two fermions and, in the second case, they can have equal masses (balanced models) or different masses (imbalanced models). The presence of time reversal symmetry in the BDI class forces the Wilson fermions to be non chiral, or, in case there are two fermions with the same mass, have opposite chirality. On the contrary, in the AIII class all fermions are chiral, what is a manifestation of the breaking of time reversal symmetry. In consequence, there are six different Wilson fermion configurations that characterize each type of topological ladder model.\\


\section{Canonical topological ladder model: the bowtie ladder}

\subsection{The bowtie ladder architecture}

From Sec.~ $3$ and Sec.~$4$, we know that a topological ladder model is characterized by two real functions of the momentum, $f_{1}(k)$ and $f_{2}(k)$, or alternatively by the real functions $\rho(k)$ and $\varphi(k)$, being these four functions related through the expression $\rho(k)e^{i\varphi(k)}=f_{1}(k)+if_{2}(k)$.
The Hamiltonian matrix of a topological ladder model can then be written in terms of such functions as:
\begin{equation}\label{eq:C5N01}
M(k)=f_{1}(k)\,\sigma_{1}+f_{2}(k)\,\sigma_{2},
\end{equation}
being $\sigma_{1}$ and $\sigma_{2}$ two anti-commuting Pauli matrices given by two orthogonal unitary vectors in the 3-dimensional real space; that is: $\sigma_{j}=\bm{n}_{j}\cdot\bm{\sigma}$ with $\bm{n}_{1}\cdot\bm{n}_{2}=0$.

The particular directions of the vectors $\bm{n}_{1}$ and $\bm{n}_{2}$ are irrelevant as they can be changed by performing a global unitary transformation. In this way, we can focus on a particular choice for $\bm{n}_{1}$ and $\bm{n}_{2}$ and study the corresponding ladder architecture and Hamiltonian. If we take $\bm{n}_{1}=\hat{x}$ and $\bm{n}_{2}=\hat{y}$ we obtain a particular realization of a topological ladder model. It constitutes a \textit{canonical} ladder model, as every topological ladder model can be obtained from it by applying the appropriate unitary transformation.\\

\begin{figure}[t]
  \centering
    \includegraphics[width=0.485\textwidth]{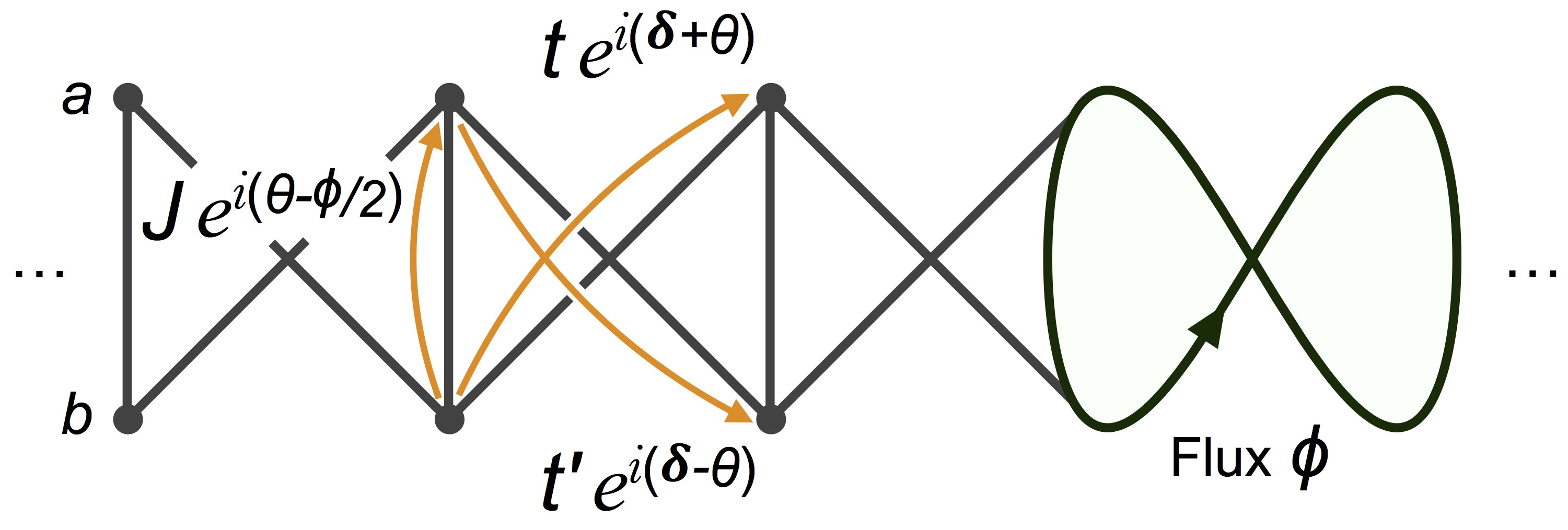}
    \caption[The bowtie ladder.]{\textbf{The bowtie ladder.} Schematic illustration of the bowtie ladder, which is a canonical ladder model, as any other ladder model can be obtained by performing the appropriate unitary transformation onto this model. The bowtie ladder consists of three different hopping terms: one vertical coupling and two diagonal couplings. This ladder geometry allows the construction of closed paths that define finite areas.  A particle that completes the elementary path $a_{n}\rightarrow b_{n+1}\rightarrow a_{n+1}\rightarrow b_{n}\rightarrow a_{n}$ gets a phase $\phi$, which is then interpreted as an effective magnetic flux penetrating the ladder.}
     \label{fig:CanonicalLadder}
\end{figure}

We start by considering the Hamiltonian matrix of such canonical ladder.
For that we need to substitute in Eq.~\ref{eq:C5N01} $\sigma_{1}$ and $\sigma_{2}$ by  $\sigma_{x}$ and $\sigma_{y}$, respectively. In this way, and using the functions $\rho(k)$ and $\varphi(k)$ instead of $f_{1}(k)$ and $f_{2}(k)$, the canonical ladder Hamiltonian matrix takes the following form:
\begin{align}
M(k)=&\rho(k)\, \left[\cos\varphi(k)\,\sigma_{x}+\sin\varphi(k)\sigma_{y}\right]=\nonumber\\
&\rho(k)\begin{pmatrix}
0 & e^{-i\varphi(k)}\\
e^{i\varphi(k)} & 0\\
\end{pmatrix},
\end{align}
so that the canonical ladder Hamiltonian in the momentum representation is:
\begin{equation}\label{eq:C5N02}
H_{\text{c}}=-\sum_{k}\rho(k)\left(e^{-i\varphi(k)}\,\hat{a}^{\dagger}_{k}\hat{b}_{k}+e^{i\varphi(k)}\,\hat{b}^{\dagger}_{k}\hat{a}_{k}\right).
\end{equation}
The Hamiltonian matrix of any topological ladder model depends on the momentum only through the two functions $\sin k$ and $\cos k$, or alternatively through $e^{ik}$ and $e^{-ik}$ (see Chapter $7$). Therefore, the most general Hamiltonian matrix corresponds to the case in which $\rho(k)e^{i\varphi(k)}$ is a general complex linear combination of the three functions $1$, $e^{ik}$ and $e^{-ik}$, that is:
\begin{equation}\label{eq:C5N03}
\rho(k)e^{i\varphi(k)}=e^{-i\theta}\left(Je^{i\phi/2}+te^{-i\delta}e^{ik}+t^{\prime}e^{i\delta}e^{-ik}\right).
\end{equation}
Being $J$, $t$ and $t^{\prime}$ three real and positive parameters and $\theta$, $\phi$ and $\delta$ three phases. We have chosen a particular parametrization of the phases which, without loss of generality, is really convenient in order to make a physical interpretation of them, as we will show in the following. In order to see how the canonical ladder Hamiltonian looks like in the position representation, we substitute $\rho(k)e^{i\varphi(k)}$ by its expression in (\ref{eq:C5N03}) in the Hamiltonian in momentum space (\ref{eq:C5N02}) and thus obtain:
\begin{align}
H_{\text{c}}=-\sum_{n}\left(Je^{i(\theta-\phi/2)}\,\hat{a}^{\dagger}_{n}\hat{b}_{n}+te^{i(\delta+\theta)}\,\hat{a}^{\dagger}_{n+1}\hat{b}_{n}+\right.\nonumber\\
\left.t^{\prime}e^{-i(\delta-\theta)}\,\hat{a}^{\dagger}_{n}\hat{b}_{n+1}+\text{h.c.}\right).
\end{align}
In this way, we can identify the three parameters $J$, $t$ and $t^{\prime}$ as the amplitudes of a vertical and two diagonal couplings in the ladder, respectively. Therefore, due to its ladder architecture, we call this model the \textit{bowtie} ladder model (see Fig.~\ref{fig:CanonicalLadder}).

In order to see which role plays each parameter in the bowtie ladder Hamiltonian, we go back to the momentum representation and write the Hamiltonian as:
\begin{equation}\label{eq:CanonicalLadderHamiltonianMR}
H_{\text{c}}=-\sum_{k}
\begin{pmatrix}
\hat{a}_{k}^{\dagger} & \hat{b}_{k}^{\dagger}
\end{pmatrix}
R_{z}(\theta)\,M_{\text{c}}(k)\,R_{z}^{\dagger}(\theta)
\begin{pmatrix}
\hat{a}_{k}\\
\hat{b}_{k}
\end{pmatrix},
\end{equation}
where $R_{z}(\theta)=e^{i\theta\sigma_{z}/2}$ is a rotation of an angle $\theta$ around the $z$-axis and:
\begin{align}\label{eq:CanonicalLadderHamiltonianMatrix}
M_{\text{c}}(k)=&\left[J\cos\frac{\phi}{2}+\left(t+t^{\prime}\right)\cos(k-\delta)\right]\sigma_{x}+\nonumber\\
&\left[J\sin\frac{\phi}{2}+\left(t-t^{\prime}\right)\sin(k-\delta)\right]\sigma_{y}.
\end{align}
By inspecting the bowtie ladder Hamiltonian matrix $M_{\text{c}}(k)$, as well as the bowtie ladder geometry in Fig.~\ref{fig:CanonicalLadder}, we can identify three different phases in the model, with three distinct physical meanings. These are:
\begin{itemize}
\item
\textit{Effective magnetic flux $\phi$}

The phase $\phi$ determines the total phase accumulated by a particle completing a closed path in the ladder. A non-vanishing accumulated phase along a closed path corresponds to an effective magnetic flux penetrating the area defined by the path. Therefore, we can identify the phase $\phi$ as an effective magnetic flux. All closed paths that can be defined in the canonical ladder are obtained as a combination of the elementary path $a_{n}\rightarrow b_{n+1}\rightarrow a_{n+1}\rightarrow b_{n}\rightarrow a_{n}$. As we show in the next sections, the phase $\phi$ affects the symmetries of the Hamiltonian and the properties of the edge states.
\item
\textit{Shift $\delta$ in the momentum-isospin correspondence.}

The phase $\delta$ produces a shift in the Hamiltonian matrix with respect to the momentum [see Eq.(\ref{eq:CanonicalLadderHamiltonianMatrix})]. This produces a shift in the correspondence between the momentum of each eigenstate and its associated vector in the Bloch sphere, which makes them be genuinely different from those eigenstates corresponding to the case in which there is no phase $\delta$. This phase also affects the symmetries of the Hamiltonian and the properties of the edge states, as we show in the following sections and Chapter $10$.

\item
\textit{Irrelevant phase $\theta$.}

The phase $\theta$ produces a rotation around the $z$-axis [see Eq.(\ref{eq:CanonicalLadderHamiltonianMR})], which is a global unitary operation and does not affect the symmetries of the system. In the following we neglect this phase for simplicity being all possible rotations around the $z$-axis included in our analysis by adding the phase $\theta$ as shown in Fig.~\ref{fig:CanonicalLadder}.
\end{itemize}

%
%
\subsection{Symmetry properties of the bowtie ladder}
\subsubsection{The $(\bm{n}_{0}, \bm{n}_{c}, \bm{n}_{s})$-decomposition}
In order to know what are the symmetries of the bowtie ladder Hamiltonian and how they depend on the different parameters present in the model, we decompose its Hamiltonian matrix as $M_{\text{c}}(k)=(\bm{n}_{0}+\bm{n}_{c}\cos k+\bm{n}_{s}\sin k)\cdot\bm{\sigma}$, being:
\begin{equation}
\begin{cases}
\bm{n}_{0}=J\cos(\phi/2)\,\hat{x}+J\sin(\phi/2)\,\hat{y},\\
\bm{n}_{c}=(t+t^{\prime})\cos\delta\,\hat{x}-(t-t^{\prime})\sin\delta\,\hat{y},\\
\bm{n}_{s}=(t+t^{\prime})\sin\delta\,\hat{x}+(t-t^{\prime})\cos\delta\,\hat{y}.
\end{cases}
\end{equation}
\textit{a) Chiral symmetry}\\

It is clear that the bowtie ladder Hamiltonian has chiral symmetry, as the three vectors $\bm{n}_{0}$, $\bm{n}_{c}$ and $\bm{n}_{s}$ lie in the $xy$-plane. The Hamiltonian matrix is a linear combination of $\sigma_{x}$ and $\sigma_{y}$, Eq.~(\ref{eq:CanonicalLadderHamiltonianMatrix}) and therefore $\sigma_{z}M_{\text{c}}(k)\sigma_{z}=-M_{\text{c}}(k)$; so that the chiral symmetry condition, Eq.~(\ref{eq:ConditionS}), is always fulfilled with $U_{S}=\sigma_{z}$.\\

\noindent \textit{b) Time reversal and charge conjugation symmetries}\\

\begin{figure}[t]
  \centering
    \includegraphics[width=0.485\textwidth]{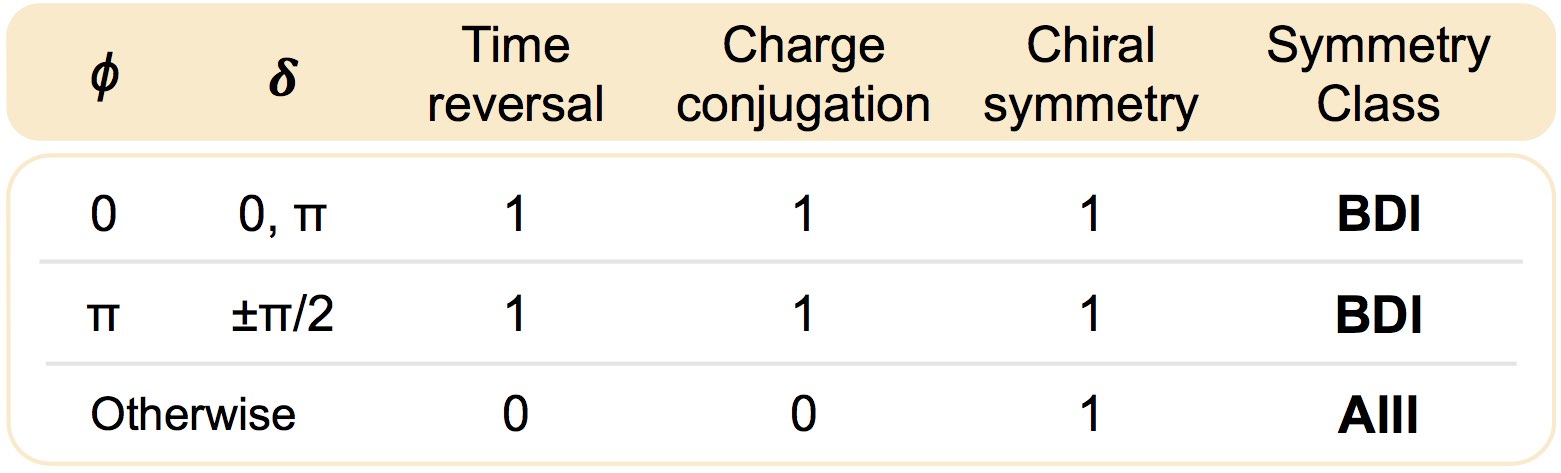}
    \caption[Bowtie ladder symmetry class]{\textbf{Bowtie ladder symmetry class}. Presence (1) or absence (0) of time reversal, charge conjugation and chiral symmetries in the bowtie ladder Hamiltonian. The symmetries of the bowtie ladder Hamiltonian are determined by the phases of its couplings, namely: $\delta$ and $\phi$. Only four particular configurations of parameters correspond to a Hamiltonian that belongs to the BDI symmetry class, being in the AIII class otherwise.}
    \label{fig:CLClass}
\end{figure}

There are two conditions that a ladder model has to fulfil in order to be time reversal symmetric, which are: $\bm{n}_{c}\cdot\bm{n}_{s}=0$ and $\bm{n}_{0}\cdot\bm{n}_{s}=0$ (see Chapter $6$). These two conditions applied to the bowtie ladder imply:
\begin{align}
&\bm{n}_{c}\cdot\bm{n}_{s}=0\quad\rightarrow\quad tt^{\prime}\sin2\delta=0,\label{eq:CanonicalLadderTR1}\\
&\bm{n}_{0}\cdot\bm{n}_{s}=0\quad\rightarrow\quad t\sin\left(\delta+\frac{\phi}{2}\right)+t^{\prime}\sin\left(\delta-\frac{\phi}{2}\right)=0.\label{eq:CanonicalLadderTR2}
\end{align}
The first time reversal condition, Eq.~(\ref{eq:CanonicalLadderTR1}), implies that $\delta$ can only take the values $0$, $\pm\pi/2$ and $\pi$. When $\delta=0$ or $\pi$, the second time reversal condition, Eq.~(\ref{eq:CanonicalLadderTR2}), is fulfilled only if $\phi=0$, whereas it is satisfied for $\phi=\pi$ when $\delta=\pm\pi/2$. Therefore there are four different configurations of the bowtie ladder parameters whose corresponding Hamiltonian exhibits time reversal symmetry. In the presence of chiral symmetry, either both time reversal and charge conjugation symmetries are present in the model, or none of them is. Consequently, these four configurations belong to the BDI symmetry class, whereas any other configuration of parameters corresponds to the AIII symmetry class.
As a result, we obtain Fig.~\ref{fig:CLClass}, where we show the symmetries of the bowtie ladder Hamiltonian and their dependence on its parameters.

\subsubsection{Hamiltonian matrix structure}
Alternatively, we can obtain the symmetries of the canonical ladder by analasyng its Hamiltonian matrix.

There are two ways in which the Hamiltonian matrix of a ladder with chiral symmetry can be written. On one hand, using two anticommuting Pauli matrices, $\sigma_{1}$ and $\sigma_{2}$, and two functions of the momentum $f_{1}(k)$ and $f_{2}(k)$. This representation is easier to deal with as it uses the orthonormal basis $\{\bm{n}_{1}, \bm{n}_{2}\}$. However, these two vectors are not unique, since we can choose any other orthogonal basis within the plane in which the Hamiltonian matrix lives; therefore, the two functions $f_{1}(k)$ and $f_{2}(k)$ are also not unique and there is no systematic way in which we can read the symmetry class of the Hamiltonian by looking at these functions. 

On the other hand, we can separate the component of the Hamiltonian matrix that goes with $\sin k$ and the one with $\cos k$, what is more convenient in order to read the symmetries of the Hamiltonian. Nevertheless, this second representation uses two Pauli matrices, $\sigma_{c}$ and $\sigma_{s}$, whose anticommutator is in general different from zero. In this way, we can write in general:
\begin{equation}
M(k)=\left(\alpha+\beta\cos k\right)\sigma_{c}+\left(\eta+\gamma\sin k\right)\sigma_{s}.
\end{equation}
Once we have expressed the Hamiltonian matrix of a ladder model using this representation it is straightforward to read its symmetries. \textit{The model belongs to the BDI class if and only if $\{\sigma_{c}, \sigma_{s}\}=0$ and $\eta=0$, being in the AIII class otherwise}. In other words, only in the BDI case the odd and even components of the Hamiltonian matrix can be separated in orthogonal directions (see Chapter $6$).\\

The bowtie ladder Hamiltonian matrix is written using the first representation in Eq.(\ref{eq:CanonicalLadderHamiltonianMatrix}). If we write it using the second representation we get:
\begin{widetext}
\begin{align}
M_{\text{c}}(k)=&\sqrt{t^2+t^{\prime\,2}+2tt^{\prime}\cos2\delta}\left(\frac{J\cos\delta\cos(\phi/2)}{t+t^{\prime}}-\frac{J\sin\delta\sin(\phi/2)}{t-t^{\prime}}+\cos k\right)\sigma_{c}\,+\nonumber\\
+\,&\sqrt{t^2+t^{\prime\,2}-2tt^{\prime}\cos2\delta}\left(\frac{J\sin\delta\cos(\phi/2)}{t+t^{\prime}}+\frac{J\cos\delta\sin(\phi/2)}{t-t^{\prime}}+\sin k\right)\sigma_{s}
\end{align}
\end{widetext}
where the two Pauli matrices $\sigma_{c}$ and $\sigma_{s}$ are:
\begin{align}
\sigma_{c}=\frac{\left(t+t^{\prime}\right)\cos\delta\,\sigma_{x}-\left(t-t^{\prime}\right)\sin\delta\,\sigma_{y}}{\sqrt{t^2+t^{\prime2}+2tt^{\prime}\cos2\delta}},\\
\sigma_{s}=\frac{\left(t+t^{\prime}\right)\sin\delta\,\sigma_{x}+(\left( t-t^{\prime}\right)\cos\delta\,\sigma_{y}}{\sqrt{t^2+t^{\prime2}-2tt^{\prime}\cos2\delta}}.
\end{align}
Therefore, the two conditions that must be satisfied in order for the Hamiltonian to be time reversal symmetric are:
\begin{align}
&\{\sigma_{c}, \sigma_{s}\}=\frac{4\,tt^{\prime}\sin2\delta\,\mathbb{I}}{\sqrt{(t^{2}+t^{\prime\,2})^{2}-(2tt^{\prime}\cos2\delta)^{2}}}=0\\
&\eta=J\sqrt{t^2+t^{\prime\,2}-2tt^{\prime}\cos2\delta}\,\Bigg(\frac{\sin\delta\cos(\phi/2)}{t+t^{\prime}}+\Bigg.\nonumber\\
&\Bigg.\frac{\cos\delta\sin(\phi/2)}{t-t^{\prime}}\Bigg)=0.
\end{align}
The first condition implies that $\delta=0$, $\pm\pi/2$ or $\pi$. When $\delta=0$ or $\pi$, $\eta\propto\sin(\phi/2)$ and thus $\phi=0$. In the case $\delta=\pm\pi/2$, $\eta\propto\cos(\phi/2)$ and therefore $\phi=\pi$. In this way, we obtain the same result as before, using the three vectors $\bm{n}_{0}$, $\bm{n}_{c}$ and $\bm{n}_{s}$ (see Fig.~\ref{fig:CLClass}).
%
%
\begin{figure}[t]
  \centering
    \includegraphics[width=0.425\textwidth]{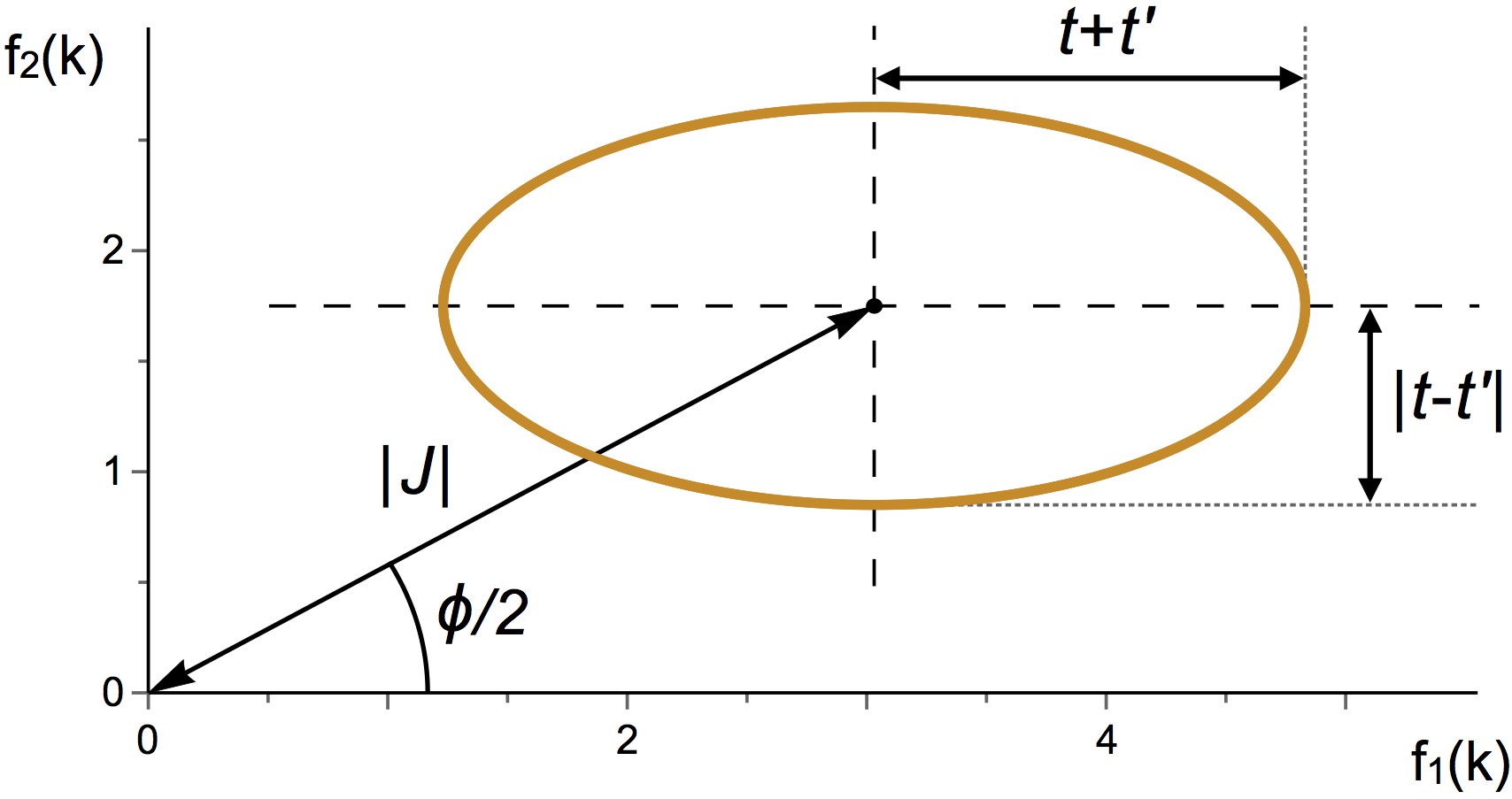}
    \caption[Bowtie ladder Hamiltonian matrix ellipse.]{\textbf{Bowtie ladder Hamiltonian matrix ellipse.} , The curve described by the bowtie ladder Hamiltonian matrix is an ellipse centred at $(J\cos(\phi/2), J\sin(\phi/2))$ and with axes $t+t^{\prime}$ and $||t|-|t^{\prime}||$. From the corresponding ellipse equation we can derive the condition that the parameters need to fulfil for the Hamiltonian to be topologically non-trivial [Eq.~(\ref{eq:TopologyCondition})].}
    \label{fig:GeneralEllipse}
\end{figure}
\subsection{Bowtie ladder topological phase.}

The simplest way of writing the bowtie ladder Hamiltonian matrix is using the form $M_{\text{c}}(k)=f_{1}(k)\sigma_{x}+f_{2}(k)\sigma_{y}$.
From Eq.~(\ref{eq:CanonicalLadderHamiltonianMatrix}) we can identify the functions $f_{1}(k)$ and $f_{2}(k)$ and see that they satisfy the following ellipse equation:
\begin{equation}\label{eq:EllipseEquation}
\left[\frac{f_{1}(k)-J\cos(\phi/2)}{t+t^{\prime}}\right]^{2}+\left[\frac{f_{2}(k)-J\sin(\phi/2)}{t-t^{\prime}}\right]^{2}=1,
\end{equation}
so that the curve described by the Hamiltonian matrix in the complex plane is an ellipse centred at the point $(J\cos(\phi/2), J\sin(\phi/2))$ and whose horizontal and vertical axes are $t+t^{\prime}$ and $|t-t^{\prime}|$, respectively (see Fig.~\ref{fig:GeneralEllipse}). Form this ellipse equation we can obtain the condition that defines the region in the parameter space in which the system exhibits a topologically non-trivial nature. Such condition is determined by imposing that the ellipse encloses the origin and thus the Zak phase takes the value $\pi$. This condition is:
\begin{equation}\label{eq:TopologyCondition}
J\sqrt{t^2+t^{\prime\,2}-2tt^{\prime}\cos\phi}<|t^{2}-t^{\prime\,2}|.
\end{equation}

It is important to mention that we exclude the case in which $t=t^{\prime}$ from our analysis, as it corresponds to a trivial Hamiltonian. From Eq.~(\ref{eq:CanonicalLadderHamiltonianMatrix}) we see that the component that goes with $\sin(k-\delta)$ in the Hamiltonian matrix vanishes in this situation and thus the Hamiltonian matrix ellipse collapses into a line [see Eq.~(\ref{eq:EllipseEquation})]. Therefore, the corresponding winding number is always zero and the model cannot access a topologically non-trivial phase.

%
%
%
\begin{figure}[t]
  \centering
    \includegraphics[width=0.485\textwidth]{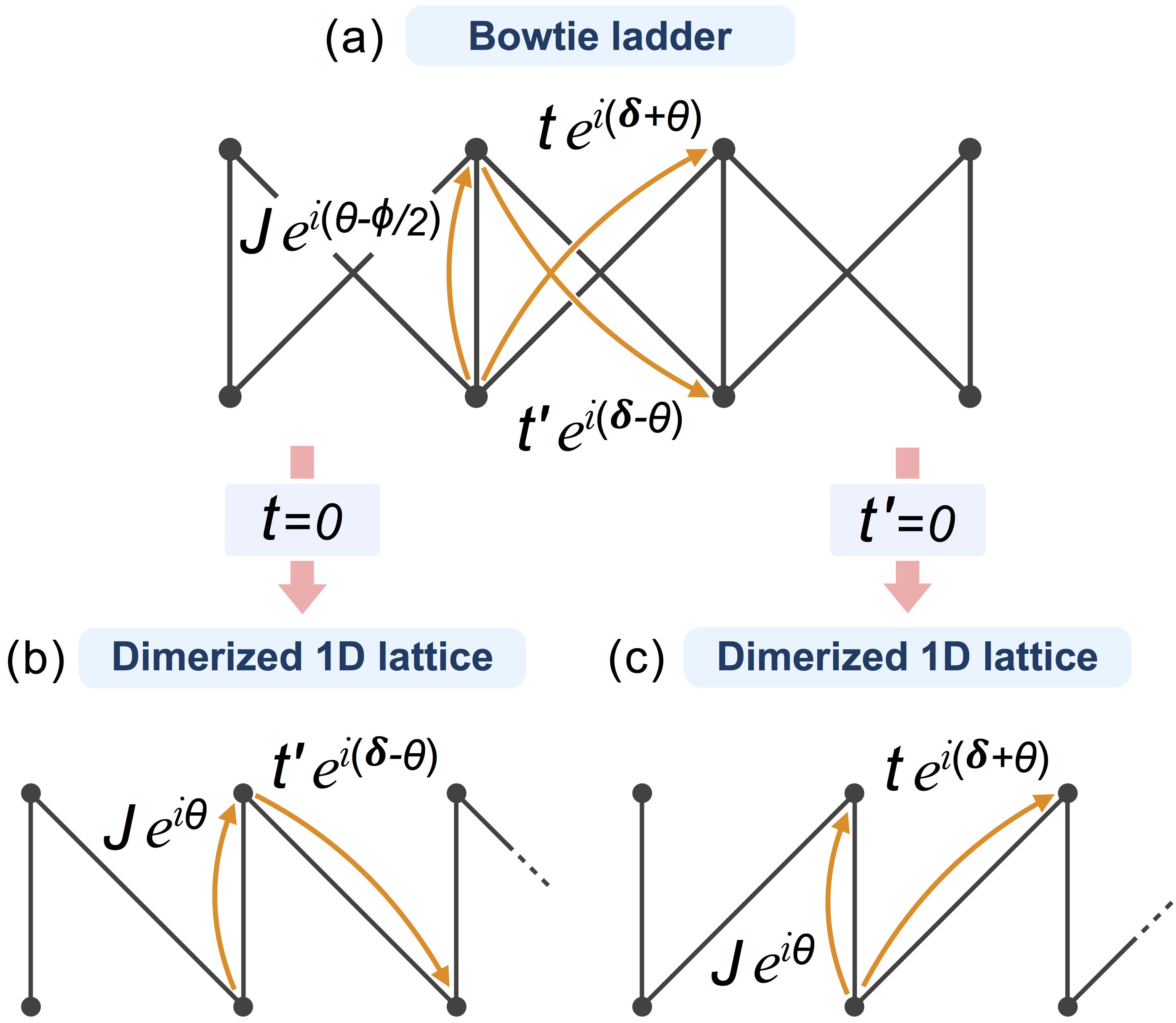}
    \caption[1D dimerized lattice from the canonical ladder.]{\textbf{1D dimerized lattice from the canonical ladder.} The canonical ladder (a) can collapse into a one-dimensional dimerized lattice by setting one of the two diagonal coupling amplitudes to zero. That is, making $t^{\prime}=0$ (b) or $t=0$ (c).}
    \label{fig:ChainCanonicalLadder}
  \end{figure}  
\subsection{Bowtie ladder in the BDI symmetry class}
  
Among the four different parameter configurations of the bowtie ladder Hamiltonian that belong to the BDI symmetry class (see Fig.~\ref{fig:CLClass}), the bowtie ladder model can explore all different types of BDI ladder models, namely: the SSH model, the balanced BDI model and the imbalanced BDI model.\\
  
\textbf{\textit{i)} Bowtie ladder realizing the SSH model}\\

The bowtie ladder collapses into a one-dimensional dimerized lattice if we set one of the two diagonal coupling amplitudes to zero; that is, $t=0$ or $t^{\prime}=0$. In both cases the system becomes a chain with just two alternate couplings (see Fig.~\ref{fig:ChainCanonicalLadder}) and thus the phase $\phi$ becomes redundant and can be neglected, as in a one-dimensional chain no closed path defining a finite area can be constructed (apart from the path formed by the whole system when considering closed boundary conditions). Evaluating the time reversal conditions, Eq.~(\ref{eq:CanonicalLadderTR1}) and Eq.~(\ref{eq:CanonicalLadderTR2}), for $t=0$ or $t^{\prime}=0$ we see that the corresponding Hamiltonian belongs to the BDI symmetry class only if $\delta=0$ or $\delta=\pi$, being in the AIII class otherwise.
  
\begin{figure}[t]
  \centering
    \includegraphics[width=0.485\textwidth]{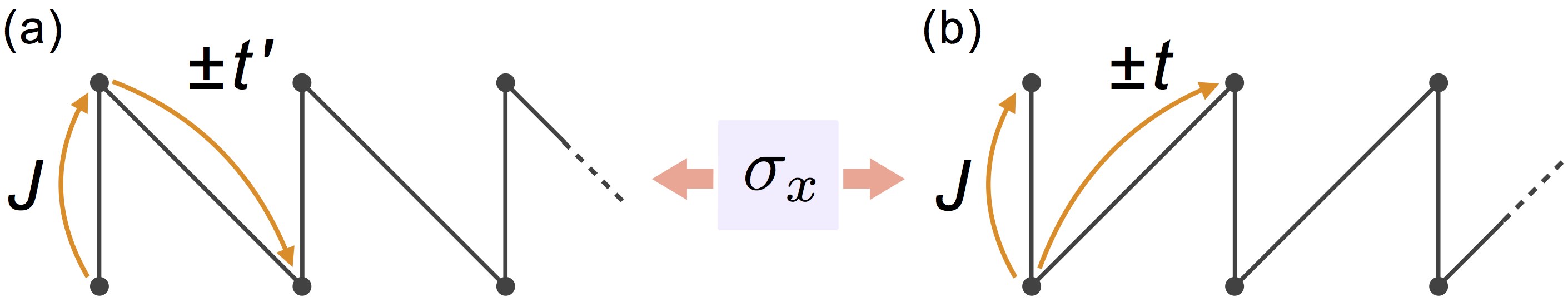}
    \caption[SSH model from the canonical ladder.]{\textbf{SSH model from the canonical ladder.} The SSH model is obtained from the canonical ladder Hamiltonian by removing one of the two diagonal couplings, that is, taking $t=0$ (a) or $t^{\prime}=0$ (b), as well as removing all phases in the tunneling amplitudes. This two possibilities correspond to different Hamiltonians which, nevertheless, are connected by applying the unitary $\sigma_{x}$.}
    \label{fig:SSHCanonicalLadder}
  \end{figure}

\begin{figure*}
  \centering
    \includegraphics[width=0.85\textwidth]{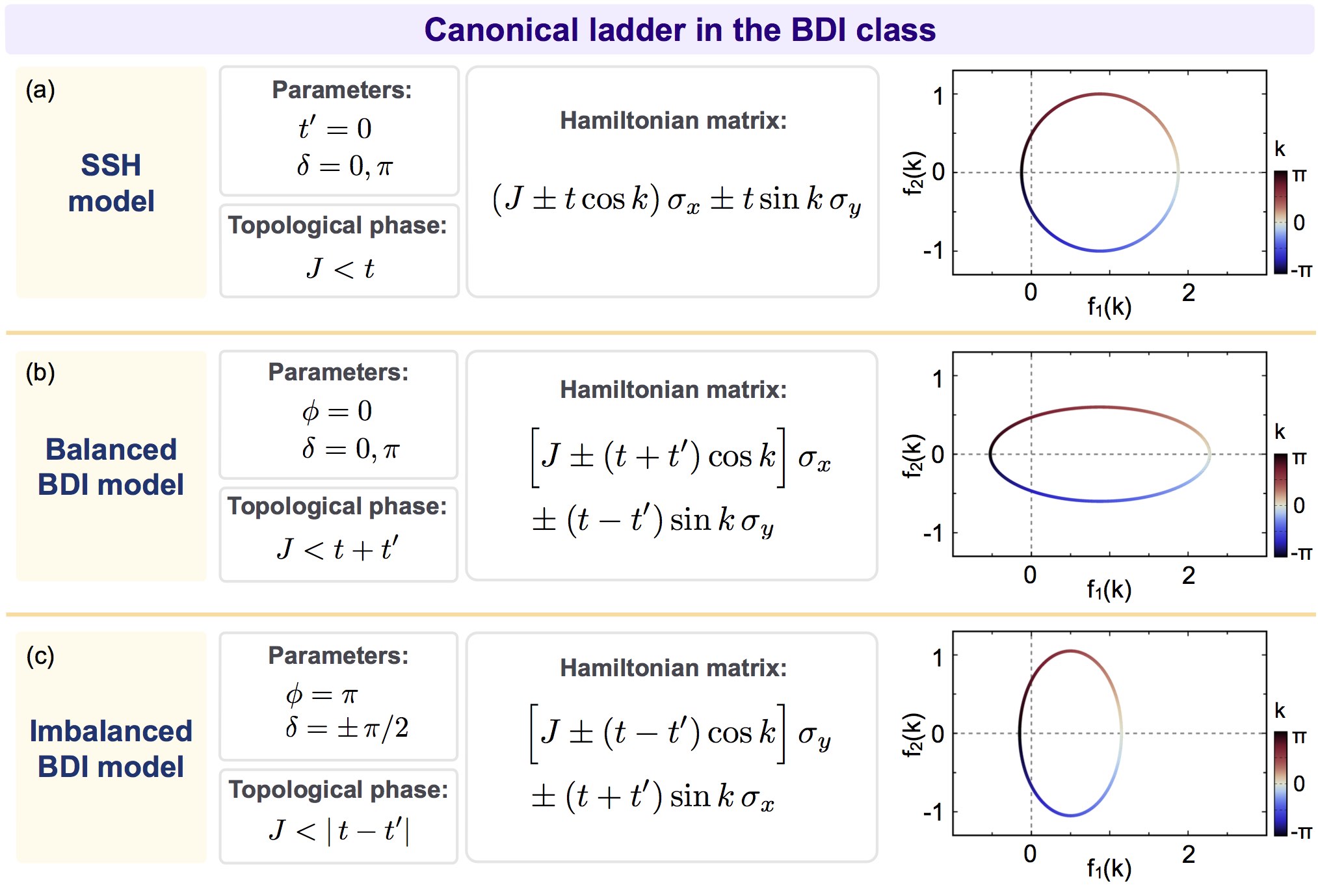}
    \caption[Bowtie ladder in the BDI class.]{\textbf{Bowtie ladder in the BDI class.} The bowtie ladder serves as a canonical ladder model and allows us to explore and realize all types of topological ladder models. In particular, the three distinct types of BDI ladder models can be achieved by choosing the appropriate parameters configuration. Taking $t^{\prime}=0$ and $\delta=0$ or $\delta=\pi$, it realizes the SSH model (a). For $\phi=0$ and $\delta=0$ or $\delta=\pi$, it corresponds to the balanced BDI model (b). Finally, for $\phi=\pi$ and $\delta=\pi/2$ or $\delta=-\pi/2$, we obtain a realization of the imbalanced BDI model.}
    \label{fig:EllipsesClassificationBDI}
  \end{figure*}

These two different ways in which the bowtie ladder becomes a one-dimensional dimerized lattice are realizations of the SSH model. They are connected to each other by applying the unitary transformation $\sigma_{x}$, which interchanges $\hat{a}_{n}^{\dagger}$ with $\hat{b}_{n}^{\dagger}$ and consists of flipping the ladder with respect to a central horizontal line (see Fig.~\ref{fig:SSHCanonicalLadder}), and therefore we only need to consider and analyse one of them. We choose the case given by setting $t^{\prime}=0$, whose Hamiltonian matrix is:
\begin{equation}
M(k)=(J\pm t\cos k)\,\sigma_{x}\pm t\sin k\,\sigma_{y}.
\end{equation}
The $\pm$ options correspond to $\delta=0$ and $\delta=\pi$, respectively.
This Hamiltonian matrix is characterized by the fact that the ellipse in the complex plane has collapsed into a circumference [see Fig.~\ref{fig:EllipsesClassificationBDI}(a)], which is centred at the point $(J,0)$ and whose radius is $t$, so that the system is in a topologically non-trivial phase if $J<t$.\\
%
%

\textbf{\textit{ii)} Bowtie ladder realizing the balanced BDI model}\\

In the case in which the three coupling amplitudes present in the bowtie ladder Hamiltonian are non zero, there are four different configurations of parameters that correspond to a Hamiltonian in the BDI symmetry class. Two with $\phi=0$ and another two with $\phi=\pi$ (see Fig.~\ref{fig:CLClass}). The Hamiltonian matrices for the first two, with no effective magnetic flux, are the following:
\begin{equation}
M(k)=\left[J\pm(t+t^{\prime})\cos k\right]\,\sigma_{x}\pm(t-t^{\prime})\sin k\,\sigma_{y},
\end{equation}
where the $\pm$ correspond to the cases $\delta=0$ and $\delta=\pi$, respectively. These Hamiltonian matrices are characterized by the fact that the horizontal axis of the corresponding ellipse is always bigger than the vertical one [see Fig.~\ref{fig:EllipsesClassificationBDI}(b)], so that these two configurations of the bowtie ladder are realizations of the balanced BDI model (as well as the SSH-like model for a small region in the parameter space (see Fig.~\ref{fig:FiguraLadder06}). The condition for a non-trivial topology takes in this case the particular form $J<t+t^{\prime}$.\\

\textbf{\textit{iii)} Bowtie ladder realizing the imbalanced BDI model}\\

The other two configurations of the bowtie ladder that realize BDI models, with $\phi=\pi$, correspond to the Hamiltonian matrices:
\begin{equation}
M(k)=\left[J\mp(t-t^{\prime})\cos k\right]\,\sigma_{y}\pm(t+t^{\prime})\sin k\,\sigma_{x},
\end{equation}
where the $\mp$ correspond to the cases $\delta=\pm\pi/2$. The ellipses described by these Hamiltonian matrices are characterized by a horizontal axis smaller than the vertical one, considering the horizontal direction the one where the centre of the ellipse is located [see Fig.~\ref{fig:EllipsesClassificationBDI}(c)]. Consequently, these two configurations of the bowtie ladder realize the imbalanced BDI model, as well as the SSH-like model in a small region in the parameter space (see Fig.~\ref{fig:FiguraLadder06}). The condition for a non-trivial topology takes the form $J<|t-t^{\prime}|$.

\subsection{Bowtie ladder in the AIII class}

There are two ways in which the bowtie ladder can break time reversal symmetry and, thus, belong to the AIII symmetry class. On one hand, by shifting the momentum-isospin correspondence, what can be done by adding a phase $\delta$. On the other hand, by adding an effective magnetic flux per plaquette $\phi$. We analyse each case separately.
\begin{figure*}
  \centering
    \includegraphics[width=0.85\textwidth]{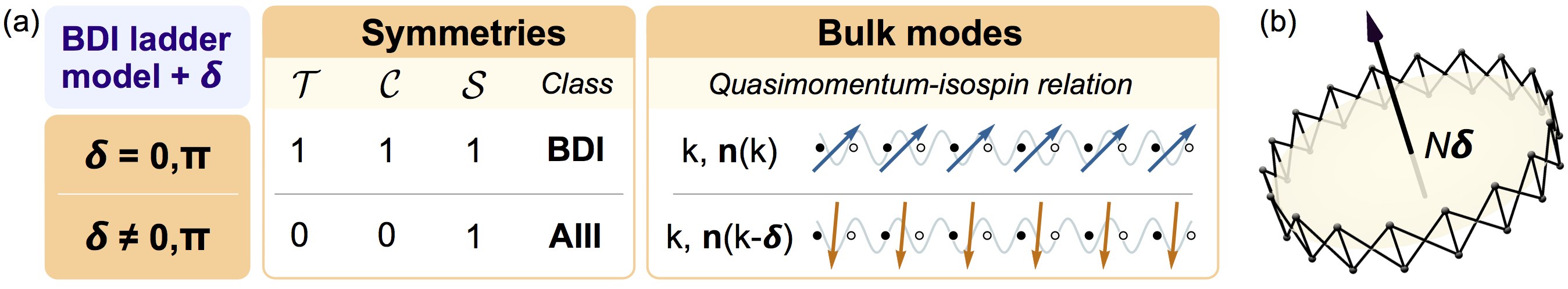}
   \caption[Shift in the momentum-isospin correspondence.]{\textbf{Shift in the momentum-isospin correspondence.} (a) Each ladder model in the BDI class can be generalized by adding a phase $\delta$. In the presence of a phase $\delta\neq0, \pi$ time reversal and charge conjugation symmetries are broken and the model belongs to the AIII class. In this case the bulk modes exhibit a different momentum-isospin relation. (b) For periodic boundary conditions and $N$ unit cells a total phase $N\delta$ is acquired by particles when completing a closed path along the whole system.}
    \label{fig:BDIFaseDelta} 
\end{figure*}
\subsubsection{Shift in the momentum-isospin correspondence}

Any model in the BDI symmetry class with Hamiltonian matrix $M(k)$ can be taken to the AIII symmetry class by adding a shift $\delta$ in the momentum-isospin relation; that is, transforming $M(k)$ into $M(k-\delta)$. In the presence of this phase $\delta$ the Hamiltonian matrix has a different form and breaks time reversal and charge conjugation symmetries. The bulk modes differ from those corresponding to the case in which there is no phase $\delta$, being characterized by a different momentum-isospin relation. 

To show these results, we start by considering a ladder model in the BDI class, whose Hamiltonian matrix, as we know from Chapter $6$, can be written as:
\begin{equation}
M(k)=(\alpha+\beta\cos k)\,\sigma_{1}+\gamma\sin k\,\sigma_{2},
\end{equation}
with $\{\sigma_{1}, \sigma_{2}\}=0$ and being $\alpha$, $\beta$ and $\gamma$ three real parameters. The Bloch eigenstates of these model are given by Eq.~(\ref{eq:BlochEigenstates}) and the corresponding energies are $E_{\pm}(k)\mp\rho(k)$, with:
\begin{equation}
\rho^2(k)=\alpha^2+\gamma^2+2\alpha\beta\cos k+(\beta^2-\gamma^2)\cos^2 k.
\end{equation}
If we introduce a shift $\delta$ with respect to the momentum, we obtain a new Hamiltonian matrix, $M(k-\delta)$, which can be decomposed as:
\begin{align}
M(k-\delta)=&\sqrt{\beta^2\cos^{2}\delta+\gamma^2\sin^{2}\delta}\left( \frac{\alpha\cos\delta}{\beta}+\cos k \right)\,\sigma_{c}\,+\nonumber\\
&\sqrt{\beta^2\sin^{2}\delta+\gamma^2\cos^{2}\delta}\left( \frac{\alpha\sin\delta}{\beta}+\sin k \right)\,\sigma_{s},
\end{align}
with $\sigma_{c}\propto\beta\cos\delta\,\sigma_{1}-\gamma\sin\delta\,\sigma_{2}$ and $\sigma_{s}\propto\beta\sin\delta\,\sigma_{1}+\gamma\cos\delta\,\sigma_{2}$.

Comparing this matrix with the general form of the Hamiltonian matrix of a ladder model with chiral symmetry, Eq.~(\ref{eq:GeneralAIIIMatrix}), we conclude that it belongs to the BDI symmetry class if and only if:
\begin{align}
&\eta=\frac{\alpha\sin\delta}{\beta}\sqrt{\beta^2\sin^{2}\delta+\gamma^2\cos^{2}\delta}=0,\\
&\{\sigma_{c}, \sigma_{s}\}\propto(\beta^2-\gamma^2)\sin\delta\cos\delta=0.
\end{align}
These two conditions are fulfilled simultaneously only if $\sin\delta=0$, that is: for $\delta=0$ and $\delta=\pi$.
The phase $\delta$ that appears in the bowtie ladder parametrization (see Fi.~\ref{fig:CanonicalLadder}) plays precisely this role, that is, introduces a shift in the Hamiltonian matrix with respect to the momentum. As a consequence, all ladder models in the BDI symmetry class come in pairs $(\delta=\delta_{0},\delta=\delta_{0}+\pi)$, whose corresponding Hamiltonian matrices have the form:
\begin{equation}
M(k)=(\alpha\pm\beta\cos k)\,\sigma_{1}\pm\gamma\sin k\,\sigma_{2}.
\end{equation}
They are particular cases of a more general one with a phase $\delta=\delta_{0}+\delta^{\prime}$. This more general model belongs to the BDI class for $\delta^{\prime}=0,\pi$, whereas it belongs to the AIII class otherwise [see Fig.~\ref{fig:BDIFaseDelta}(a)].

The Bloch eigenstates for $\delta^{\prime}\neq0$ are:
\begin{equation}
\ket{k,\delta^{\prime}}_{\pm}=\begin{pmatrix}
\hat{a}_{k}^{\dagger} & \hat{b}_{k}^{\dagger}
\end{pmatrix}
\frac{1}{\sqrt{2}}\,U\begin{pmatrix}
1\\
\pm e^{i\phi(k-\delta^{\prime})}
\end{pmatrix}\ket{0}.
\end{equation}
The phase $\delta^{\prime}$ produces a shift in the correspondence between the momentum and the isospin of each eigenstate, which are the two quantities that characterize them. As a consequence, the set of eigenmodes of the Hamiltonian for $\delta^{\prime}\neq0$ (AIII class) is genuinely different from the set of eigenmodes of the Hamiltonian for $\delta^{\prime}=0$ (BDI class) [see Fig.~\ref{fig:BDIFaseDelta}(a)].

In this way, introducing a shift in the momentum-isospin correspondence of a topological ladder Hamiltonian is a systematic way of breaking time reversal symmetry and can be donde in every ladder model. In the case of the bowtie ladder, when considering periodic boundary conditions, the phase $\delta$ also results in an effective magnetic flux $N\delta$ penetrating the ring formed by the whole system [see Fig.~\ref{fig:BDIFaseDelta}(b)].

The bowtie ladder model in the presence of a phase $\delta$ can realize each of the three different types of AIII ladder models, namely: the circular AIII model the balanced AIII model and the imbalanced AIII model.\\

\textbf{\textit{i)} Bowtie ladder realizing the circular AIII model}\\

\begin{figure*}
  \centering
    \includegraphics[width=0.85\textwidth]{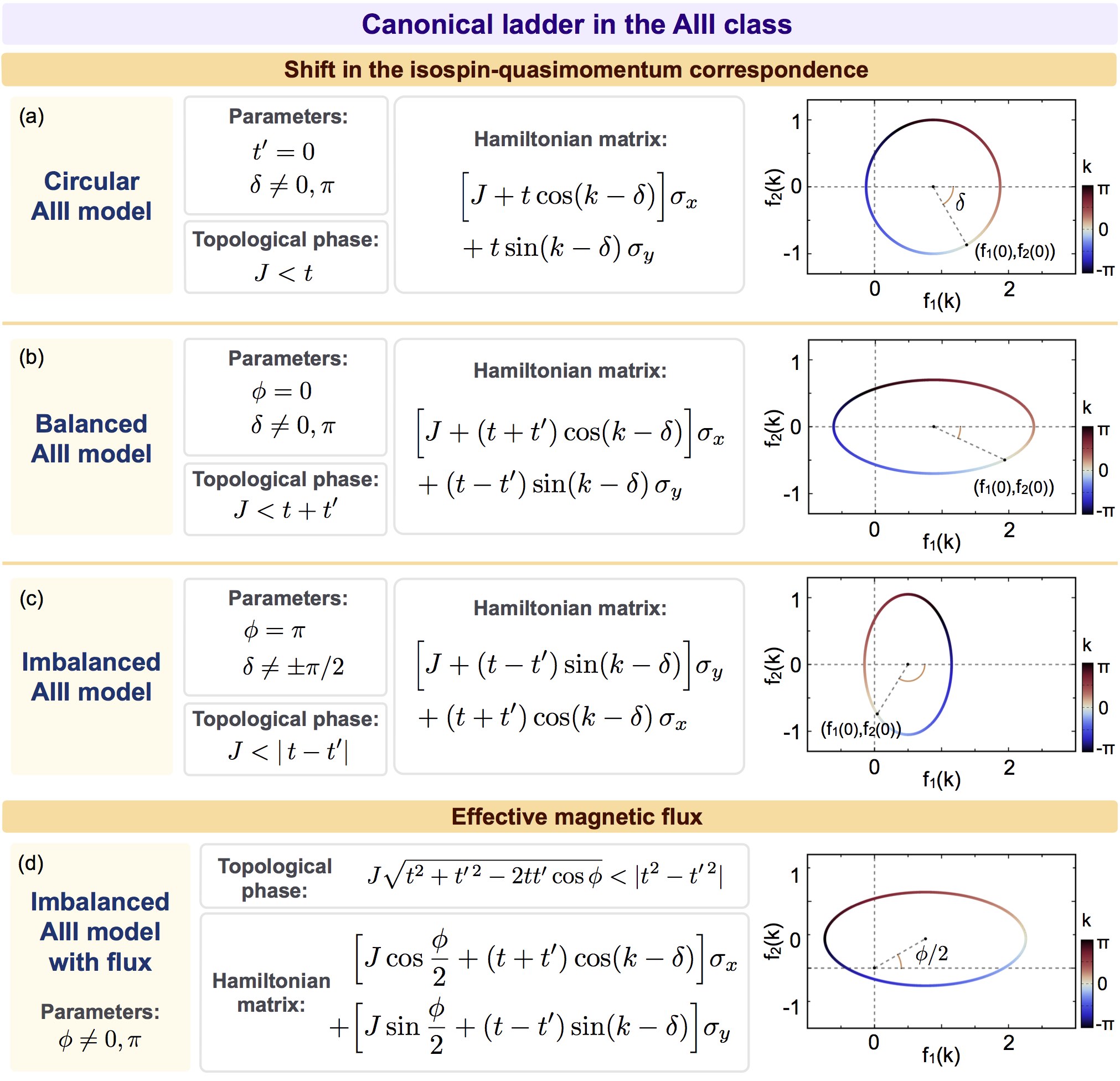}
    \caption[Bowtie ladder in the AIII class.]{\textbf{Bowtie ladder in the AIII class.} The bowtie ladder needs at least one of two ingredients in order to break time reversal symmetry and realize a model in the AIII class: a phase $\delta$ that introduces a shift in the momentum-isospin correspondence or an effective magnetic flux $\phi$. In the first case, the bowtie ladder realizes the circular AIII model (a), the balanced AIII model (b) and a particular case of the imbalance AIII model (c), in which the distance between the two energy gaps is $\pi$. In the second case, for $\phi\neq0,\pi$, the bowtie ladder realizes the most general imbalance AIII model (d), in which the distance between the two energy gaps is different from $\pi$.}
    \label{fig:FiguraLadder17}
  \end{figure*}

This case corresponds to the situation in which one of the two diagonal couplings of the bowtie ladder have been set to zero, so that the system became a dimerized one-dimensional lattice (see Fig.~\ref{fig:ChainCanonicalLadder}). Both cases, taking $t=0$ or $t^{\prime}=0$, are connected by applying a global unitary transformation. Therefore, we can consider only the case $t^{\prime}=0$, for which the Hamiltonian matrix is:
\begin{equation}
M(k)=\Big[J+t\cos(k-\delta)\Big]\,\sigma_{x}+t\sin(k-\delta)\,\sigma_{y}.
\end{equation}
It describes a circle in the complex plane [see Fig.~\ref{fig:FiguraLadder17}(a)] and corresponds to two energy bands with a single gap at momentum $q=\pi+\delta$. Therefore, it is a realization of the circular AIII model. The condition for the system to be in its topologically non trivial phase is $J<t$.\\

\textbf{\textit{ii)} Bowtie ladder realizing the balanced AIII model}\\

In case in which the three different hopping amplitudes of the model are different from zero and $\phi=0$, the bowtie ladder Hamiltonian matrix is:
\begin{equation}
M(k)=\Big[J+(t+t^{\prime})\cos(k-\delta)\Big]\,\sigma_{x}+(t-t^{\prime})\sin(k-\delta)\,\sigma_{y}.
\end{equation}
This Hamiltonian matrix corresponds to an ellipse whose horizontal axis is bigger than the vertical one [see Fig.~\ref{fig:FiguraLadder17}(b)] and two energy bands with two gaps of the same width located at $q_{1}=\delta+q$ and $q_{2}=\delta+q$ with:
\begin{equation}
q=\arccos\left[\frac{-J(t+t^{\prime})}{4tt^{\prime}}\right].
\end{equation}
Therefore, it is a realization of the balanced AIII model. The condition for a non trivial topology takes in this case the form $J<t+t^{\prime}$.\\

\textbf{\textit{iii)} Bowtie ladder realizing the imbalanced AIII model}\\

For an effective magnetic flux $\phi=\pi$, the Hamiltonian matrix is:
\begin{equation}
M(k)=\big[J-(t-t^{\prime})\cos(k-\delta^{\prime})\big]\,\sigma_{y}+(t+t^{\prime})\sin(k-\delta^{\prime})\,\sigma_{x},
\end{equation}
where $\delta^{\prime}=\delta-\pi/2$. The Hamiltonian matrix ellipse has then a horizontal axis smaller than the vertical one, considering the horizontal direction the one where the centre of the ellipse is located [see Fig.~\ref{fig:FiguraLadder17}(c)]. In consequence, the model has two gaps between the energy bands, which have different widths and are located at $q_{1}=\delta^{\prime}$ and $q_{2}=\delta^{\prime}+\pi$. Thereby, it is a realization of the imbalanced AIII model. The condition for a non-trivial topology takes the form $J<|t-t^{\prime}|$.

\subsubsection{Effective magnetic flux per plaquette}

We know that if we add a phase $\delta\neq0,\pi$ to a particular bowtie ladder parameter configuration in the BDI class, we obtain a model in the AIII class. The Hamiltonian matrix of the new model will be equal to the Hamiltonian matrix of the BDI model to which the phase $\delta$ has been added, but with a shift of $\delta$ with respect to the momentum.That is:
\begin{equation}
M_{\text{c}}(k;\delta)=M_{\text{c}}(k-\delta;0).
\end{equation}
In consequence, the Hamiltonian matrix curve will remain the same, being the momentum that corresponds to each point in the ellipse the only thing that is different. The energy bands are also shifted with respect to the momentum:
\begin{equation}
E_{\pm}(k;\delta)=E_{\pm}(k-\delta;0),
\end{equation}
so that the number of energy gaps and their widths are not affected by the phase $\delta$. However, their location in momentum space does actually change. In the SSH model there is a single energy gap at $q=0$ or $q=\pi$; after adding a phase $\delta$ to such model we obtain the circular AIII model, with a single gap at $q=\delta$ or $q=\pi+\delta$. Analogously, adding a phase $\delta$ to the balanced BDI model and the imbalanced BDI model turns them into the balanced AIII model and the imbalanced AIII model, respectively.\\

Furthermore, most AIII ladder models can only be realized by adding a phase $\delta$ to a BDI ladder model. That is the case of the circular AIII model and the balanced AIII model. If an AIII ladder model has a single gap at $q\neq0,\pi$, or two gaps of the same width, we can add a phase $\delta$ to them such that we obtain a single gap at $q=0$, for the first case, and two gaps at opposite momenta in the second case. These new configurations are in the BDI class, what means that any circular AIII model and any balanced AIII model can be taken to the BDI class by adding the appropriate phase $\delta$ to them. On the contrary, not every imbalanced AIII model can be taken to the BDI class by shifting the momentum-isospin correspondence, but only those in which the distance between the two energy gaps is exactly $\pi$, and therefore a phase $\delta$ can be added such that one gap is shifted to $q_{1}=0$ and the other to $q_{2}=\pi$.

As a result, we can conclude that a ladder model in the AIII class in which time reversal symmetry is broken by the presence of an effective magnetic field, such as the bowtie ladder with $\phi\neq0,\pi$, is a realization of the imbalanced AIII model in which the distance between the two energy gaps is different from $\pi$ . All other possible ladder models in the AIII class, namely: the circular AIII model, the balanced AIII model and the imbalance AIII model with a distance of $\pi$ between the two energy gaps, can be taken to the BDI class by adding a phase $\delta$. The only values that the effective magnetic field can take in the BDI class are $\phi=0$ and $\phi=\pi$, and the phase $\delta$ does not change the value of the effective magnetic field. Therefore, $\phi=0$ or $\phi=\pi$ for those three types of AIII ladder models. 

The bowtie ladder model with $\phi\neq0,\pi$ corresponds to the Hamiltonian matrix in Eq.~\ref{eq:CanonicalLadderHamiltonianMatrix}, which describes an ellipse centred at $(J\cos\phi/2,J\sin\phi/2)$ and with axes $t+t^{\prime}$ and $|t-t^{\prime}|$ [see Fig.~\ref{fig:FiguraLadder17}(d)]. It is a realization of the imbalanced AIII model and the condition that defines its topologically non trivial phase is the one in Eq.~\ref{eq:TopologyCondition}.


\section{Topological Ladder Edge Modes}

The topological nature of a ladder model with chiral symmetry is manifested, for open boundary conditions, in the existence of topologically protected edge modes. They are located at the edges of the system, have almost zero energy and appear when the system is in its topological phase. Here we first obtain their wave function and then present their main properties, namely: they can be localized both in position and momentum spaces, and their momentum distribution is directly related to the number and masses of the Wilson fermions described by the model, as well as to the symmetry class of the Hamiltonian.

\subsection{Edge modes wave function}

In order to derive the wave function of the edge modes of a topological ladder model, we focus on the bowtie ladder. As a canonical ladder model, any other ladder model can be obtained from it by performing a global unitary transformation. Therefore, once we obtain the edge modes of the bowtie ladder, it is straightforward to generalize the result to any ladder model. For it, we exploit the symmetries of the bowtie ladder model and also make a zero energy approximation. 

\subsubsection{Inversion-reflection-conjugation symmetry}

\begin{figure}[t]
  \centering
    \includegraphics[width=0.4\textwidth]{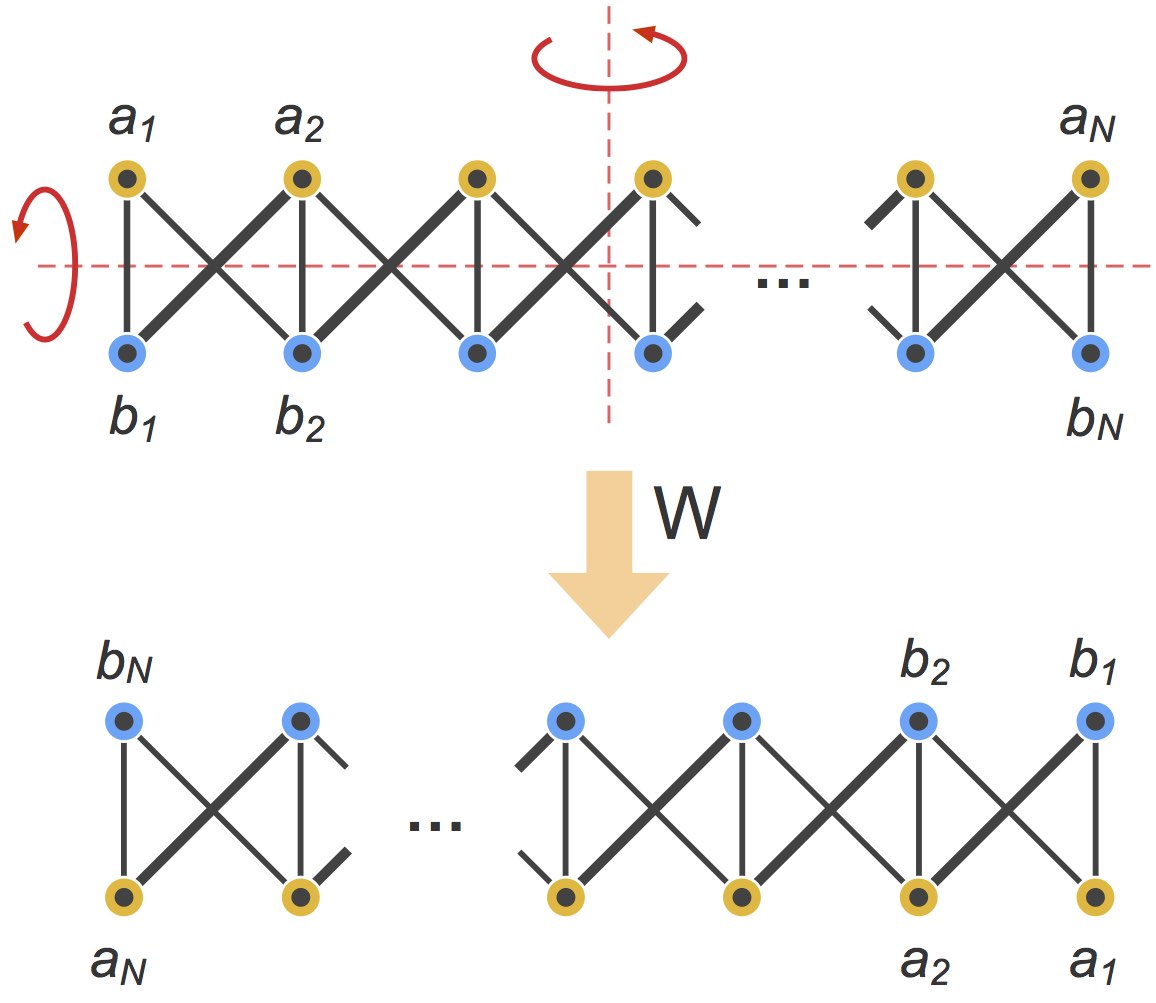}
    \caption[IRC-symmetry of the bowtie ladder.]{\textbf{IRC-symmetry of the bowtie ladder.} Schematic illustration of the unitary operator $W$, which produces a reflection and an inversion of the ladder. This transformation together with the complex conjugation leaves the bowtie ladder Hamiltonian invariant and, thus, represents a symmetry of the system.}
    \label{fig:C6N01}
\end{figure}

The bowtie ladder model has an inversion-reflection-conjugation (IRC) symmetry, which can be used in order to obtain important information about the wave function of the edge modes.
We consider the bowtie ladder and define the unitary transformation $W$ as:
\begin{equation}
W: \begin{cases}
\,\hat{a}_{n}^{\dagger}\,\longrightarrow\,\hat{b}_{N+1-n}^{\dagger}\\
\,\hat{b}_{n}^{\dagger}\,\longrightarrow\,\hat{a}_{N+1-n}^{\dagger}.
\end{cases}
\end{equation}
It is clearly a unitary transformation, as it rearranges the elements of the position basis.
This operation consists of a reflection of the ladder with respect to a central vertical line and an inversion that interchanges the $a$ and $b$ modes with each other (see Fig.~\ref{fig:C6N01}).
This unitary transformation modifies the momentum creation operators in the following way:
\begin{equation}
\hat{a}_{k}^{\dagger}=\frac{1}{\sqrt{N}}\sum_{n}e^{ikn}\,\hat{a}_{n}^{\dagger}\xrightarrow{\text{$\, W\,$}}\frac{e^{ik}}{\sqrt{N}}\sum_{n}e^{-ikn}\,\hat{b}_{n}^{\dagger}=e^{ik}\,\hat{b}_{-k}^{\dagger},
\end{equation}
where we have used that $e^{ikN}=1$. Therefore, the unitary $W$ takes the following form in the momentum representation:
\begin{equation}
W: \begin{cases}
\,\hat{a}_{k}^{\dagger}\,\longrightarrow\,e^{ik}\,(\hat{b}_{k}^{\dagger})^{*}\\
\,\hat{b}_{k}^{\dagger}\,\longrightarrow\,e^{ik}\,(\hat{a}_{k}^{\dagger})^{*}.
\end{cases}
\end{equation}
Applying this inversion-reflection transformation to the bowtie ladder Hamiltonian in Eq.~(\ref{eq:C5N02}) is equivalent to compute its complex conjugated:
\begin{align}
&WH_{\text{c}}W^{\dagger}=\nonumber\\
&-\sum_{k}\rho(k)\left[\,e^{i\varphi(k)}\,\hat{b}_{k}^{\dagger}\hat{a}_{k}+e^{-i\varphi(k)}\,\hat{a}_{k}^{\dagger}\hat{b}_{k}\,\right]^{*}=H_{\text{c}}^{*}.
\end{align}
Therefore the canonical Hamiltonian is invariant under the composition of the unitary transformation $W$ and the anti-unitary transformation $K$, the complex conjugation. Moreover, both transformations commute with each other (the unitary transformation $W$ consists in a permutation of the elements of the position basis, so that it corresponds to a real matrix in such representation with only $0$'s and $1$'s as entries), in this way:
\begin{equation}
(WH_{\text{c}}W^{\dagger})^{*}=WH_{\text{c}}^{*}W^{\dagger}=H_{\text{c}}.
\end{equation}
As a consequence, if a certain vector $\ket{e}$ is an eigenstate of the Hamiltonian with a corresponding energy $E$ the transformed state $W\ket{e}^{*}$ is also an eigenstate with the same energy.
For non degenerate states, as the edge modes, this means that the transformed state is proportional to itself. Since the transformation preserves the norm, the transformed state is equal to itself up to a phase:
\begin{equation}
V\ket{e}^{*}=e^{i\omega}\ket{e}.
\end{equation}
The phase $\omega$ can be absorbed into the state by redefining $\ket{e}=e^{i\omega/2}\ket{e}$, and thus we can consider without loss of generality that:
\begin{equation}
V\ket{e}^{*}=\ket{e}.
\end{equation}
Being $\psi_{a}(n)$ and $\psi_{b}(n)$ the two components of the edge mode wave function, that is:
\begin{equation}
\ket{e}=\sum_{n}
\begin{pmatrix}
\hat{a}_{n}^{\dagger} & \hat{b}_{n}^{\dagger}
\end{pmatrix}
\begin{pmatrix}
\psi_{a}(n)\\
\psi_{b}(n)
\end{pmatrix}\ket{0},
\end{equation}
the transformed state is:
\begin{equation}
W\ket{e}^{*}=\sum_{n}\begin{pmatrix}
\hat{a}_{n}^{\dagger} & \hat{b}_{n}^{\dagger}
\end{pmatrix}
\begin{pmatrix}
\,\psi_{b}^{*}(N+1-n)\,\\
\,\psi_{a}^{*}(N+1-n)\,
\end{pmatrix}\ket{0},
\end{equation}
and thus the IRC symmetry implies:
\begin{equation}
\psi_{b}(n)=\psi_{a}^{*}(N+1-n).
\end{equation}
As a result the edge state can be expressed in terms of just one wave function $\psi(n)$ as:
\begin{equation}\label{eq:C6N01}
\ket{e}=\frac{1}{\sqrt{2}}\sum_{n}
\begin{pmatrix}
\hat{a}_{n}^{\dagger} & \hat{b}_{n}^{\dagger}
\end{pmatrix}
\begin{pmatrix}
\,\psi(n)\,\\
\,\psi^{*}(N+1-n)\,
\end{pmatrix}\ket{0}.
\end{equation}

\subsubsection{Chiral symmetry and edge modes polarization}

The presence of chiral symmetry makes all eigenstates come in pairs of opposite energy, being the two eigenstates of each pair connected by the chiral operator $U_{S}$. That is, being the edge mode $\ket{e}$ in Eq.(\ref{eq:C6N01}) an eigenstate of the bowtie ladder Hamiltonian with some energy $E$, then the state $U_{S}\ket{e}$ is another eigenstate with energy $-E$. In the case of the bowtie ladder Hamiltonian the chiral operator is $U_{S}=\sigma_{z}$, therefore the two edge modes are:
\begin{equation}\label{eq:EdgeModesWF1}
\ket{e_{\pm}}=\frac{1}{\sqrt{2}}\sum_{n}
\begin{pmatrix}
\hat{a}_{n}^{\dagger} & \hat{b}_{n}^{\dagger}
\end{pmatrix}
\begin{pmatrix}
\,\psi(n)\,\\
\,\pm\psi^{*}(N+1-n)\,
\end{pmatrix}\ket{0},
\end{equation}
which satisfy: $U_{S}\ket{e_{\pm}}=\ket{e_{\mp}}$.

The edge modes are located at the edges of the system and the most general wave function $\psi(n)$ can be decompose into two different parts:
\begin{equation}
\psi(n)=\psi_{l}(n)+\psi_{r}(n),
\end{equation}
being $\psi_{l}(n)$ and $\psi_{r}(n)$ located at the left and right ends of the ladder, respectively. In this way, we can decompose the edge modes in the following way:
\begin{equation}\label{eq:EdgeStatesDecomp}
\ket{e_{\pm}}=\frac{1}{2}\Big(\ket{l_{\pm}}+\ket{r_{\pm}}\Big),
\end{equation}
where:
\begin{align}
&\ket{l_{\pm}}=\sum_{n}\begin{pmatrix}\hat{a}_{n}^{\dagger} & \hat{b}_{n}^{\dagger}\end{pmatrix}
\begin{pmatrix}
\,\psi_{l}(n)\,\\
\,\pm\psi_{r}^{*}(N+1-n)\,
\end{pmatrix}\ket{0},\\
&\ket{r_{\pm}}=\sum_{n}\begin{pmatrix}\hat{a}_{n}^{\dagger} & \hat{b}_{n}^{\dagger}\end{pmatrix}
\begin{pmatrix}
\,\psi_{r}(n)\,\\
\,\pm\psi_{l}^{*}(N+1-n)\,
\end{pmatrix}\ket{0},
\end{align}
so that the edge modes would be made from four different components: two located at the left end of the ladder, $\ket{l_{\pm}}$, and another two at the right edge, $\ket{r_{\pm}}$.

On the other hand, apart from being spatially located at the ends of the ladder, the edge modes are chacterized by the property of having almost zero energy. Furthermore, we can consider regions in the parameter space in which the energy of the edge modes is arbitrary close to zero so that we can make the approximation of zero energy, which is equivalent to consider the size of the system to be infinite compared to the spatial extension of the edge modes wave packets. In this situation the edge modes are annihilated by the Hamiltonian:
$H_{\text{c}}\ket{e_{\pm}}=0$.
However, the Hamiltonian is a local operator, as it connects each site in the ladder with itself and the sites within the same and the adjacent unit cells. As a consequence, the edge modes can be eigenstates of the Hamiltonian with zero energy only if the two components from which they are formed are also annihilated by the Hamiltonian, since they are separated by an arbitrary long distance. That is:
\begin{equation}
H_{\text{c}}\ket{e_{\pm}}=0\implies H_{\text{c}}\ket{l_{\pm}}=H_{\text{c}}\ket{r_{\pm}}=0,
\end{equation}
which would imply that the four states $\ket{l_{+}}$, $\ket{l_{-}}$ ,$\ket{r_{+}}$ and $\ket{r_{-}}$ are zero energy eigenstates of the Hamiltonian. In this situation the two original edge modes of the system would split into four different edge modes, what has no sense. The origin of this contradiction is found in the assumption that the wavefunction $\psi(n)$ has two components located at each edge of the ladder. Therefore, we conclude that it must be located at one particular edge.
For instance, let us consider the case in which it is localized at the left side of the system. Then, the edge modes would be written as:
\begin{equation}
\ket{e_{\pm}}=\frac{1}{\sqrt{2}}\Big(\ket{l}\pm\ket{r}\Big),
\end{equation}
where $\ket{l}$ and $\ket{r}$ are states localized at the left and right edges of the system, respectively:
\begin{align}
&\ket{l}=\sum_{n}\,\psi(n)\,\hat{a}_{n}^{\dagger}\ket{0}\label{eq:EdgeModeSpatialL}\\
&\ket{r}=\sum_{n}\,\psi^{*}(N+1-n)\,\hat{b}_{n}^{\dagger}\ket{0}\label{eq:EdgeModeSpatialR}.
\end{align}

\begin{figure}[t]
  \centering
    \includegraphics[width=0.485\textwidth]{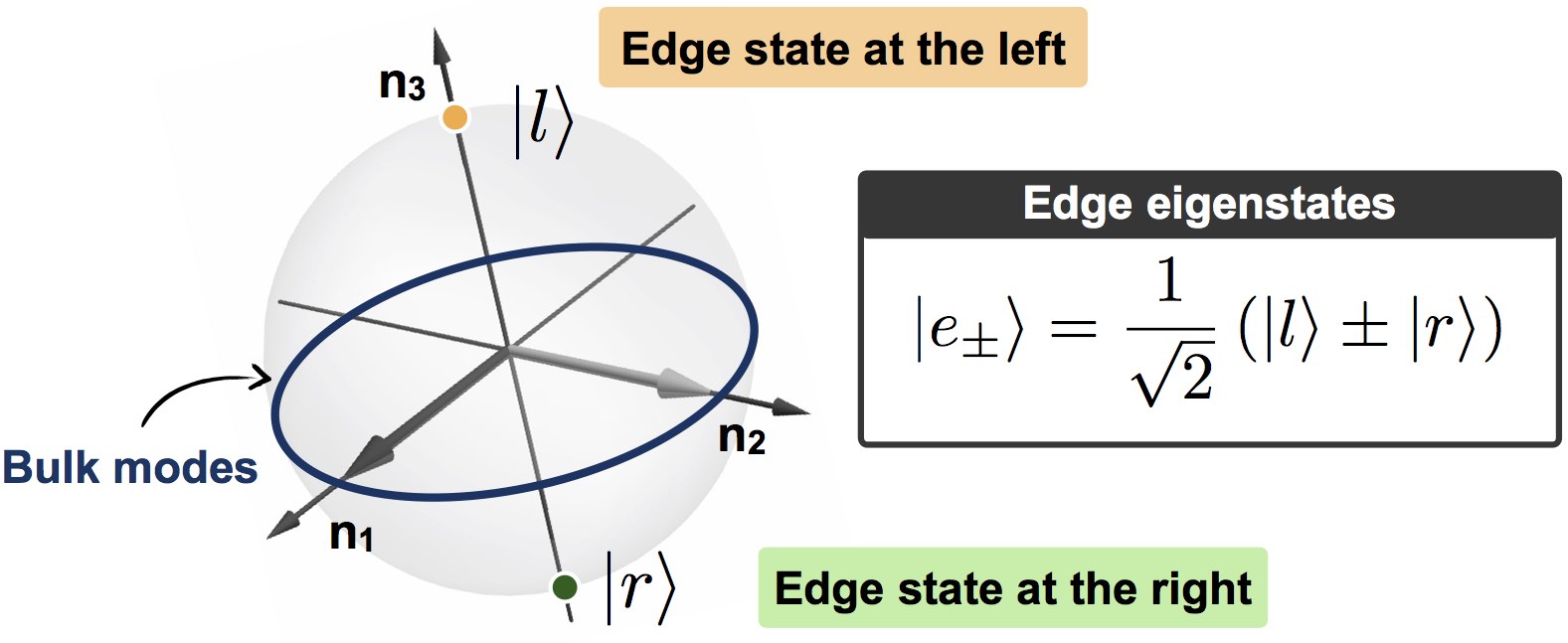}
    \caption[Edge modes polarization]{\textbf{Edge modes polarization.} The edge eigenstates of the Hamiltonian, $\ket{e_{\pm}}$, consist of the symmetric and antisymmetric superpositions of the edge states $\ket{l}$ and $\ket{r}$, which are localized at the left and right edges of the system. The isospin state of these two edge states are the ones given by the eigenvectors of the chiral operator. This correlation between the isospin of each edge modes and the edge where it is located is what we call the edge modes polarization. If we place all eigenstates of the Hamiltonian in the Bloch sphere, the bulk modes are on the equator and the edge modes are at the poles; being the equator the circumference formed by the intersection of the sphere and the plane where the Hamiltonian matrix lives, so that the poles correspond to the direction of the chiral operator $U_{S}=\bm{n}_{3}\cdot\bm{\sigma}$.}
    \label{fig:FiguraLadder19}
\end{figure}

In conclusion, the edge states which are exact eigenstates of the Hamiltonian, $\ket{e_{\pm}}$, are composed of two edge states, $\ket{l}$ and $\ket{r}$, which are localized at a single edge of the system. In the zero energy approximation, and due to the local nature of the Hamiltonian, the edge modes $\ket{l}$ and $\ket{r}$ can be considered to be also eigenstates of the system. The presence of chiral symmetry implies that the two edge exact eigenstates $\ket{e_{\pm}}$ must be transformed by the chiral operator into each other. This means that the subspace spanned by the edge eigenstates $\ket{e_{\pm}}$, which is the same as the one spanned by the two single-edge modes $\ket{l}$ and $\ket{r}$, is invariant under the chiral operator. Since the single-edge modes are separated by an arbitrary long distance and the chiral operator acts locally, the only possible situation is that they are eigenstates of the chiral operator. Let us choose the edge mode at the left size of the system to be the one corresponding to the eigenvalue $+1$ and the one at the right to the eigenvalue $-1$, that is:
\begin{align}
&U_{S}\,\ket{l}=\ket{l},\\
&U_{S}\,\ket{r}=-\ket{r}.
\end{align}
So that the isospin state of each single-edge mode, i.e. the vector that corresponds to the $a-b$ internal degree of freedom, corresponds to one of the two orthogonal eigenvectors of the chiral operator, which stands for a \textit{polarization} property of the edge modes (see Fig.~\ref{fig:FiguraLadder19}).
In the case of the bowtie ladder, for which $U_{S}=\sigma_{z}$, this means that the left edge mode occupies only $a$-sites in the ladder, whereas the right edge mode occupies only $b$-sites.
The edge eigenstates of the Hamiltonian are then formed as the symmetric and antisymmetric superpositions of the edge modes $\ket{l}$ and $\ket{r}$, see Eq.~\ref{eq:EdgeStatesDecomp}, and fulfil the relation $U_{S}\,\ket{e_{\pm}}=\ket{e_{\mp}}$.

Finally, we could ask ourselves what decides the particular polarization of the edge modes. That is, when does the edge mode located at the left end of the system occupy $a$-modes and when $b$-modes? It turns out that it depends on the two diagonal couplings of the bowtie ladder. For $t>t^{\prime}$ the edge mode at the left occupies $a$-modes and the one at the right $b$-modes, whereas the polarization is the opposite for $t<t^{\prime}$. In the first situation, when $t>t^{\prime}$, we can get arbitrary close to the limit case $t^{\prime}=0$. At that point the ladder becomes a dimerized $(a-b)$ chain, where the left edge is an $a$-site and the right edge a $b$-site. In contrast, for $t<t^{\prime}$, the ladder can be continuously connected to an $(b-a)$ chain, where the left edge is a $b$-site and the right one an $a$-site. Both situations cannot be continuously connected without a phase transition, as the point $t=t^{\prime}$ corresponds to a trivial topology, for which there are no edge modes (see Sec.$5$).

\subsubsection{Zero energy approximation}

A very good approximation of the edge modes wave function can be obtained by considering them to be zero energy eigenstates of the Hamiltonian.

We start by separating the bowtie ladder Hamiltonian for periodic boundary conditions, $H_{\text{prdc}}$, into two parts: the bowtie ladder Hamiltonian for open boundary conditions, $H_{\text{open}}$, and an edge Hamiltonian, $H_{\text{edge}}$. That is:
\begin{equation}
H_{\text{prdc}}=H_{\text{open}}+H_{\text{edge}},
\end{equation}
being:
\begin{equation}
H_{\text{edge}}=-te^{i\delta}\,\hat{a}_{1}^{\dagger}\hat{b}_{N}^{}-t^{\prime}e^{i\delta}\,\hat{b}_{1}^{\dagger}\hat{a}_{N}^{}+\text{H.c.}
\end{equation}
We consider the edge modes in Eq.~(\ref{eq:EdgeModesWF1}) and impose that they are zero energy eigenstates of the Hamiltonian for open boundary conditions:
\begin{equation}
H_{\text{open}}\ket{e_{\pm}}\approx0,
\end{equation}
so that, in terms of the Hamiltonian for periodic boundary conditions and the edge Hamiltonian, we have:
\begin{equation}\label{eq:EdgeModesHperiodicHedge}
H_{\text{prdc}}\ket{e_{\pm}}-H_{\text{edge}}\ket{e_{\pm}}\approx0.
\end{equation}
First, we compute the result of applying the Hamiltonian for periodic boundary conditions to the edge modes, $H_{\text{prdc}}\ket{e_{\pm}}$. For that we write the edge modes in the momentum representation, which is more convenient inasmuch as the Hamiltonian for periodic boundary conditions is easily written in such basis. Defining the function $F(k)$ as:
\begin{equation}
F(k)=\frac{1}{\sqrt{N}}\sum_{n=1}^{N}e^{-ikn}\psi(n),
\end{equation}
the edge modes can be written as:
\begin{equation}
\ket{e_{\pm}}=\frac{1}{\sqrt{2}}\sum_{k}
\begin{pmatrix}
\hat{a}_{k}^{\dagger} & \hat{b}_{n}^{\dagger}
\end{pmatrix}
\begin{pmatrix}
\,F(k)\,\\
\,\pm e^{-ik}\,F^{*}(k)\,
\end{pmatrix}\ket{0}.
\end{equation}
The bowtie ladder Hamiltonian for periodic boundary condition is:
\begin{equation}
H_{\text{prdc}}=-\sum_{k}\rho(k)
\begin{pmatrix}
\hat{a}_{k}^{\dagger} & \hat{b}_{n}^{\dagger}
\end{pmatrix}
\begin{pmatrix}
0  & e^{-i\varphi(k)}\\
e^{i\varphi(k)} & 0
\end{pmatrix}
\begin{pmatrix}
\,\hat{a}_{k}\,\\
\,\hat{b}_{k}\,
\end{pmatrix},
\end{equation}
so that the result of applying it to the edge modes is:
\begin{align}\label{eq:EdgeModesHperiodic}
&H_{\text{prdc}}\ket{e_{\pm}}=\nonumber\\
&-\frac{1}{\sqrt{2}}\sum_{k}\rho(k)
\begin{pmatrix}
\hat{a}_{k}^{\dagger} & \hat{b}_{n}^{\dagger}
\end{pmatrix}
\begin{pmatrix}
\,\pm e^{-ik}e^{-i\varphi(k)}F^{*}(k)\,\\
\,e^{i\varphi(k)}\,F(k)\,
\end{pmatrix}\ket{0}.
\end{align}
Secondly, we compute the result of applying the edge Hamiltonian to the edge modes, $H_{\text{edge}}\ket{e_{\pm}}$. For that we need to take into account the polarization of the edge modes, that is, the edge mode at the left size of the system occupies only $a$-modes, whereas the one at the right occupies only $b$-modes. Therefore, only two sites are affected by the edge Hamiltonian, $a_{1}$ and $b_{N}$, and we get:
\begin{equation}
H_{\text{edge}}\ket{e_{\pm}}=-\frac{t}{\sqrt{2}}\Big[ e^{-i\delta}\,\psi(1)\,\hat{b}_{N}^{\dagger}\pm e^{i\delta}\,\psi^{*}(1)\,\hat{a}_{1}^{\dagger}\Big]\ket{0},
\end{equation}
which in momentum representation takes the form:
\begin{equation}\label{eq:EdgeModesHedge}
H_{\text{edge}}\ket{e_{\pm}}=\frac{t}{\sqrt{2N}}\sum_{k}
\begin{pmatrix}
\hat{a}_{k}^{\dagger} & \hat{b}_{n}^{\dagger}
\end{pmatrix}
\begin{pmatrix}
\,\pm e^{-ik}e^{i\delta}\psi^{*}(1)\,\\
\,e^{-i\delta}\,\psi(1)\,
\end{pmatrix}\ket{0}.
\end{equation}
From (\ref{eq:EdgeModesHperiodicHedge}), (\ref{eq:EdgeModesHperiodic}) and (\ref{eq:EdgeModesHedge}) we obtain:
\begin{equation}
F(k)=-\frac{1}{\sqrt{N}}\,t\,e^{-i\delta}\,\psi(1)\,\frac{e^{-i\varphi(k)}}{\rho(k)},
\end{equation}
and thus the edge modes are easily written in momentum representation as:
\begin{align}
&\ket{l}=\frac{1}{\sqrt{\kappa}}\sum_{k}\frac{e^{-i\varphi(k)}}{\rho(k)}\,\hat{a}_{k}^{\dagger}\ket{0},\label{eq:EdgeModeL}\\
&\ket{r}=\frac{1}{\sqrt{\kappa}}\sum_{k}\frac{e^{-ik}e^{i\varphi(k)}}{\rho(k)}\,\hat{b}_{k}^{\dagger}\ket{0},\label{eq:EdgeModeR}
\end{align}
being $\kappa$ a normalization constant.

\subsection{Edge states general properties}

Once we have obtained a quite good approximation for the edge modes wave functions, Eq.~(\ref{eq:EdgeModeL}) and Eq.~(\ref{eq:EdgeModeR}), we can derive their more relevant properties. These are:

\begin{itemize}
\item
The edge states are localized in momentum space at the positions of the energy gaps between the two energy bands.

\item
The edge states are located in position space at the ends of the ladder.

\item
The edge states can be simultaneously well localized in momentum and position spaces.
\item
The symmetry class of a topological ladder model is manifested through the edge states momentum distribution.
\end{itemize}

\subsubsection{Edge modes in momentum space}

The momentum density distribution of the edge states can be easily obtained from their wave functions, Eq.(\ref{eq:EdgeModeL}) and Eq.(\ref{eq:EdgeModeR}). Both states have the same momentum density distribution:
\begin{equation}
\langle \hat{n}_{k}\rangle=\frac{1}{\kappa\,\rho^{2}(k)},\label{eq:MomentumDistributionGeneral}
\end{equation}
where $\hat{n}_{k}=\hat{a}_{k}^{\dagger}\hat{a}_{k}+\hat{b}_{k}^{\dagger}\hat{b}_{k}$. Due to the polarization property of the edge states we know that, in the particular case of the bowtie ladder, the edge state located at the left end of the system occupies only $a$-modes, whereas the one at the right end occupies only $b$-modes. Therefore:
\begin{align}
&\langle l|\,\hat{a}_{k}^{\dagger}\hat{a}_{k}\ket{l}=\langle r|\,\hat{b}_{k}^{\dagger}\hat{b}_{k}\ket{r}=\frac{1}{\kappa\,\rho^{2}(k)}\\
&\langle r|\,\hat{a}_{k}^{\dagger}\hat{a}_{k}\ket{r}=\langle l|\,\hat{b}_{k}^{\dagger}\hat{b}_{k}\ket{l}.
\end{align}
In general, for an arbitrary topological ladder model, the edge state located at each end of the system occupies one of the two orthogonal superpositions of $a$-modes and $b$-modes corresponding to each eigenvector of the chiral operator $U_{S}$. Nevertheless, the momentum distribution $\langle \hat{n}_{k}\rangle$, understood as the probability for a particle that occupies a certain state to be found with the momentum value $k$, regardless of its isospin, depends only on the function $\rho(k)$, that is, the energy bands of the system. In this way, we only need to analyse Eq.~(\ref{eq:MomentumDistributionGeneral}) in order to learn about the properties of the edge states in momentum space.

The maxima of the momentum density distribution of the edge modes, $\langle \hat{n}_{k}\rangle$, will be located at the minima of the function $\rho(k)$. Being $q_{j}$ a minimum of $\rho(k)$, which correspond to the location of an energy gap between the two bands, we can make a second order expansion of $\rho(k)$ around that minimum:
\begin{equation}\label{eq:2OrderExpansionRho}
\rho(k)\approx\frac{E_{j}}{2}+\frac{(k-q_{j})^{2}}{2m_{j}},
\end{equation}
begin $E_{j}$ the gap width and $m_{j}$ the mass of the Wilson fermion associated to that gap, which is given by:
\begin{equation}
\frac{1}{m_{j}}=\left.\frac{d^{2}\rho(k)}{dk^{2}}\right |_{k=q_{j}}.
\end{equation}
In this way we can approximate the edge states momentum distribution around $q_{j}$ by:
\begin{equation}\label{eq:MomentumDensityDistributionApprox}
\langle \hat{n}_{k}\rangle\approx \frac{4}{\kappa\,E_{j}^{2}}\left[ 1+\Big(\frac{k-q_{j}}{\xi_{j}}\Big)^{2} \right]^{-2},
\end{equation}
being $\xi_{j}=\sqrt{m_{j}\,E_{j}}$. As we see, the edge states momentum distribution takes the form of the square of a Cauchy distribution around each energy gap and, therefore, we can conclude that the edge states are strongly localized in momentum space at the position of the energy gaps of the system.

The edge states momentum distribution shows a peak for each energy gap between the two bands, being $4/\kappa\,E_{j}^{2}$ the height of each peak and $\delta k_{j}=2\xi_{j}\sqrt{\sqrt{2}-1}$ its corresponding FWHM. We can obtain the probability associated to each peak by making an approximation and integrating over a continuum momentum
\begin{equation}
\text{prob}_{j}\approx\int_{-\infty}^{\infty}dk\frac{4}{\kappa\,E_{j}^{2}}\left[ 1+\Big(\frac{k-q_{j}}{\xi_{j}}\Big)^{2} \right]^{-2}=\frac{2\pi}{\kappa}\sqrt{\frac{m_{j}}{E_{j}^{3}}}.
\end{equation}
As a consequence, taking into account the fact that the energy bands correspond to the distance from the origin to the points of an ellipse, in case there are two different energy gaps between the bands, and thus two different peaks in the edge states momentum distribution, the probability associated to the peak placed at the smallest energy gap momentum will be larger.

As a result, we can establish a correspondence between the edge states present in a topological ladder model and the Wilson fermions described by such model, according to which:
\begin{itemize}
\item
The number of Wilson fermions is equal to the number of momentum components that constitute the edge modes.
\item
The momenta at which the Wilson fermions arise correspond to the location of the edge modes momentum components.
\item
The masses of the Wilson fermions correspond to the relative weights of the edge modes momentum components, in such a way that:\\
\begin{enumerate}[label=(\roman*)]
\item
If two Wilson fermions have the same masses, the corresponding momentum components contribute equally to the edge modes.
\item
If two Wilson fermions have different masses, the momentum peak associated to the lightest fermion is better defined, so that its corresponding component is predominant in the edge states wave function.
\end{enumerate}
\end{itemize}

\subsubsection{Edge modes in position space}

In the case of the position space, each edge sate shows a different density distribution. However, they are not independent from each other, as the wave function of the edge state $\ket{r}$ can be obtained by complex conjugating and reflecting the wave function of the edge state $\ket{l}$, see Eq.~(\ref{eq:EdgeModeSpatialL}) and Eq.~(\ref{eq:EdgeModeSpatialR}). We have:
\begin{align}
&\langle l|\,\hat{n}_{x}\ket{l}=|\psi(x)|^{2}\label{eq:PositionDenDistL}\\
&\langle r|\,\hat{n}_{x}\ket{r}=|\psi(N+1-x)|^{2},\label{eq:PositionDenDistR}
\end{align}
being $\hat{n}_{x}=\hat{a}_{x}^{\dagger}\hat{a}_{x}+\hat{b}_{x}^{\dagger}\hat{b}_{x}$ the particle number operator at position $x$, regardless of the isospin state. In the same way that occurs in momentum space, in position space the two edge states occupy different orthogonal isospin states, given by the chiral operator. Nevertheless, the position density distributions in Eq.~(\ref{eq:PositionDenDistL}) and Eq.~(\ref{eq:PositionDenDistR}) are valid for any ladder model.

We first consider the edge state located at the left side of the system, Eq.~(\ref{eq:EdgeModeL}). As we already know, it is mostly located in momentum space around the energy gap momenta and, therefore, we can decompose such edge mode in several momentum components. For that, we need to make an expansion of the function $e^{-i\varphi(k)}$ around $q_{j}$, being $q_{j}$ a minimum of $\rho(k)$. The value of $\varphi(k)$, for $k$ close to $q_{j}$, will be equal to $\varphi(q_{j})$ plus some phase $\delta\varphi_{j}$, and thus:
\begin{equation}
e^{-i\varphi(k)}=e^{-i\left[\varphi(q_{j})+\delta\varphi_{j}\right]}=e^{-i\varphi(q_{j})}(\cos\delta\varphi_{j}-i\sin\delta\varphi_{j})
\end{equation}
\begin{figure}[t]
  \centering
    \includegraphics[width=0.485\textwidth]{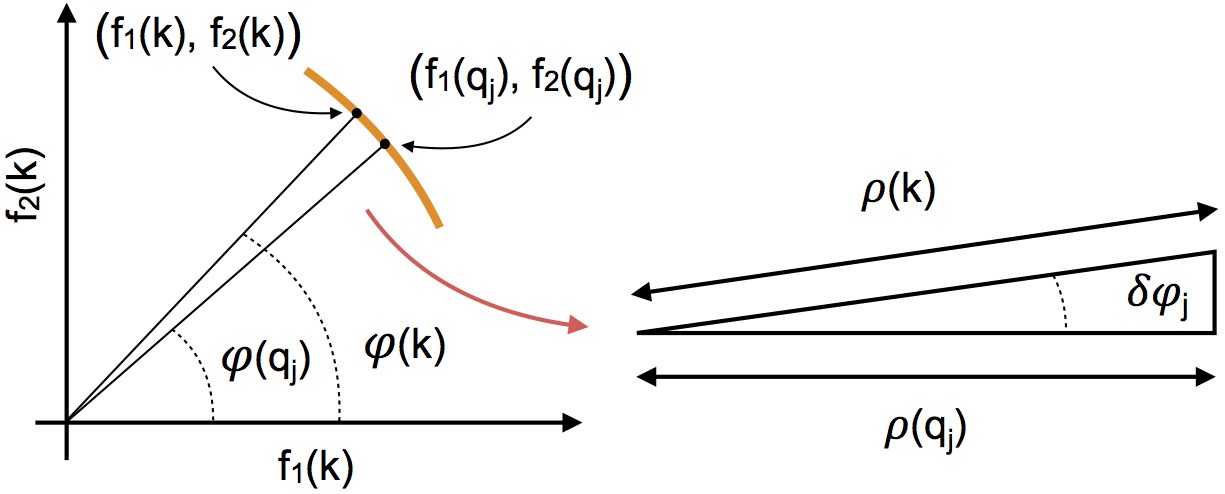}
   \caption[Function $\varphi(k)$ around the momentum of an energy gap.]{\textbf{Function $\varphi(k)$ around the momentum of an energy gap.} For a momentum $k$ close to $q_{j}$, being $q_{j}$ the location of an energy gap between the two bands can be written as $\varphi(q_{j})$ plus some variation $\delta\varphi_{j}$, which can be computed in terms of $\rho(k)$ and $\rho(q_{j})$ (see text).}
    \label{fig:C6N02} 
\end{figure}
From Fig.~\ref{fig:C6N02} we can compute first and second order approximations for the sine and cosine of the variation of $\varphi(k)$ with respect to $\varphi(q_{j})$. Taking into account that $\rho(q_{j})=E_{j}/2$ and using the second order expansion of $\rho(k)$ around $q_{j}$ in Eq.~(\ref{eq:2OrderExpansionRho}) we obtain:
\begin{align}
&\cos\delta\varphi_{j}\approx1-\left(\frac{k-q_{j}}{\xi_{j}}\right)^{2}\\
&\sin\delta\varphi_{j}\approx\sqrt{2}\,\left(\frac{k-q_{j}}{\xi_{j}}\right)
\end{align}

By using this expansion of $\varphi(k)$ at each energy gap momentum we can write the edge mode $\ket{l}$ as a sum of different momentum components:
\begin{equation}
\ket{l}=\sum_{j}\ket{l_{j}},
\end{equation}
running $j$ over the number of energy gaps of the system and being:
\begin{equation}\label{eq:EdgeModeComponentMomentum}
\ket{l_{j}}\approx\frac{2e^{-i\varphi(q_{j})}}{\sqrt{\kappa}}\sum_{k}\frac{1-\left(\frac{k-q_{j}}{\xi_{j}}\right)^{2}-i\sqrt{2}\left(\frac{k-q_{j}}{\xi_{j}}\right)}{E_{j}+\left(\frac{k-q_{j}}{\xi_{j}}\right)^{2}}\,\hat{a}^{\dagger}_{k}\ket{0}.
\end{equation}
We compute the inverse Fourier transform of this momentum wave function and thus obtain the component $\ket{l_{j}}$ in the position representation:
\begin{equation}
\ket{l_{j}}\approx\frac{\sqrt{2}(\sqrt{2}-1)}{\sqrt{\kappa}}\sqrt{\frac{m_{j}}{E_{j}}}\,e^{-i\varphi(q_{j})}\sum_{n}\,e^{iq_{j}n}\,e^{-\xi_{j}n}\,\hat{a}_{n}^{\dagger}\,\ket{0}.
\end{equation}
As we see, the spatial wave function consists of two factors: $e^{iq_{j}n}$, which means that the momentum distribution is centred at the value $q_{j}$, as we already new, and $e^{-\xi_{j}n}$, which is a decaying exponential whose maximum value corresponds to the position $n=1$. The spatial density distribution of such state is then:
\begin{equation}\label{eq:PositionDensityDistributionApproxL}
\bra{l_{j}}\hat{n}_{x}\ket{l_{j}}\approx\frac{2(3-2\sqrt{2})}{\kappa}\,\frac{m_{j}}{E_{j}}\,e^{-2\xi_{j}n},
\end{equation}
Analogously, the edge mode $\ket{r}$ can also be decomposed in different momentum modes:
\begin{equation}
\ket{r}=\sum_{j}\ket{r_{j}},
\end{equation}
being:
\begin{align}
\ket{r_{j}}\approx\frac{\sqrt{2}(\sqrt{2}-1)}{\sqrt{\kappa}}\sqrt{\frac{m_{j}}{E_{j}}}\,e^{i\varphi(q_{j})}\,e^{-iq_{j}(N+1)}\nonumber\\
\sum_{n}\,e^{iq_{j}n}\,e^{-\xi_{j}(N+1-n)}\,\hat{b}_{n}^{\dagger}\,\ket{0},
\end{align}
since we know that the spatial wave function of $\ket{r}$ can be obtained by complex conjugating and reflecting the wave function of $\ket{l}$, see Eq.~(\ref{eq:EdgeModeSpatialL}) and Eq.~(\ref{eq:EdgeModeSpatialR}). The spatial density distribution of the component $\ket{r_{j}}$ is then:
\begin{equation}\label{eq:PositionDensityDistributionApproxR}
\bra{r_{j}}\hat{n}_{x}\ket{r_{j}}\approx\frac{2(3-2\sqrt{2})}{\kappa}\,\frac{m_{j}}{E_{j}}\,e^{-2\xi_{j}(N+1-n)}.
\end{equation}
In this way, we conclude that the edge states $\ket{l}$ and $\ket{r}$ are localized at the left and right edges of the system, decaying their spatial density distribution exponentially towards the bulk. Each edge mode component $\ket{l_{j}}$ and $\ket{r_{j}}$ has a FWHM of $\delta x_{j}=\log 2/(2\xi_{j})$.

\subsubsection{Simultaneous momentum-position localization}

From the approximated momentum and spatial density distributions of the edge states, Eq.~(\ref{eq:MomentumDensityDistributionApprox}), Eq.~(\ref{eq:PositionDensityDistributionApproxL}) and Eq.~(\ref{eq:PositionDensityDistributionApproxR}), and their corresponding localization lengths, $\delta k_{j}=2\xi_{j}\sqrt{\sqrt{2}-1}$ and $\delta x_{j}=\log 2/(2\xi_{j})$, we know that the larger an energy gap is, the better defined the position of the corresponding edge states component is and the worse defined its momentum is. However, there is a region in the parameter space for which both the momentum and the position of the edge modes can be simultaneously well defined.


\subsubsection{Symmetry class correspondence}

There exists a relation between the symmetry class of the Hamiltonian and the momentum of the symmetry protected edge modes that the system exhibits when it is found to be in its topologically non-trivial phase.

In this context, if a Hamiltonian $H$ presents timer reversal symmetry there is a global unitary transformation $U_{T}$ such that:
\begin{equation}
U_{T}\,H\,U^{\dagger}_{T}=H.
\end{equation}
Therefore, if a certain state $\ket{e}$ is an eigenstate of the Hamiltonian with some energy $E$, then the transformed state $U_{T}\ket{e}^{*}$ is also an eigenstate of the Hamiltonian with the same energy. In case of a non degenerate state, as the edge states, this means that this transformation leaves the state invariant up to a phase, which can be absorbed in the unitary operator $U_{T}$. That is:
\begin{equation}
U_{T}\ket{e}^{*}=\ket{e},
\end{equation}
and consequently, opposite momentum modes are on average equally occupied:
\begin{align}
\langle e|\,\hat{n}_{k}\,|e\rangle=&\langle e^{*}|\,\hat{n}_{-k}\,|e^{*}\rangle=\nonumber\\
&\langle e^{*}|\,U^{\dagger}_{T}\,\hat{n}_{-k}\,U_{T}\,|e^{*}\rangle=\langle e|\,\hat{n}_{-k}\,|e\rangle.
\end{align}
As a result, the momentum density distribution of the edge modes of a time reversal symmetric Hamiltonian is an even function of the momentum and, thus, such states have a zero average momentum.
In this way, there is a correspondence between the symmetry class of a topological ladder model and the momentum distribution of the edge modes that such model exhibits. The BDI class is characterized by edge modes with a symmetric momentum distribution, and thus with a zero average momentum, whereas the AIII class corresponds to edge modes with an asymmetric momentum distribution, or a nonzero average momentum.


\begin{figure*}
    \includegraphics[width=0.85\textwidth]{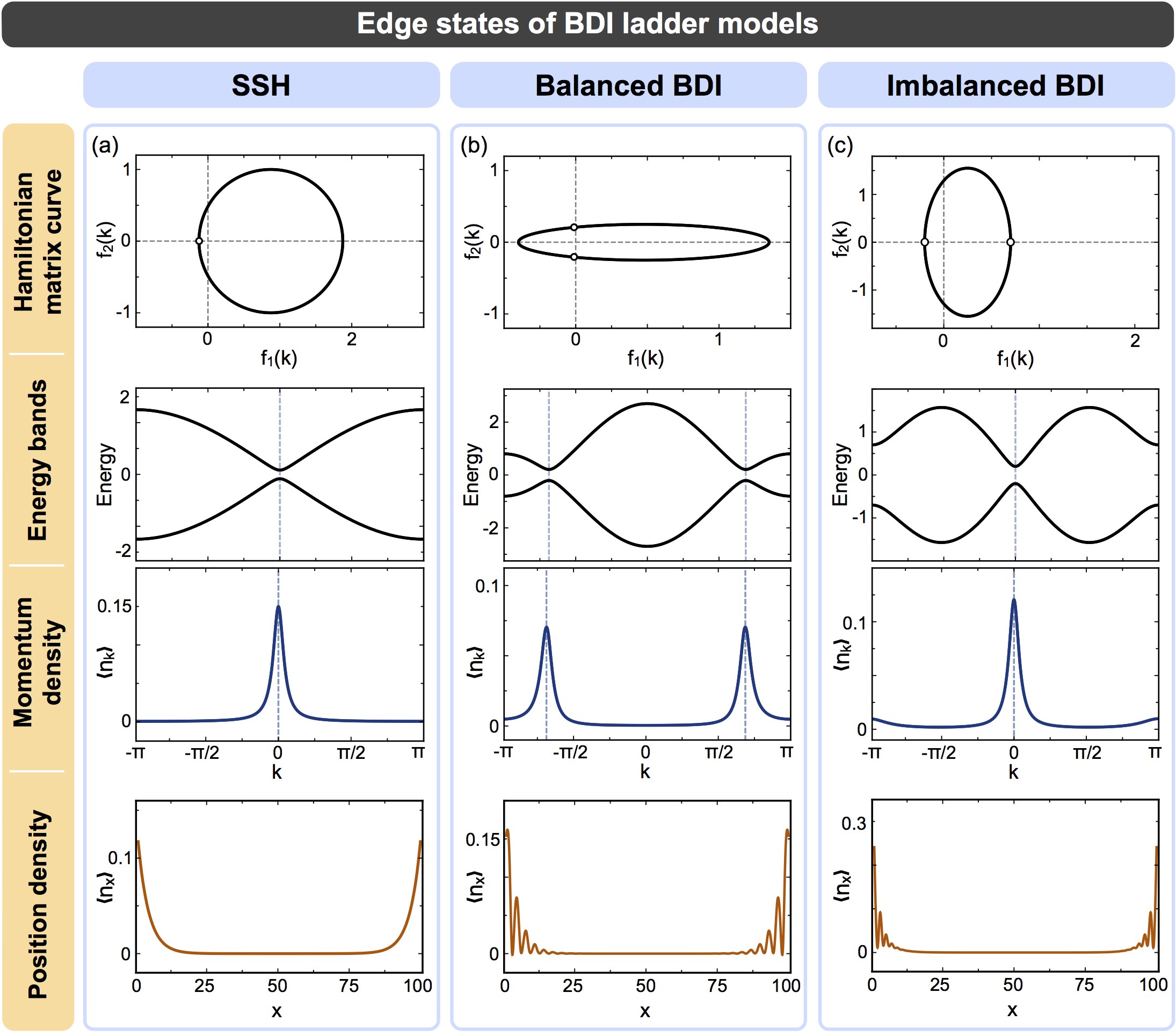}
    \caption[Edge states in the BDI class.]{\textbf{Edge states in the BDI class.} In this figure we show the Hamiltonian matrix curve, the energy bands, and the edges states momentum and position density distributions for each of the three different types of topological ladder models in the BDI class, namely: (a) the SSH model, (b) the balanced BDI model, and (c) the imbalanced BDI model. Each of them is characterized by the edge states momentum density distribution, which is related to the number and masses of the Wilson fermions described by the model. In this way, the edge states momentum density distribution shows a single peak at momentum $0$ or $\pi$ for the SSH model, two peaks of the same height located at opposite momenta for the balanced BDI model, and two peaks of different height located at momenta $0$ and $\pi$ for the imbalanced BDI model. The momentum and position density distributions we show here have been obtained by exact numerical diagonalization of the bowtie ladder Hamiltonian for parameters: $J=0.875$, $t=1$, $t^{\prime}=0$, $\delta=\pi$ and $\phi=0$ (SSH model); $J=0.95$, $t=1$, $t^{\prime}=0.75$, $\delta=0$ and $\phi=0$ (balanced BDI model); and $J=0.25$, $t=1$, $t^{\prime}=0.55$, $\delta=\pi/2$ and $\phi=\pi$ (imbalanced BDI model).}
    \label{fig:FiguraLadder21} 
\end{figure*}
\begin{figure*}
  \centering
    \includegraphics[width=0.85\textwidth]{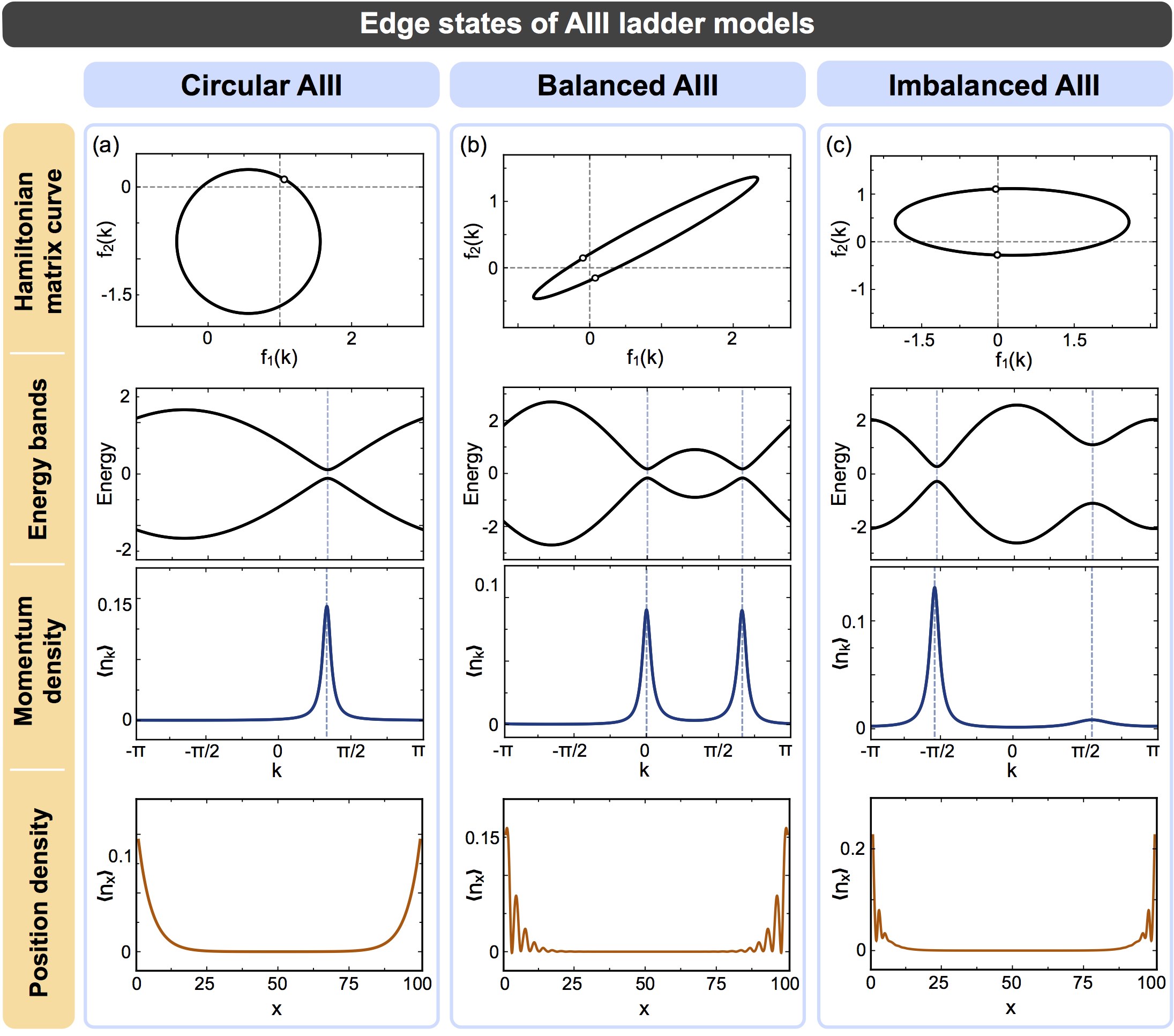}
   \caption[Edge states in the AIII class.]{\textbf{Edge states in the AIII class.}In this figure we show the Hamiltonian matrix curve, the energy bands, and the edges states momentum and position density distributions for each of the three different types of topological ladder models in the AIII class, namely: (a) the circular AIII model, (b) the balanced AIII model, and (c) the imbalanced  model. Each of them is characterized by the edge states momentum density distribution, which is related to the number and masses of the Wilson fermions described by the model. In this way, the edge states momentum density distribution shows a single peak at any momentum different from $0$ and $\pi$ for the cricular AIII model, two peaks of the same height located at not opposite momenta for the balanced AIII model, and two peaks of different height located at any momenta, excluding the situation in which one is located at momentum $0$ and the other at momentum $\pi$, for the imbalanced AIII model. The momentum and position density distributions we show here have been obtained by exact numerical diagonalization of the bowtie ladder Hamiltonian for parameters: $J=0.875$, $t=1$, $t^{\prime}=0$, $\delta=-2\pi/3$ and $\phi=0$ (circular AIII model); $J=0.9$, $t=1$, $t^{\prime}=0.8$, $\delta=-2\Pi/3$ and $\phi=0$ (balanced AIII model); and $J=0.5$, $t=1.5$, $t^{\prime}=0.8$, $\delta=0$ and $\phi=5\pi/8$ (imbalanced AIII model).}
    \label{fig:FiguraLadder22} 
\end{figure*}
\begin{figure}[h]
  \centering
    \includegraphics[width=0.485\textwidth]{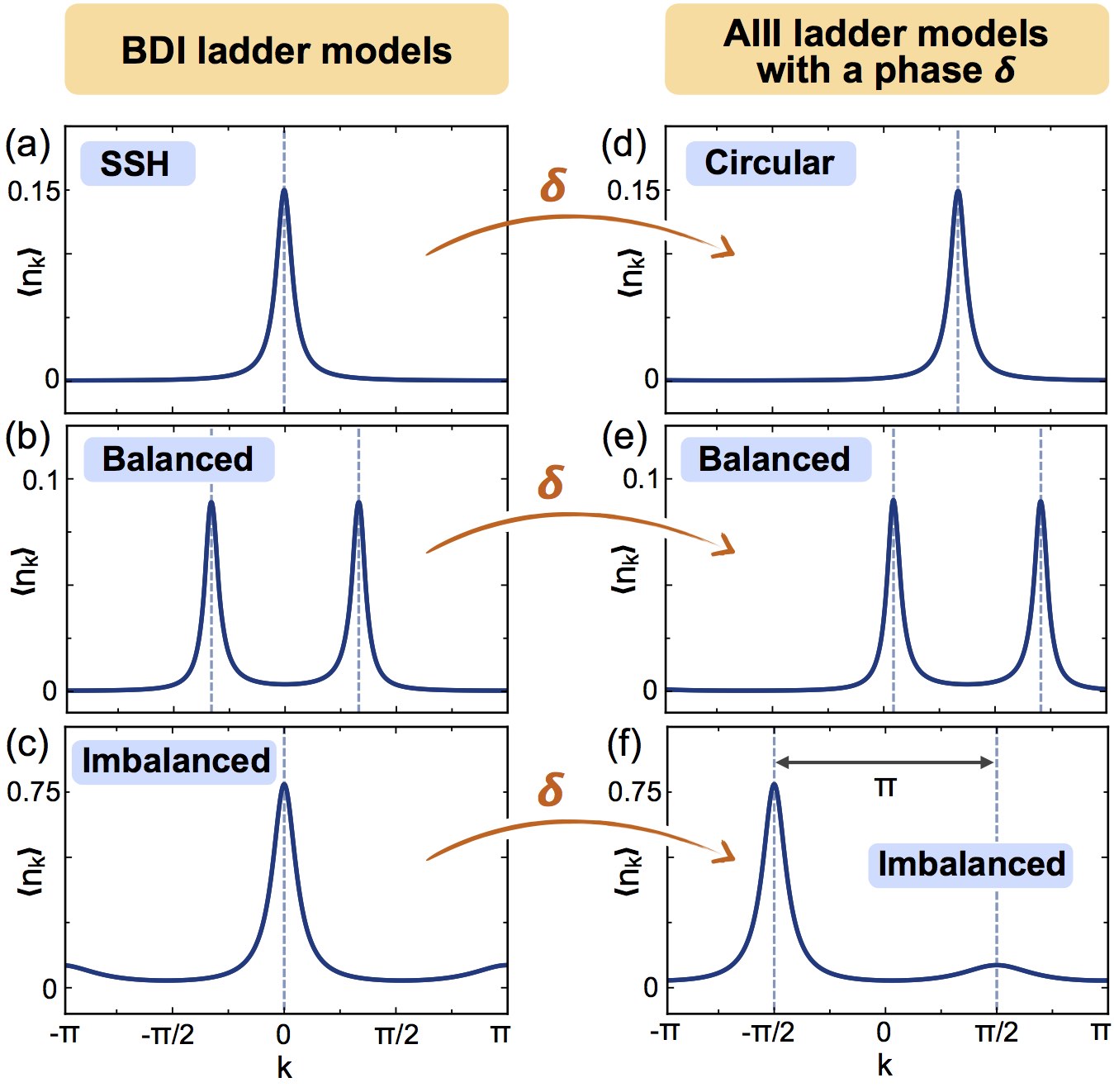}
    \caption[Connection between BDI and AIII ladder models.]{\textbf{Connection between BDI and AIII ladder models.} Adding a phase $\delta$ to ladder models in the BDI class, makes the model enter the AIII class and affects the edge states by shifting their momentum density distribution. In this way, the SSH model (a), the balanced BDI model (b) and the imbalanced BDI model (c) are connected in a one-to-one correspondence to the circular AIII model (d), the balanced AIII model (e) and a particular case of the imbalanced AIII model (f), respectively. The most general imbalanced AIII model, in which the edge states show a momentum distribution with two peaks separated by a distance different from $\pi$, has no correspondence in the BDI class and is realized by ladder configurations with an effective magnetic flux. Here we show the edge states momentum density distributions for different parameter configurations of the bowtie ladder Hamiltonian, corresponding each to one of the cases mentioned above. This parameters configurations are: (a) $J=0.875$, $t=1$, $t^{\prime}=0$, $\delta=\pi$ and $\phi=0$; (b) $J=0.9$, $t=1$, $t^{\prime}=0.8$, $\delta=\pi$ and $\phi=0$; (c) $J=0.2$, $t=1$, $t^{\prime}=0.55$, $\delta=\pi/2$ and $\phi=\pi$; (d) $J=0.875$, $t=1$, $t^{\prime}=0$, $\delta=-2\pi/3$ and $\phi=0$; (e) $J=0.9$, $t=1$, $t^{\prime}=0.8$, $\delta=-5\pi/8$ and $\phi=0$; and (f) $J=0.2$, $t=1$, $t^{\prime}=0.55$, $\delta=0$ and $\phi=\pi$. All momentum density distributions have been obtained by exact numerical diagonalization of the Hamiltonian.}
    \label{fig:FiguraLadder23} 
\end{figure}
\begin{figure*}
  \centering
    \includegraphics[width=0.8\textwidth]{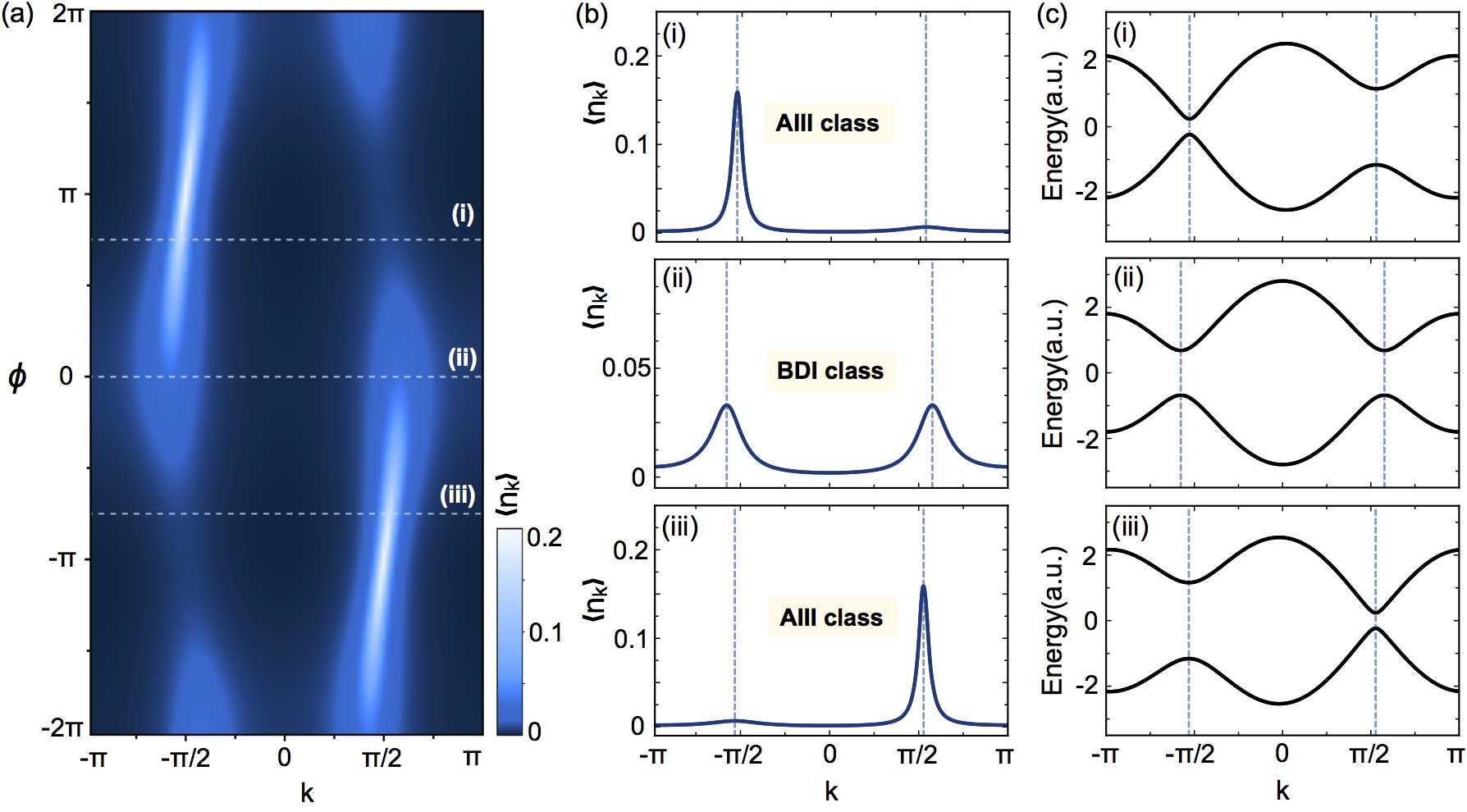}
    \caption[Edge states under an effective magnetic flux.]{\textbf{Edge states under an effective magnetic flux.} (a) Momentum density distribution of the edge states of the bowtie ladder Hamiltonian with parameters $J=0.5$, $t=1.5$, $t^{\prime}=0.8$ and $\delta=0$, as a function of the effective magnetic flux $\phi$. (b) Edge states momentum density distribution and (c) energy bands of the bowtie ladder Hamiltonian for the same parameter configurations and (i) $\phi=3\pi/4$, (ii) $\phi=0$, and (iii) $\phi=-3\pi/4$. The edge states momentum density distributions have been obtained by exact numerical diagonalization of the bowtie ladder Hamiltonian.}
    \label{fig:FiguraLadder24} 
\end{figure*}

\subsection{6 types of topological edge modes}

To conclude this chapter, we know that the edge states of a topological ladder model are determined by the two functions $\rho(k)$ and $\varphi(k)$, which are given by the Hamiltonian matrix curve, as well as by the chiral operator $U_{S}$. The exact edge eigenstates of a topological ladder Hamiltonian are $\ket{e_{\pm}}=(\ket{l}\pm\ket{r})/\sqrt{2}$, being:
\begin{align}
&\ket{l}=\frac{1}{\sqrt{\kappa}}\sum_{k}\frac{e^{-i\varphi(k)}}{\rho(k)}\,\hat{\psi}_{k}^{\dagger}\,\hat{u}_{+}\ket{0},\label{eq:EdgeModeGeneralChiralOperatorL}\\
&\ket{r}=\frac{1}{\sqrt{\kappa}}\sum_{k}\frac{e^{-ik}e^{i\varphi(k)}}{\rho(k)}\,\hat{\psi}_{k}^{\dagger}\,\hat{u}_{-}\ket{0},\label{eq:EdgeModeGeneralChiralOperatorR}
\end{align}
where $\hat{u}_{\pm}$ are the eigenvectors of the chiral operator with eigenvalues $\pm1$, expressed using the basis formed by the eigenvectors of $\sigma_{z}$, so that for $U_{S}=\sigma_{z}$ we have that $\hat{\psi}_{k}^{\dagger}\,\hat{u}_{+}=\hat{a}_{k}^{\dagger}$ and $\hat{\psi}_{k}^{\dagger}\,\hat{u}_{-}=\hat{b}_{k}^{\dagger}$.

The states $\ket{l}$ and $\ket{r}$ are localized at the left and right edges of the ladder, respectively, decaying their spatial density distributions exponentially towards the bulk. Their spatial localization depends on the quantities $\xi_{j}$, defined for each energy gap, but does not take into account the locations of the gaps. In consequence, we cannot tell any significant difference between the BDI class and the AIII class or between different types of ladder models, when we look at the edge states spatial density distribution. In contrast, it is through their momentum distribution how both symmetry classes can be distinguished, as well as different types of ladder models. The edge states are localized in momentum space at the positions of the energy gaps, being the relative weight of each momentum component determined by the corresponding gap width and the mass of the associated Wilson fermion. In this way, each of the six different types of topological ladder models corresponds to a distinct type of edge states, characterized by their momentum density distribution.\\

There are three different types of edge modes that can exist in a BDI topological ladder, corresponding each to a different way of obtaining a symmetric momentum density distribution, imposed by the presence of chiral symmetry. The SSH model has edge states with a single momentum component, which is located at $q=0$ or $q=\pi$, the only two possibilities for a zero average momentum [see Fig.~\ref{fig:FiguraLadder21}(a)]. The balanced BDI model has edge states of two equally weighted momentum components, therefore they are located at opposite momenta so that the average momentum is zero [see Fig.~\ref{fig:FiguraLadder21}(b)]. Finally, the imbalanced BDI model has edge states with two differently weighted momentum components. The presence of time reversal symmetry makes them be located at $q_{1}=0$ and $q_{2}=\pi$ [see Fig.~\ref{fig:FiguraLadder21}(c)].

In the AIII symmetry class there are three distinct types of edge states, one for each type of AIII ladder model. The circular AIII model has edge states with just one momentum component, located at momentum $q\neq0,\pi$, so that time reversal symmetry is broken [see Fig.~\ref{fig:FiguraLadder22}(a)]. The balanced AIII model has edge states with two equally weighted components, located at not opposite momenta, so that their average momentum is different from zero [see Fig.~\ref{fig:FiguraLadder22}(b)]. At last, the imbalanced AIII model has edge states with two components of different weights, which can be located at any momentum values, excluding the case in which one component has momentum $0$ and the other $\pi$ [see Fig.~\ref{fig:FiguraLadder22}(c)].\\

In Sec~$5$ we analysed how the bowtie ladder Hamiltonian can break time reversal symmetry and, thus, enter the AIII class. We concluded that the model needs at least one of the following two ingredients: a shift in the momentum-isospin correspondence introduced by the presence of a phase $\delta$, and an effective magnetic flux $\phi$. Furthermore, we know what type of AIII ladder models are obtained after adding a phase $\delta$ or an effective magnetic flux $\phi$ to each of the three types of BDI ladder models. That is, adding a phase $\delta$ to the SSH model, the balanced BDI model and the imbalanced BDI model results into the circular AIII model, the balanced AIII model and the imbalanced AIII model, respectively. However, the imbalanced AIII model that can be achieved by this means is just a particular case, in which the distance in momentum space between the two energy gaps remains constant and equal to $\pi$. The most general imbalanced AIII model corresponds to the case in which there is an effective magnetic flux $\phi$ penetrating the ladder.

This correspondence between BDI and AIII ladder models is quite evident when we look at the edge states. Introducing a shift in the momentum-isospin correspondence in a ladder model means changing the Hamiltonian matrix $M(k)$ into $M(k-\delta)$. Therefore, the two functions $\rho(k)$ and $\varphi(k)$ are transformed into $\rho(k-\delta)$ and $\varphi(k-\delta)$. 

By looking at the edge states in the momentum representation, Eq.~(\ref{eq:EdgeModeGeneralChiralOperatorL}) and Eq.~(\ref{eq:EdgeModeGeneralChiralOperatorR}), we can easily deduce that their momentum density distributions will be also shifted. That is:
\begin{equation}
\langle\hat{n}_{k}\rangle_{\delta}=\langle\hat{n}_{k-\delta}\rangle_{\delta=0}.
\end{equation}
In Fig.~\ref{fig:FiguraLadder23} we show the relation between the edge states of the three types of BDI ladder models and the edge states of the three types of AIII ladder models.

In addition, we show in Fig.~\ref{fig:FiguraLadder24} how the edge states momentum distribution depends on the value of an effective magnetic flux $\phi$. For $\phi=0$, the edge states show two equally weighted peaks in their momentum density distribution located at opposite momenta. This configuration corresponds to the balanced BDI model. For a non zero effective magnetic flux, the relative weight of each momentum component changes, as well as their location in momentum space. In this case the system realizes the imbalanced AIII model. For $\phi=\pm\pi$, the distance between the two momentum peaks is exactly $\pi$, and therefore there model could be brought to the BDI class by introducing a phase $\delta=\pm\pi/2$ that would shift the momentum peaks to $q_{1}=0$ and $q_{2}=\pi$. That situation would correspond to the imbalance BDI model.

\newpage
\section{Ladder geometries}

Here we present all ladder geometries that realize a model for a topological insulator. For that, we first consider the most general ladder model and impose the time reversal and chiral symmetry conditions. After that, we obtain all possible parameter configurations that correspond to a topological model. Each of them will lead to a different ladder geometry. Finally, we analyse a couple of ladder models that we consider to be more relevant. 

\subsection{General ladder and symmetry conditions}

\begin{figure}[t]
  \centering
    \includegraphics[width=0.485\textwidth]{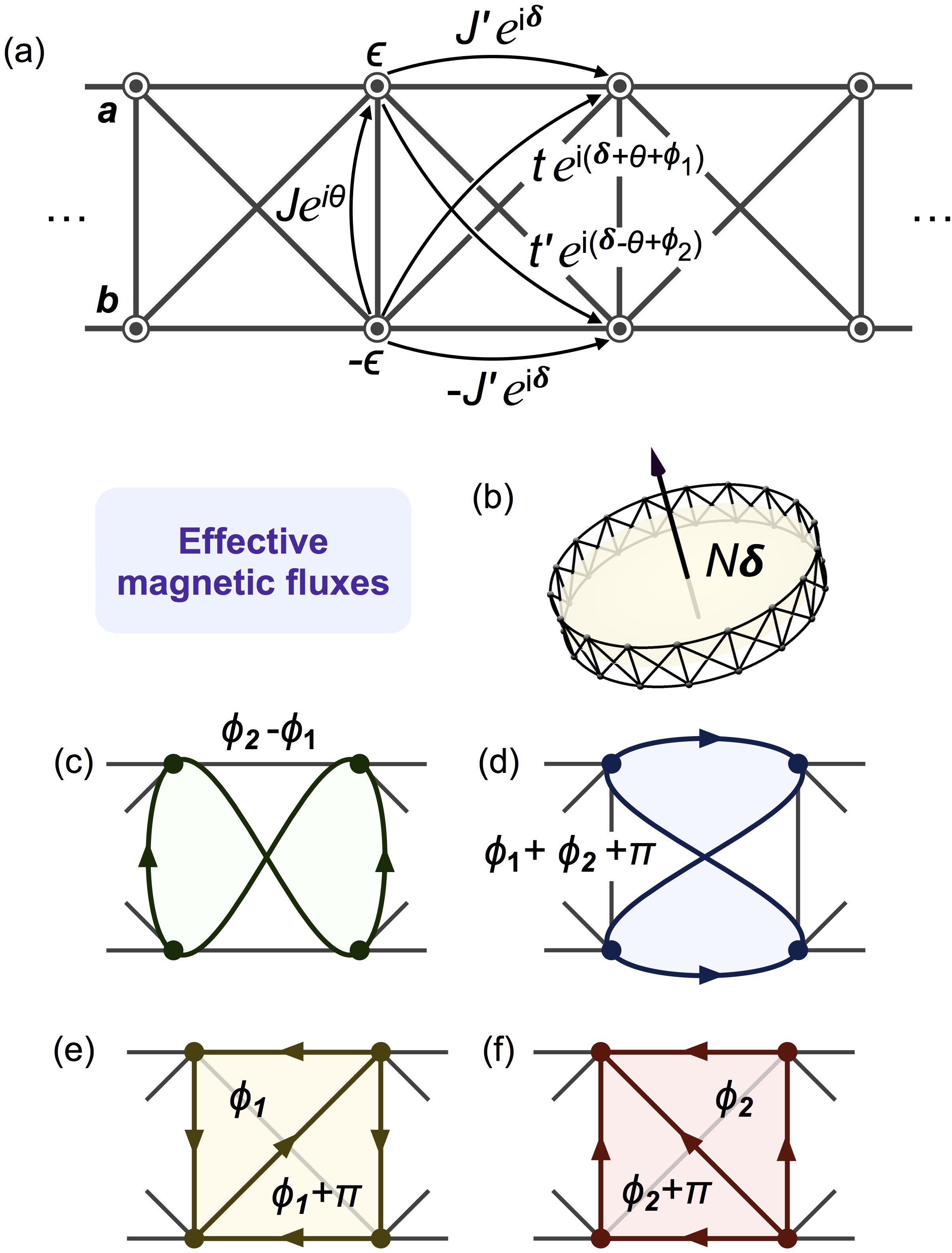}
    \caption[General topological ladder model.]{\textbf{General topological ladder model.} (a) Schematic illustration of the most general ladder model that can exhibit chiral symmetry. The phases of the tunneling amplitudes are parametrized, without loss of generality, so that they have a direct physical interpretation. The phase $\theta$ produces a general rotation on the Hamiltonian matrix and is irrelevant. The phase $\delta$ changes the correspondence between the momentum and the isospin associated to each eigenstate, as well as leads to a phase $N\delta$ hat particles pick up every time they complete a closed path along the whole system when there are periodic boundary conditions (b). Finally, the two phases $\phi_{1}$ and $\phi_{2}$ are related to the different effective magnetic fluxes that the ladder can exhibit (c), (d), (e) and (f).}
    \label{fig:LadderFluxes} 
\end{figure}

We consider the most general ladder geometry, which contains two horizontal tunneling amplitudes, one for each leg of the ladder, two diagonal tunneling amplitudes, one vertical tunneling amplitude that connects the two sites in the same lattice cell, and two on-site energy terms. All these parameters are contained in the matrices $C$ and $T$, being the most general ladder Hamiltonian written in terms of these matrices as in Eq.~(\ref{eq:GeneralLadderHamiltonian}). Our purpose is to obtain all different ladder architectures that realize a model for a topological insulator. Therefore, we look for ladder geometries that fulfil the two chiral symmetry conditions: Eq.~(\ref{eq:ChiralCondition1}) and Eq.~(\ref{eq:ChiralCondition2}).
According to the first one, the Hamiltonian matrix must have no component proportional to the identity. This implies that the two on-site energy terms and the two horizontal couplings, one for each leg of the ladder, must be opposite to each other.
Therefore, the matrices $C$ and $T$ for the most general ladder model that satisfies the first condition for chiral symmetry are:
\begin{equation}
C=\begin{pmatrix}
\epsilon & Je^{i\theta}\\
\\
Je^{-i\theta} & -\epsilon
\end{pmatrix}
,\quad
T=e^{i\delta}\begin{pmatrix}
J^{\prime} & te^{i(\phi_1+\theta)}\\
\\
t^{\prime}e^{i(\phi_{2}-\theta)} & -J^{\prime}
\end{pmatrix}.
\end{equation}
Where $\epsilon$ is a real parameter that corresponds to the on-site energies, $J$, $J^{\prime}$, $t$ and $t^{\prime}$ are real parameters that correspond to the vertical, horizontal and diagonal coupling amplitudes and $\theta$, $\delta$, $\phi_{1}$ and $\phi_{2}$ are real parameters that correspond to the four independent phases in the coupling terms [see Fig.~\ref{fig:LadderFluxes}(a)].

We have chosen a parametrization of the phases which, without loss of generality, is convenient in order to make a physical interpretation of them, as we show in the following. 

The Hamiltonian can be written in momentum representation as:
\begin{equation}\label{eq:HamiltonianoFaseTheta}
H=-\sum_{k}\psi^{\dagger}_{k}\,R_z(\theta)\,M(k)\,R_z^{\dagger}(\theta)\,\psi^{}_{k},
\end{equation}
where $R_z(\theta)=e^{i\theta\sigma_z/2}$ is a rotation of an angle $\theta$ around the $z$-axis and:
\begin{equation}\label{eq:GeneralLadderMatrixM}
M(k)=\begin{pmatrix}
\epsilon+2J\cos(k-\delta)& z^{*}(k-\delta)\\
\\
z(k-\delta) & -\epsilon-2J\cos(k-\delta)
\end{pmatrix},
\end{equation}
\noindent being $z(k)=J+te^{i(k-\phi_1)}+t^{\prime}e^{-i(k-\phi_2)}$.
From the form of this Hamiltonian matrix we can distinguish three kind of phases in the ladder parameters:
\begin{itemize}
\item
\textit{Effective magnetic fluxes $\phi_{1}$ and $\phi_{2}$}

The two parameters $\phi_1$ and $\phi_2$ determine the total phase accumulated by a particle completing a closed path in the ladder and, thus, they can be identified as effective magnetic fluxes penetrating the ladder. All closed paths that can be defined in the ladder are obtained as a combination of four elementary paths [see Fig.~\ref{fig:LadderFluxes}(c), (d), (e) and (f)]. These phases $\phi_1$ and $\phi_2$ affect the symmetries of the Hamiltonian and the properties of the edge states.\\

\item
\textit{Shift $\delta$ in the momentum-isospin correspondence.}

The phase $\delta$ produces a shift in the Hamiltonian matrix with respect to the momentum, Eq.~(\ref{eq:GeneralLadderMatrixM}), so that it introduces a shift in the correspondence between the momentum of each eigenstate and its associated vector in the Bloch sphere, which makes them be genuinely different from those eigenstates corresponding to the case in which there is no phase $\delta$. This phase is a generalization of the the phase $\delta$ that appears in the bowtie ladder.

\item
\textit{Irrelevant phase $\theta$.}

The phase $\theta$ produces a rotation around the $z$-axis, Eq.~(\ref{eq:HamiltonianoFaseTheta}), which is a global unitary operation and does not affect the symmetries of the system. Therefore we neglect this phase for simplicity and each model we study includes all possible rotations around the $z$-axis by adding the phase $\theta$ as shown in Fig.~\ref{fig:LadderFluxes}(a).
\end{itemize}

This analysis of the phases present in the most general topological ladder Hamiltonian is analogous to the one we did for the bowtie ladder in Chapter $8$: one phase can be removed by applying a global unitary transformation and, thus, it can be neglected; another one produces a shift in the isospin-quasimomentum relation; and the rest correspond to effective magnetic fluxes penetrating the ladder.
In fact, the bowtie ladder parametrized as in Fig.~\ref{fig:CanonicalLadder} corresponds to the particular case of the general ladder in Fig.~\ref{fig:LadderFluxes} in which $\epsilon=0$, $J^{\prime}=0$, $\theta=-\phi/2$, $\phi_{1}=\phi/2$ and $\phi_{2}=-\phi/2$.

Our purpose is to find all parameter configurations that correspond to a topological model in the BDI class or in the AIII class. For that, we decompose the Hamiltonian matrix in Eq.~(\ref{eq:GeneralLadderMatrixM}) as $M(k)=\left(\bm{n}_{0}+\bm{n}_{c}\cos k+\bm{n}_{s}\sin k \right)\cdot\bm{\sigma}$, being:
\begin{align}
&\bm{n}_{0}=J\,\hat{x}+\epsilon\,\hat{z}\nonumber\\
&\bm{n}_{c}=(t\,c_1+t^{\prime}c_2)\,\hat{x}-(t\,s_1-t^{\prime}s_2)\,\hat{y}+2J^{\prime}\cos\delta\,\hat{z}\nonumber\\
&\bm{n}_{s}=(t\,s_1+t^{\prime}s_2)\,\hat{x}+(t\,c_1-t^{\prime}c_2)\,\hat{y}+2J^{\prime}\sin\delta\,\hat{z},
\end{align}
\noindent where $c_i=\cos(\phi_i+\delta)$ and $s_i=\sin(\phi_i+\delta)$. In order for the Hamiltonian to be chiral symmetric, the three vectors $\bm{n}_{0}$, $\bm{n}_{c}$ and $\bm{n}_{s}$ must be linear dependent, Eq.~(\ref{eq:ChiralCondition2}), which means that the determinant of the matrix formed by their components must be zero. Imposing this requirement we obtain the condition for chiral symmetry in terms of the ladder parameters:
\begin{align}\label{eq:ChiralConditionParameters}
\epsilon\left( t^{2}-t^{\prime2} \right)-2JJ^{\prime}\left( t\cos\phi_{1}-t^{\prime}\cos\phi_{2} \right)=0.
\end{align}
The conditions for time reversal symmetry are $\bm{n}_{c}\cdot\bm{n}_{s}=0$ and $\bm{n}_{0}\cdot\bm{n}_{s}=0$, Eq.~(\ref{eq:GeneralTRCondition1}) and Eq.~(\ref{eq:GeneralTRCondition2}). Imposing them we obtain the conditions that the parameters of the model must fulfil for it to have time reversal symmetry:
\begin{align}
&2J^\prime\epsilon\sin\delta+Jt\sin(\delta+\phi_1)+Jt^{\prime}\sin(\delta+\phi_2)=0 \label{eq:TRConditionParameters1}\\
&2tt^{\prime}\sin(2\delta+\phi_1+\phi_2)+2J^{\prime2}\sin2\delta=0. \label{eq:TRConditionParameters2}
\end{align}

\newpage
\subsection{Finding all topological ladder configurations}

\subsubsection{Imposing chiral symetry}

In order to obtain all ladder geometries realizing a model for a topological insulator we first remark on two ingredients which are needed for a ladder geometry to be chiral symmetric:
\begin{enumerate}
\item
\textit{Opposite horizontal couplings.}

The chiral symmetry implies that the Hamiltonian matrix has no component proportional to the identity. As a consequence, the two horizontal tunneling amplitudes must have the same modulus and opposite sing. That is, either both terms appear in the ladder configuration or none of them does.
\item
\textit{Non-vanishing diagonal couplings.}

If a model has no diagonal couplings, that is $t=t^{\prime}=0$, the condition for chiral symmetry (\ref{eq:ChiralConditionParameters}) is fulfilled independently of the rest of the parameters. However, such a model will always be in a trivial phase.
To see this we consider the most general ladder model with no diagonal couplings, see Fig.~\ref{fig:SquareLadder}, whose corresponding Hamiltonian matrix is:
$M(k)=\left[\epsilon+2J^{\prime}\cos(k-\delta)\right]\,\sigma_{z}+J\,\sigma_{x}$.
There is no component with $\sin k$ and therefore the curve described in the complex plane is a line, whose winding number is 0.
\begin{figure}[h]
\centering
    \includegraphics[width=0.3\textwidth]{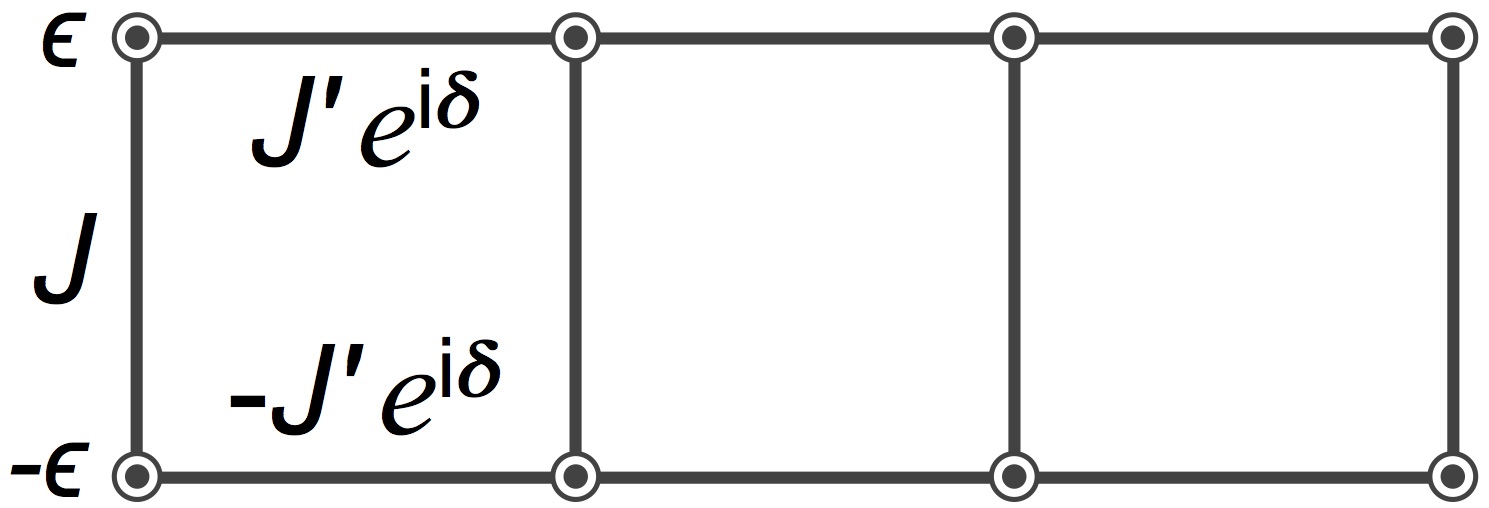}
   \caption[Square ladder]{\textbf{Square ladder.} Schematic illustration of the most general ladder model with no diagonal couplings. Despite of being chiral symmetric, it is always in a trivial phase.}
    \label{fig:SquareLadder} 
\end{figure}
\end{enumerate}
In this way, when looking for topological ladder models we have to discard all configurations with only one horizontal coupling and consider only those with at least one diagonal coupling. To find them we first consider those models with just one diagonal coupling and secondly those with two.\\

\noindent \textit{\textbf{i) Models with one diagonal coupling.}}

\begin{itemize}
\item \textit{Models with $J=0$ or $J^{\prime}=0$.}

First we consider ladder geometries with just two couplings, being one of them diagonal. In this situation the chiral symmetry condition, Eq.~(\ref{eq:ChiralConditionParameters}), implies that $\epsilon=0$, so that there are no on-site energy terms in the Hamiltonian. This case includes four ladder geometries (see Fig.~\ref{fig:1DiagonalTermLadders1}). 
\begin{figure}[t]
\centering
    \includegraphics[width=0.485\textwidth]{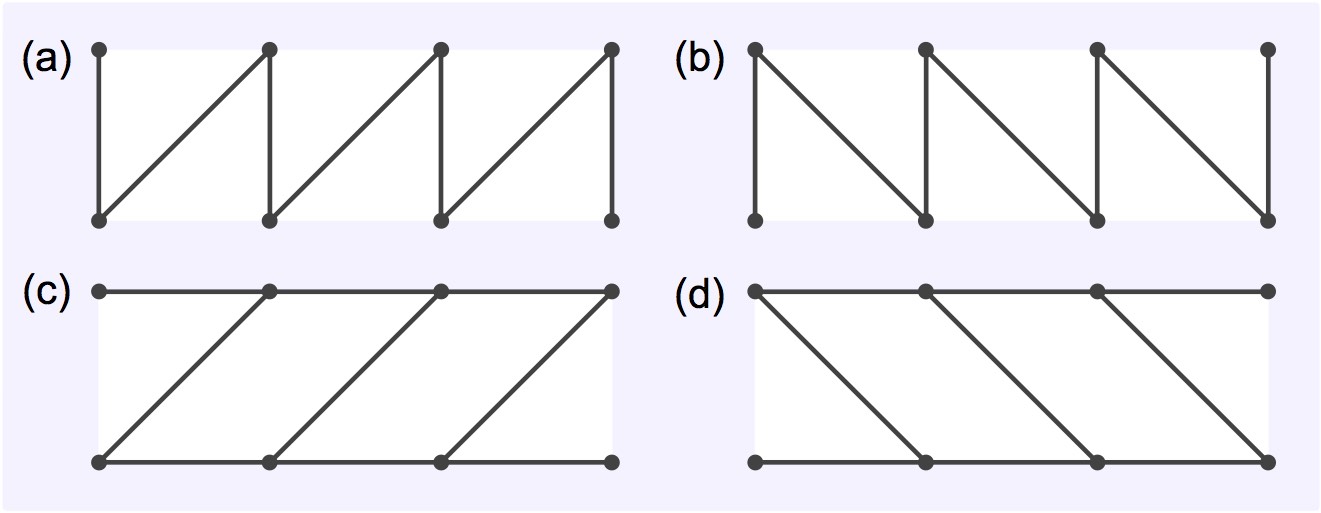}
    \caption[Ladder geometries with chiral symmetry (i).]{\textbf{Ladder geometries with chiral symmetry (i).} Topological ladder geometries with one diagonal coupling and no on-site energy.}
     \label{fig:1DiagonalTermLadders1} 
  \end{figure}
\item \textit{Models with vertical and horizontal couplings $J$ and $J^{\prime}$.}

In this situation the condition for chiral symmetry, Eq.~(\ref{eq:ChiralConditionParameters}), constitutes a constraint to the on-site energy $\epsilon=(2JJ^{\prime}/t)\cos\phi$,
where $t$ is the diagonal coupling in the model and $\phi$ can be $\phi_{1}$ or $\phi_{2}$. This case includes two ladder geometries (see Fig.~\ref{fig:1DiagonalTermLadders2}).
\begin{figure}[b]
  \centering
    \includegraphics[width=0.485\textwidth]{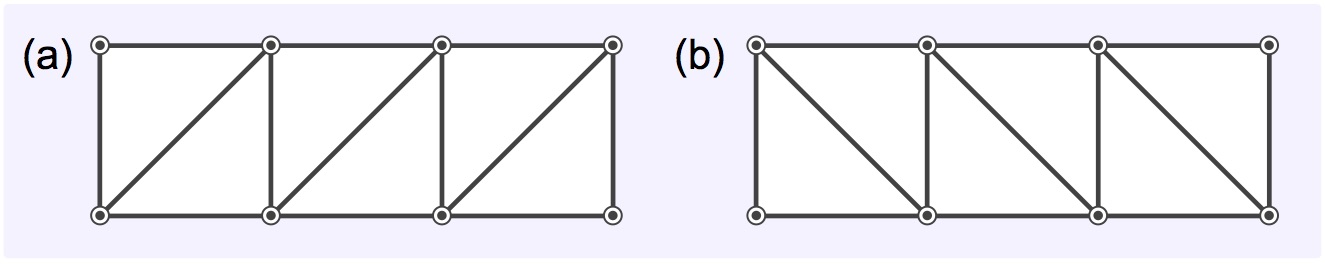}
    \caption[Ladder geometries with chiral symmetry (ii).]{\textbf{Ladder geometries with chiral symmetry (ii).} Topological ladder geometries with one diagonal coupling and constrained on-site energy.}
    \label{fig:1DiagonalTermLadders2} 
\end{figure}
\end{itemize}

\noindent \textit{\textbf{ii) Models with two diagonal couplings.}}

\begin{itemize}
\item \textit{Models with $J=0$ or $J^{\prime}=0$.}

In this case the second term in the chiral condition, Eq.~(\ref{eq:ChiralConditionParameters}), vanishes and therefore $\epsilon(t^{2}-t^{\prime 2}) =0$, so that $\epsilon=0$ or $t=t^{\prime}$.
Then, there are two configurations for each ladder geometry in this situation: one with no on-site energy and two independent diagonal couplings and another one with on-site energy and two diagonal couplings of the same amplitude. 
There are three ladder geometries of this kind. One with just the two diagonal couplings [Fig.~\ref{fig:2DiagonalTermLadders1}(a)], another having, in addition, the vertical tunneling $J$ [Fig.~\ref{fig:2DiagonalTermLadders1}(b)], and finally the one with the horizontal coupling $J^{\prime}$ [Fig.~\ref{fig:2DiagonalTermLadders1}(c) and (d)]. The first two correspond to a trivial model when the two diagonal couplings are the same, which means that the only possibility is having $\epsilon=0$ and $t\neq t^{\prime}$. The last case, in contrast, corresponds to a topological model in both situations. In this way we obtain four different ladder models.
\begin{figure}[t]
  \centering
    \includegraphics[width=0.485\textwidth]{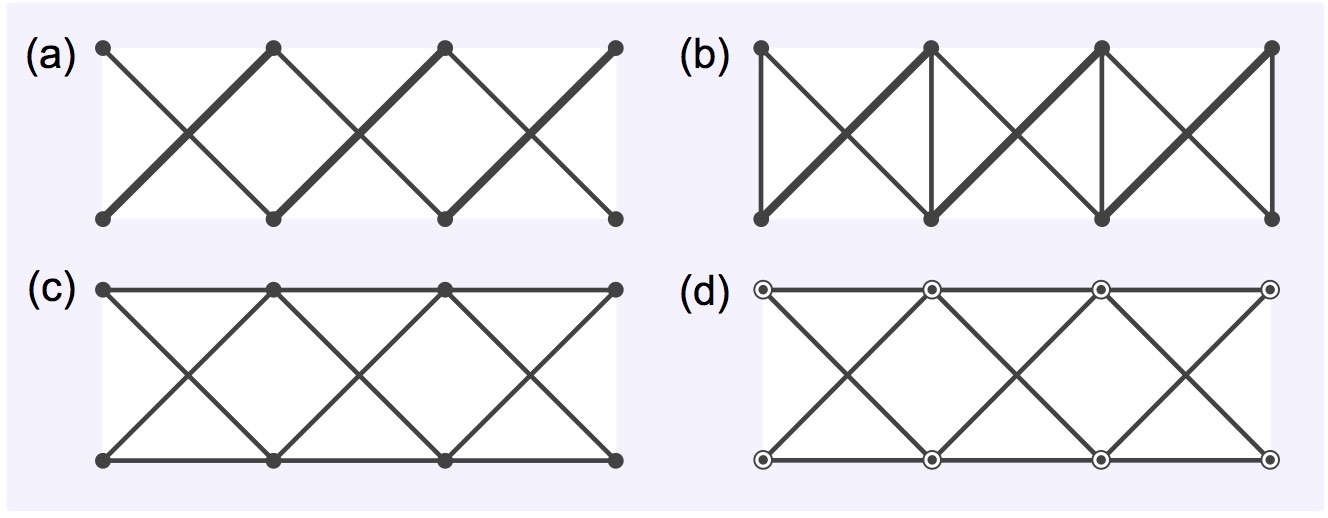}
    \caption[Ladder geometries with chiral symmetry (iii).]{\textbf{Ladder geometries with chiral symmetry (iii).} Topological ladder geometries with two diagonal terms and $J=0$ or $J^{\prime}=0$.}
    \label{fig:2DiagonalTermLadders1} 
\end{figure}

\item \textit{Models with vertical and horizontal couplings $J$ and $J^{\prime}$.}

In this case we can distinguish two situations:
\begin{enumerate}
\item
If the two diagonal couplings are the same, $t=t^{\prime}$, the first term in the chiral condition, Eq.~(\ref{eq:ChiralConditionParameters}), vanishes and, thus, the on-site energy can take any value. The chiral condition is then fulfilled if $\cos\phi_{1}=\cos\phi_{2}$. This gives us two models, one with $\phi_{1}=\phi_{2}$ and another with $\phi_{1}=-\phi_{2}$. However, the second possibility corresponds to a trivial topology so there is only one model with this configuration [Fig.~\ref{fig:2DiagonalTermLadders2}(a)].
\item
If the two diagonal couplings are different, $t\neq t^{\prime}$, the chiral condition implies a constraint to the on-site energy $\epsilon=2JJ^{\prime}\left(t\cos\phi_{1}-t^{\prime}\cos\phi_{2}\right)/(t^{2}-t^{\prime 2})$.
This corresponds to the last model [Fig.~\ref{fig:2DiagonalTermLadders2}(b)].
\begin{figure}[b]
  \centering
    \includegraphics[width=0.485\textwidth]{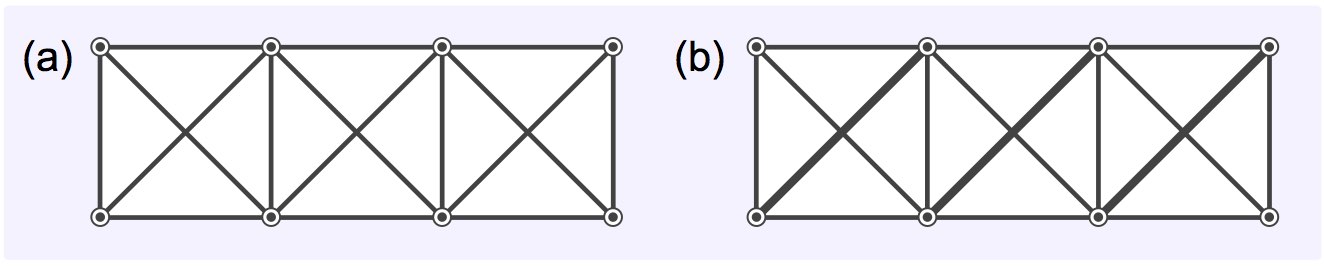}
    \caption[Ladder geometries with chiral symmetry (iv).]{\textbf{Ladder geometries with chiral symmetry (iv).} Topological ladder geometries with two diagonal terms plus both vertical and horizontal couplings.}
    \label{fig:2DiagonalTermLadders2} 
\end{figure}
\end{enumerate}
\end{itemize}

In this way we have obtained twelve topological ladder geometries, Fig.~\ref{fig:1DiagonalTermLadders1}, Fig.~\ref{fig:1DiagonalTermLadders2}, Fig.~\ref{fig:2DiagonalTermLadders1} and Fig.~\ref{fig:2DiagonalTermLadders2}. However, some of them correspond to a Hamiltonian matrix of the form:
$M(k)=\beta\cos k\,\sigma_{c}+\gamma\sin k\,\sigma_{s}$,
which means that they are always in a topological phase and have no phase transition, as the curve described by the Hamiltonian matrix is an ellipse centred at the origin whose winding number is always $1$. This special behaviour occurs because they are the particular case in which $J=0$ of a more general geometry. Therefore, we do not need to consider them.
These ladder geometries are the ones in Fig.\ref{fig:1DiagonalTermLadders1}(c) and (d), Fig.~\ref{fig:2DiagonalTermLadders1}(a) and Fig.\ref{fig:2DiagonalTermLadders1}(c), which are particular cases of the models in Fig.\ref{fig:1DiagonalTermLadders2}(a) and (b), Fig.~\ref{fig:2DiagonalTermLadders1}(b) and Fig.\ref{fig:2DiagonalTermLadders2}(b), respectively. 
\begin{figure*}
  \centering
    \includegraphics[width=1\textwidth]{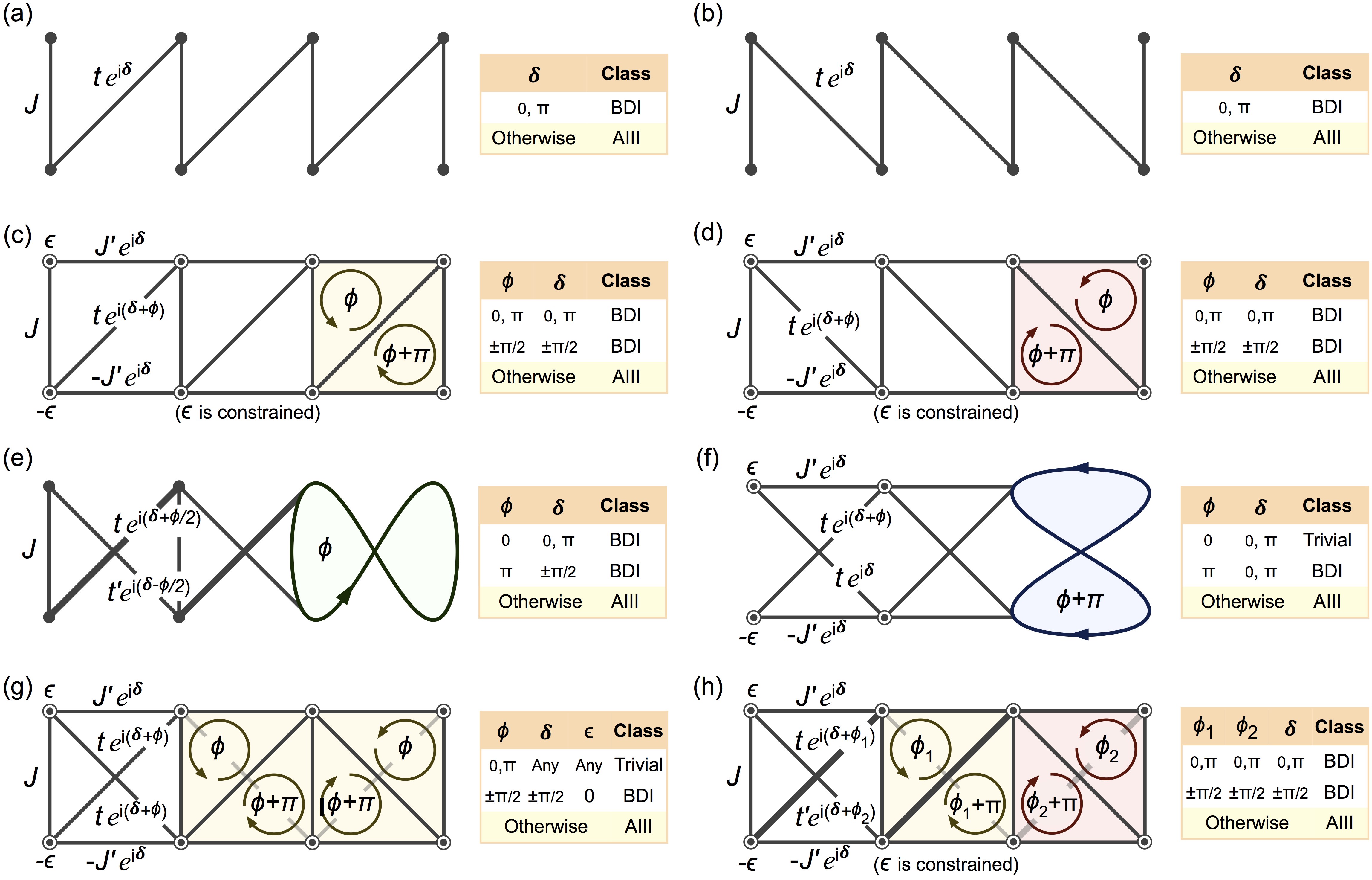}
    \caption[Topological ladder models.]{\textbf{Topological ladder models.} Here we show all topological ladder models. For each of them, we specify in a table which parameter configurations correspond to a model in the BDI symmetry class and which to a model in the AIII symmetry class. Ladder models (a) and (b) are in fact one-dimensional dimerized lattice models, which can only realize the SSH model and the circular AIII model. The rest of the ladder models preset in this figure can access all types of topological ladder models. In ladder models (e), the bowtie ladder, and (h), one of the diagonal couplings is represented with a thicker line. This indicates that the two diagonal tunneling amplitudes must have different values, otherwise the model would be topologically trivial. A double circle in every site indicates the presence of an on-site energy term in the Hamiltonian. The chiral symmetry condition imposes a constraint to the on-site energy in ladder models (c) and (d), for which $\epsilon=(2JJ^{\prime}/t)\cos\phi$, and in model (h), for which $\epsilon=2JJ^{\prime}(t\cos\phi_{1}-t^{\prime}\cos\phi_{2})/(t^2-t^{\prime 2})$.}
    \label{fig:FiguraLadder31} 
\end{figure*}

\subsubsection{Imposing time reversal symmetry}

In order to find all ladder models in the BDI class and in the AIII class we need to impose the two time reversal symmetry conditions, Eq.~(\ref{eq:TRConditionParameters1}) and Eq.~(\ref{eq:TRConditionParameters2}), to the topological ladder geometries we have obtained beforehand.
When they are fulfilled we have a realization of the BDI class, whereas it corresponds to the AIII class otherwise.
By inspecting those conditions we see that they are trivially fulfilled if all phases vanish, which is obvious as this case corresponds to a real Hamiltonian. Nevertheless, there are other solutions to the time reversal symmetry conditions that correspond to complex Hamiltonians. In order to obtain them, we have to evaluate the time reversal symmetry conditions for each particular geometry and find the values of $\delta$, $\phi_{1}$ and $\phi_{2}$ for which they are satisfied.

As an example, we consider the ladder geometry in Fig.~\ref{fig:2DiagonalTermLadders1}(b). The two time reversal symmetry conditions for this particular case are:
\begin{align}
&J(t+t^{\prime})\sin\delta\cos\frac{\phi}{2}+J(t-t^{\prime})\cos\delta\sin\frac{\phi}{2}=0\label{eq:TRConditionModeloParticular1}\\
&4tt^{\prime}\sin\delta\cos\delta=0,\label{eq:TRConditionModeloParticular2}
\end{align}
being $\phi$ the only effective magnetic flux in this ladder. The solutions to these equations are: $\phi=0,\,\delta=0,\pi$ or $\phi=\pi,\,\delta=\pm\pi/2$,
which correspond to four different BDI models with this particular ladder geometry [see Fig.~\ref{fig:FiguraLadder31}(c)].

Analogously, the time reversal symmetry conditions can be easily analysed for all other ladder geometries. As a result, we present in Fig.~\ref{fig:FiguraLadder31} all ladder models with chiral symmetry and detail the parameter configurations for which they belong to the BDI class and to the AIII class. As we see, all parameter configurations that correspond to a model in the BDI class come in pairs, corresponding the two configurations within each pair to two different values of the phase $\delta$ that differ in $\pi$. The reason for that is that any BDI model can be generalized by introducing a shift in the momentum-isospin correspondece, which makes the model break time reversal symmetry and enter the AIII class. Only when such shift is equal to $0$ or $\pi$ the model is still in the BDI class. We explain this in detail in Sec.~$V$ for the bowtie ladder Hamiltonian, however, the analysis can be applied to any ladder model.\\

Among the eight different topological ladder models we present in Fig.~\ref{fig:FiguraLadder31}, the first two, (a) and (b), consist of a one-dimensional dimerized lattice. The third ladder model, (e), is actually the bowtie ladder, which we have studied in Sec.~$V$. Furthermore, we know that it serves as a canonical ladder model, as the bowtie ladder Hamiltonian is the most general ladder Hamiltonian we can imagine up to a global unitary transformation. Therefore, all other ladder models in Fig.~\ref{fig:FiguraLadder31} can be obtain from the bowtie ladder by performing the appropriate rotation.

Nevertheless, we highlight and show in more detail two particular ladder geometries, (f) and (g), and indicate how to connect them to the canonical ladder.
Models (c), (d) and (h) have a constraint in their on-site energy $\epsilon$, which has to take a very particular value that depends on all other parameters, including the effective magnetic fluxes. The moment the on-site energy $\epsilon$ has not the appropriate value, chiral symmetry is broken and the model would not be even topological. This would imply a difficulty from an experimental point of view, and therefore we consider these particular ladder geometries of less interest.

\subsection{The hourglass ladder}

\begin{figure}[t]
  \centering
    \includegraphics[width=0.4\textwidth]{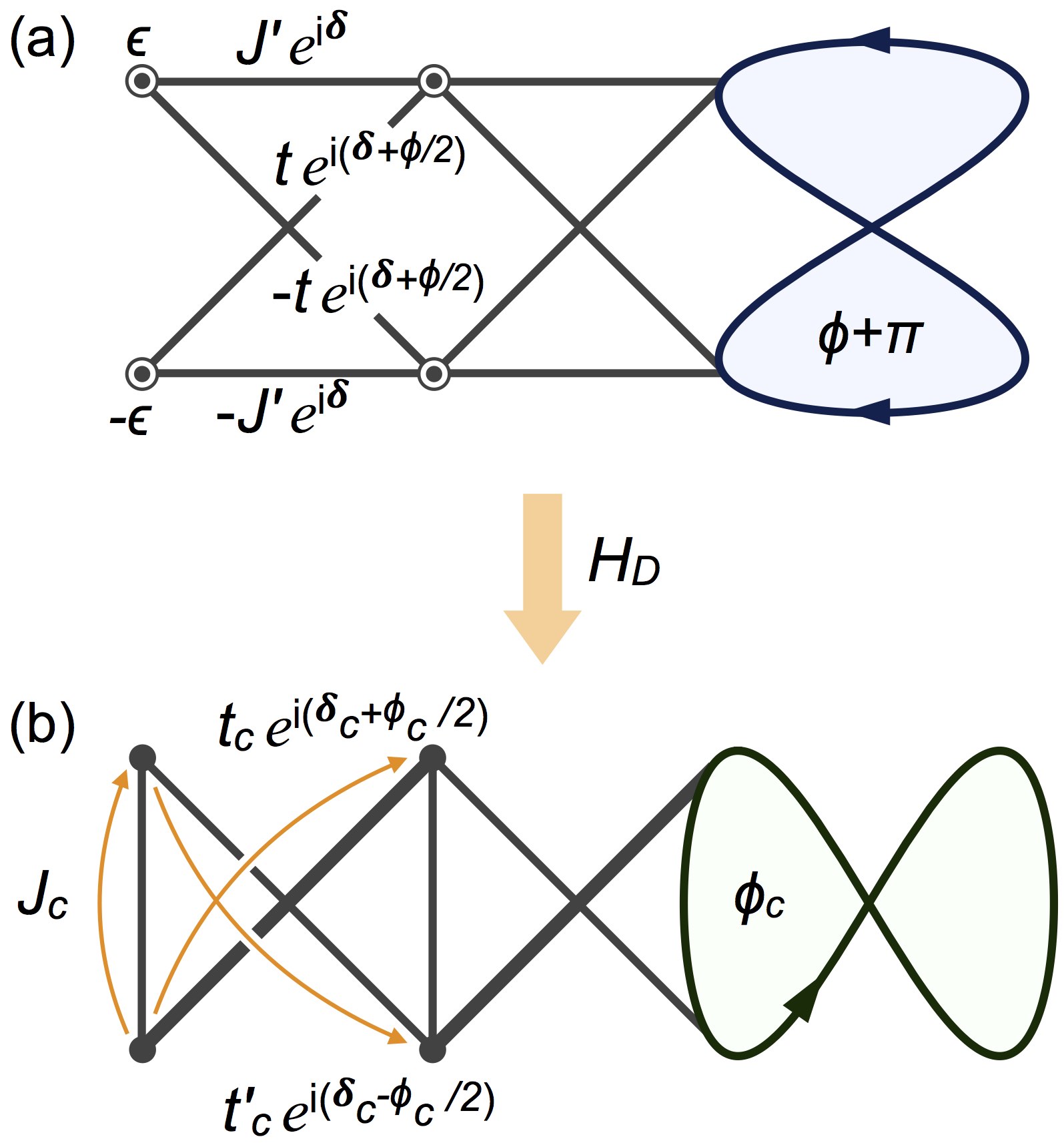}
    \caption[The hourglass ladder.]{\textbf{The hourglass ladder.} Schematic illustration of the hourglass ladder. Its Hamiltonian matrix lives on the $yz$-plane and is connected to the bowtie ladder Hamiltonian matrix by applying the unitary $H_{D}=(\sigma_{x}+\sigma_{z})/\sqrt{2}$.}
    \label{fig:LadderX} 
\end{figure}

As the first remarkable example of a topological ladder model different from the bowtie ladder, we consider ladder (a) in Fig.~\ref{fig:LadderX}, which we call the \textit{hourglass} ladder and whose Hamiltonian is:
\begin{align}
&H_{\text{hg}}=\nonumber\\
&-\sum_{n}\epsilon\,(\hat{a}^{\dagger}_{n}\hat{a}_{n}-\hat{b}^{\dagger}_{n}\hat{b}_{n})+\left[J^{\prime}e^{i\delta}\,(\hat{a}^{\dagger}_{n+1}\hat{a}_{n}-\hat{b}^{\dagger}_{n+1}\hat{b}_{n})+\right. \nonumber\\
&\left.t\,e^{i(\delta+\phi/2)}\,(\hat{a}^{\dagger}_{n+1}\hat{b}_{n}-\hat{b}^{\dagger}_{n+1}\hat{a}_{n})+\text{h.c.}\right].
\end{align}
By changing to the momentum representation we obtain the corresponding Hamiltonian matrix, which is:
\begin{equation}
M_{\text{hg}}(k)=\left[\epsilon+2J^{\prime}\cos(k-\delta)\right]\,\sigma_{z}+2t\sin(k-\delta-\phi/2)\,\sigma_{y}.
\end{equation}
As we can see, the Hamiltonian matrix lives on the $yz$-plane, so that it can be connected to the bowtie ladder Hamiltonian matrix by performing a rotation around the $x$ axis (see Fig.~\ref{fig:LadderX}). That is:
\begin{equation}
H_{D}\,M(k)\,H_{D}^{\dagger}=\begin{pmatrix}
0 & w^{*}(k)\\
w(k) & 0\end{pmatrix},
\end{equation}
where $H_{D}=(\sigma_{x}+\sigma_{z})/2$ is the Hadamard transformation and:
\begin{equation}
w(k)=\epsilon+2J^{\prime}\cos(k-\delta)-i\,2t\sin(k-\delta-\phi/2).
\end{equation}
Now that the Hamiltonian matrix has the canonical form, in which only the off-diagonal terms are different from zero, it can be directly related to the canonical ladder Hamiltonian matrix, which is:
\begin{equation}
M_{c}(k)=\begin{pmatrix}
0 & z^{*}(k)\\
z(k) & 0\end{pmatrix},
\end{equation}
with:
\begin{equation}\label{eq:ZCanonicalLadder}
z(k)=J_{c}+t_{c}e^{i(k-\delta_{c}-\phi_{c}/2)}+t_{c}^{\prime}e^{-i(k-\delta_{c}+\phi_{c}/2)}.
\end{equation}
Both models are identified by taking $w(k)=z(k)$, what establishes a relation between the parameters of the first model, $\epsilon$, $J^{\prime}$, $t$, $\phi$ and $\delta$, and the ones of the canonical ladder: $J_{c}$, $t_{c}$, $t_{c}^{\prime}$, $\phi_{c}$ and $\delta_{c}$.

\subsection{The box ladder}

\begin{figure}[t]
  \centering
    \includegraphics[width=0.4\textwidth]{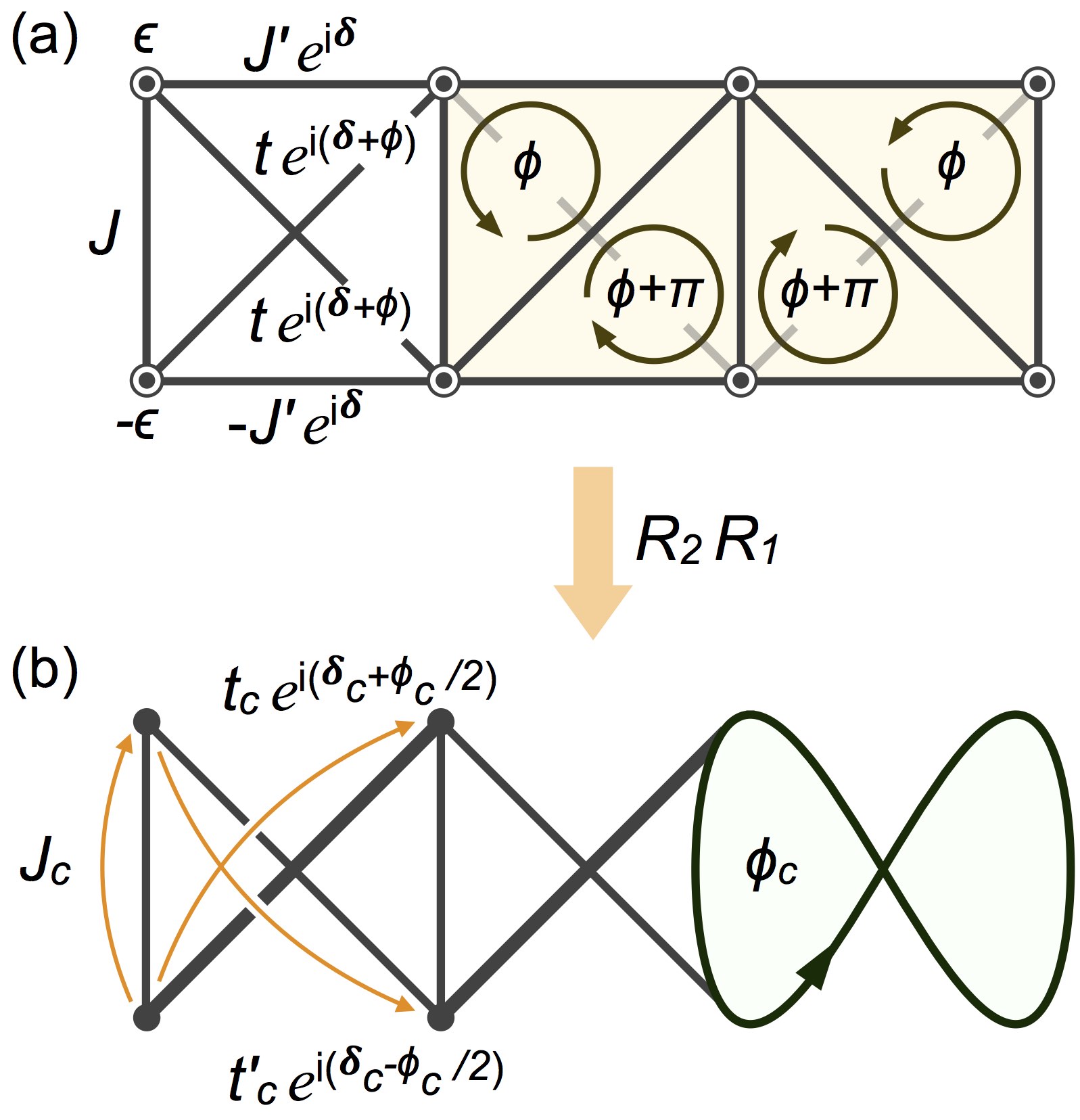}
    \caption[The box ladder model.]{\textbf{The box ladder model.} Schematic illustration of the box ladder. Its Hamiltonian matrix lives on the $xz$-plane and is connected to the canonical ladder Hamiltonian matrix by applying the unitary transformations $R_{1}$ and $R_{2}$ (see text).}
    \label{fig:LadderIX} 
\end{figure}

Another remarkable topological ladder model is ladder (a) in Fig.~\ref{fig:LadderIX} which we call the box ladder and whose Hamiltonian is:
\begin{align}
&H_{\text{box}}=\nonumber\\
&-\sum_{n}\epsilon\,(\hat{a}^{\dagger}_{n}\hat{a}_{n}-\hat{b}^{\dagger}_{n}\hat{b}_{n})+\left[J^{\prime}e^{i\delta}\,(\hat{a}^{\dagger}_{n+1}\hat{a}_{n}-\hat{b}^{\dagger}_{n+1}\hat{b}_{n})+\right.\nonumber\\
&\left.t\,e^{i(\delta+\phi)}\,(\hat{a}^{\dagger}_{n+1}\hat{b}_{n}+\hat{b}^{\dagger}_{n+1}\hat{a}_{n})+J\,\hat{a}^{\dagger}_{n}\hat{b}_{n}+\text{h.c.}\right].
\end{align}
The corresponding Hamiltonian matrix is:
\begin{align}
M_{\text{box}}(k)=&\left[\epsilon+2J^{\prime}\cos(k-\delta)\right]\,\sigma_{z}+\nonumber\\
&\left[J+2t\cos(k-\delta-\phi)\right]\,\sigma_{x}.
\end{align}
It is on the $xz$-plane and is connected to the canonical ladder Hamiltonian matrix by applying two combined rotations, which are: $R_{1}=e^{i\pi(\sigma_{x}+\sigma_{y}+\sigma_{z})/3\sqrt{3}}$ and $R_{2}=e^{i\Theta\sigma_{z}/2}$, where $\Theta=\arctan(J/\epsilon)$. In this way:
\begin{equation}
R_{2}R_{1}\,M_{\text{box}}(k)\,R_{1}^{\dagger}R_{2}^{\dagger}=\begin{pmatrix}
0 & v^{*}(k)\\
v(k) & 0\end{pmatrix},
\end{equation}
being:
\begin{align}
v(k)=&\sqrt{\epsilon^{2}+J^{2}}+2J^{\prime}\cos(k-\delta)e^{-i\Theta}+\nonumber\\
&i2t\sin(k-\delta-\phi)e^{-i\Theta}.
\end{align}
The condition $v(k)=z(k)$ [see Eq.~(\ref{eq:ZCanonicalLadder})] establishes the relation between the parameters of the box ladder model and the ones of the canonical ladder.

There are three ladder models studied in previous works which are particular cases of this ladder configuration. The first one is the Creutz ladder \cite{Creutz1999}, which corresponds to ladder (a) in Fig.~\ref{fig:LadderIX} for $\epsilon=0$, $\delta=-\pi/2$ and $\phi=\pi/2$. The second one is the imbalance Creutz ladder \cite{Juenemann2017}, which corresponds to ladder (a) in Fig.~\ref{fig:LadderIX} for $J=0$, $\delta=\pi/2$, $\phi=\pi/2$ and $J^{\prime}=t$. Finally, the third one is the shifted Creutz ladder \cite{Gholizadeh2018}, which corresponds to ladder (a) in Fig.~\ref{fig:LadderIX} for $\epsilon=0$, $\delta=0$ and $\phi=-\pi/2$.\\

This research was funded by the Deutsche Forschungsgemeinschaft (DFG, German Research Foundation) via Research Unit FOR 2414 under project number 277974659.

%
%

%
%
%
%
\appendix
\section{Chiral symmetry and time reversal conditions}

In order to prove the conditions for chiral symmetry, \ref{eq:ChiralCondition1} and  \ref{eq:ChiralCondition2}, and for $T$ symmetry,  \ref{eq:GeneralTRCondition1} and \ref{eq:GeneralTRCondition2}, we first show two useful results.
We consider a general $2\times2$ hermitian matrix $H$ and a genaral $2\times2$ unitary matrix $U$, which can be written as:
\begin{align}
&H=\alpha\mathbb{I}+\beta\,\bm{n}\cdot\bm{\sigma}\label{Hgeneral}\\
&U=e^{i\varphi}\left(\cos\frac{\theta}{2}\mathbb{I}+i\sin\frac{\theta}{2}\bm{m}\cdot\bm{\sigma}\right),
\end{align}
where $\bm{n}$ and $\bm{m}$ are unit vectors in $\mathbb{R}^3$, $\alpha$ and $\beta$ are two real numbers, $\theta$ an angle and $\varphi$ a global phase.
We compute the transformation of $H$ under $U$ and get:
\begin{equation}
UHU^{\dagger}=\alpha\mathbb{I}+\beta\,\bm{n'}\cdot\bm{\sigma},\label{eq:HTransformadoPorU}
\end{equation}
with
$\bm{n'}=\cos\theta\,\bm{n}-\sin\theta\,\bm{m}\times\bm{n}+2\sin^{2}\frac{\theta}{2}(\bm{m}\cdot\bm{n})\,\bm{m}$.
We are interested in those unitary transformations that leave $H$ invariant and those that produce a change in its sign. In the second case we want that $UHU^{\dagger}=-H$. Comparing \ref{Hgeneral} and \ref{eq:HTransformadoPorU} we see that this implies: 
$\alpha=0$ and $\bm{n'}=-\bm{n}$.
If $\bm{m}$ is codirectional to $\bm{n}$, $\bm{n'}=\bm{n}$ and there is no solution. Otherwise the set $\set{\bm{n},\bm{m},\bm{m}\times\bm{n}}$ constitutes a basis of $\mathbb{R}^3$ and we can get the solution by comparing coefficients, getting three equations:
\begin{align}
&\cos\theta=-1\\
&2\sin^{2}\frac{\theta}{2}\,\bm{m}\cdot\bm{n}=0\\
&\sin\theta=0.
\end{align}
The only solution to them is $\theta=\pi$ and $\bm{m}\cdot\bm{n}=0$, so we get the following result.
\textit{Result 1}: given a $2\times2$ hermitian matrix $H$, there exists a unitary transformation $U$ such that $UHU^{\dagger}=-H$ if and only if $H=\beta\,\bm{n}\cdot\bm{\sigma}$. Moreover, the unitary transformation $U$ has the form $U=e^{i\varphi}\,\bm{m}\cdot\bm{\sigma}$ for any $\varphi$ and being $\bm{m}$ such that $\bm{m}\cdot\bm{n}=0$.

On the other hand, for $H$ to be invariant under $U$ we need that $\mathbf{n'}=\mathbf{n}$, which is the case if $\bm{m}$ is codirectional to $\bm{n}$. Otherwise we have that:
\begin{align}
&\cos\theta=1\\
&2\sin^{2}\frac{\theta}{2}\,\mathbf{m}\cdot\mathbf{n}=0\\
&\sin\theta=0,
\end{align}
whose only solution is $U=e^{i\varphi}\mathbb{I}$. As a consequence we obtain the second result.
\textit{Result 2}: given a $2\times2$ hermitian matrix $H=\alpha\mathbb{I}+\beta\,\bm{n}\cdot\bm{\sigma}$, the only unitary transformations $U$ such that $UHU^{\dagger}=H$ are those with the form $U=e^{i\varphi}\left(\cos\frac{\theta}{2}\mathbb{I}+i\sin\frac{\theta}{2}\bm{n}\cdot\bm{\sigma}\right)$.\\

Now we can easily derive the chiral conditions \ref{eq:ChiralCondition1} and \ref{eq:ChiralCondition2} and the $T$ conditions \ref{eq:GeneralTRCondition1} and \ref{eq:GeneralTRCondition2}. The Hamiltonian matrix of a general ladder model is:
\begin{equation}
M(k)=\lambda(k)\mathbb{I}+(\bm{n}_{0}+\bm{n}_{c}\cos k+\bm{n}_{s}\sin k)\cdot\bm{\sigma},
\end{equation}
Using \textit{result 1} we see that the Hamiltonian matrix fulfils the chiral condition (\ref{eq:ConditionS}) only if $\lambda(k)=0$ and there exists a vector $\bm{a}$ such that $\bm{a}\cdot(\bm{n}_{0}+\bm{n}_{c}\cos k+\bm{n}_{s}\sin k)=0\quad\forall k$.
Therefore, the model will be chiral symmetric if the vectors $\bm{n}_{0}$, $\bm{n}_{c}$ and $\bm{n}_{s}$ lie in a common plane.

The model has $T$ symmetry (condition \ref{eq:ConditionT}) if there is a unitary transformation $U_{T}$ such that
$U_{T}\,M(-k)^{\dagger}U_{T}^{\dagger}=M(-k)$.
Using $\bm{\sigma}^{*}=-\sigma_{y}\bm{\sigma}\sigma_{y}$ and defining $\widetilde{U}_{T}=U_{T}\sigma_{y}$, this implies:
\begin{align}
&\widetilde{U}_{T}\left(\bm{n}_{0}\cdot\bm{\sigma}\right)\widetilde{U}_{T}^{\dagger}=-\bm{n}_{0}
\cdot\bm{\sigma}\\
&\widetilde{U}_{T}\left(\bm{n}_{c}\cdot\bm{\sigma}\right)\widetilde{U}_{T}^{\dagger}=-\bm{n}_{c}
\cdot\bm{\sigma}\\
&\widetilde{U}_{T}\left(\bm{n}_{s}\cdot\bm{\sigma}\right)\widetilde{U}_{T}^{\dagger}=\bm{n}_{s}
\cdot\bm{\sigma}.
\end{align}
For the first two equations to be fulfilled, using \textit{result 1}, the unitary transformation $\widetilde{U}_{T}$ must have the form $\widetilde{U}_{T}=\bm{b}\cdot\bm{\sigma}$ with $\bm{b}\cdot\bm{n}_{0}=\bm{b}\cdot\bm{n}_{c}=0$. 
However, from the last equation, using \textit{result 2}, $\bm{b}$ must be proportional to $\bm{n}_{s}$, therefore the condition for time reversal symmetry is fulfilled if:
$\bm{n}_{s}\cdot\bm{n}_{0}=\bm{n}_{s}\cdot\bm{n}_{c}=0$. 
In other words, the even and odd components of the Hamiltonian matrix must be perpendicular to each other for the system to exhibit time reversal symmetry.\\

\end{document}